%% file: bare_conf_post_submission2.tex
\documentclass[journal]{IEEEtran}
\usepackage[nolist]{acronym}
\usepackage{color}

\ifCLASSINFOpdf
   \usepackage[pdftex]{graphicx}
\else
   \usepackage[dvips]{graphicx}
\fi

\usepackage[cmex10]{amsmath}
\usepackage{url}
\usepackage{subfigure}
\begin{document}
\title{Exploring the Practical Limits of Cooperative Awareness in Vehicular Communications}

\author{Mate Boban and Pedro M. d'Orey%

\thanks{Copyright (c) 2016 IEEE. Personal use of this material is permitted. However, permission to use this material for any other purposes must be obtained from the IEEE by sending a request to pubs-permissions@ieee.org. 

The research leading to these results has received funding from the European Union's Seventh Framework Programme (FP7/2007-2013) under grant agreement n. 270410, DRIVE C2X. This work was partially supported by the Funda\c{c}\~{a}o para a Ci\^{e}ncia e a Tecnologia (FCT) under grant UID/EEA/50008/2013.
We would like to thank Francesco Alesiani for proofreading the paper and for providing valuable comments. 
M. Boban and P. M. d'Orey carried out this work while at NEC Laboratories Europe.
}
\thanks{M. Boban is with Huawei Technologies Duesseldorf GmbH, European Research Center, Riesstrasse 25, Munich, Germany (e-mail:mate.boban@huawei.com).

 P. M. d'Orey is with the Instituto de Telecomunica\c{c}\~{o}es, Universidade do Porto, Rua do Campo Alegre, 1021/1055, Porto, Portugal (e-mail: pedro.dorey@dcc.fc.up.pt).}}


\maketitle
\input{acronyms}

\begin{abstract}

We perform an extensive study of cooperative awareness in vehicular communication based on periodic message exchange. We start by analyzing measurements collected on four test sites across Europe. To measure cooperative awareness, we use three metrics: 1) neighborhood awareness ratio; 2) ratio of neighbors above range; and 3) packet delivery rate. Using the collected data, we define a simple model for calculating  neighborhood awareness given packet delivery ratio for a given environment. Finally, we perform realistic, large-scale simulations to explore the achievable performance of cooperative awareness   under realistic transmit power and transmit rate constraints. Our measurements and simulation results show that: i) above a certain threshold, there is little benefit in increasing  
cooperative message rate to improve the awareness; higher transmit power and fewer messages transmissions are a better approach, since message delivery is dominated by shadowing.  ii) the efficacy of cooperative awareness varies greatly in different environments on both large scale (e.g., 90\% awareness is achievable up to 200~m in urban and over 500~m in highway) and small scale (e.g., vehicles in nearby streets can have significantly different awareness); iii) V2V and V2I communication have distinct awareness  patterns; iv) each location has a distinct transmit power that achieves high awareness;  and v) achieving high awareness levels results in increased reception of potentially unwanted messages; therefore, a balance needs to be found between awareness and interference, depending on the specific context. We hope our results will serve as a starting point for designing more effective periodic message exchange services for cooperative awareness.
\end{abstract}



\section{Introduction} \label{sec:Introduction}
\input{intro}

\section{Related Work} \label{sec:RelWork}
\input{relatedWork}

\section{Measurement-based Evaluation Of Cooperative Awareness}\label{sec:Evaluation}
\input{experimental}

\subsection{Results}\label{sec:Results}
\input{results}

\section{Modeling the Relationship Between Cooperative Awareness and Packet Delivery Rate}\label{sec:Model}
\input{modeling}

\section{Large Scale Simulation of Cooperative Awareness} \label{sec:Simulation}
\input{simulation}

\section{Conclusions} \label{sec:Conclusions}
Periodic broadcast of single-hop cooperative messages is the basis for future cooperative ITS systems in the EU, US, Japan, and other markets. Through measurements and simulations, we analyzed the ability of periodic message exchange to enable cooperative awareness as well as their impact on channel load. First, we empirically evaluated the performance of cooperative awareness using measurements collected in four test sites in Europe within the scope of DRIVE-C2X project. The measurements were performed in three distinct environments (urban, suburban and highway) and between vehicles (\ac{V2V} communication) and vehicles and infrastructure (\ac{V2I} communication). Next, we developed a simple model to estimate cooperative awareness for an environment, provided that \ac{PDR} information is available. The model can be used to define upper and lower bound of achievable awareness and to get insight into the performance of cooperative awareness in a given environment. Finally, we performed large-scale simulations with thousands of vehicles in urban and highway environments to explore the limits of cooperative awareness, given the practical limitations in terms of transmit power and rate of periodic messages.

Our results demonstrate that cooperative awareness is strongly dependent on link quality and propagation conditions. The propagation environment where vehicles move determines the maximum achievable communication range and neighborhood awareness: the more complex the environment, the lower the awareness. With respect to the link type, the results show that the advantageous positions of RSUs improve the awareness levels for \ac{V2I} communications when compared with \ac{V2V} communications. Furthermore, higher effective transmit power can, while increasing awareness levels, also (prohibitively) increase the interference by far-away nodes; this effect is especially evident for \ac{V2V} communication in highway scenarios and \ac{V2I} communication in general. Furthermore, both measurements and simulations showed that increasing transmit power has a much more significant impact on awareness than transmit rate. In fact, irrespective of the environment, above a certain transmit rate per observed time period (upper-bounded by three to four messages), increasing the rate results in minimal improvement of awareness, while at the same time increasing the channel load.

With regards to the application performance, our results show that applications requiring high awareness levels (e.g., 90\%) up to 100~m can be satisfied in virtually all environments. For larger distances, high awareness is possible in certain types of environments (e.g., highway), whereas in others the awareness is limited by the harsh propagation environment and regulatory limits on transmit power level (this is the case above 200 m in typical urban environments). Furthermore, transitions between environments incur a significant difference in awareness; therefore, it is beneficial for applications to dynamically detect and adapt the parameters (e.g., transmit power) according to the current surroundings. We hope our results regarding the benefits and  practical limitations of cooperative awareness message exchange will help in the design of future safety and efficiency ITS protocols and applications.
 

\bibliographystyle{IEEEtran}
\bibliography{draftIII_tex}

\begin{IEEEbiography}[{\includegraphics[width=1in,height=1.25in,clip,keepaspectratio]{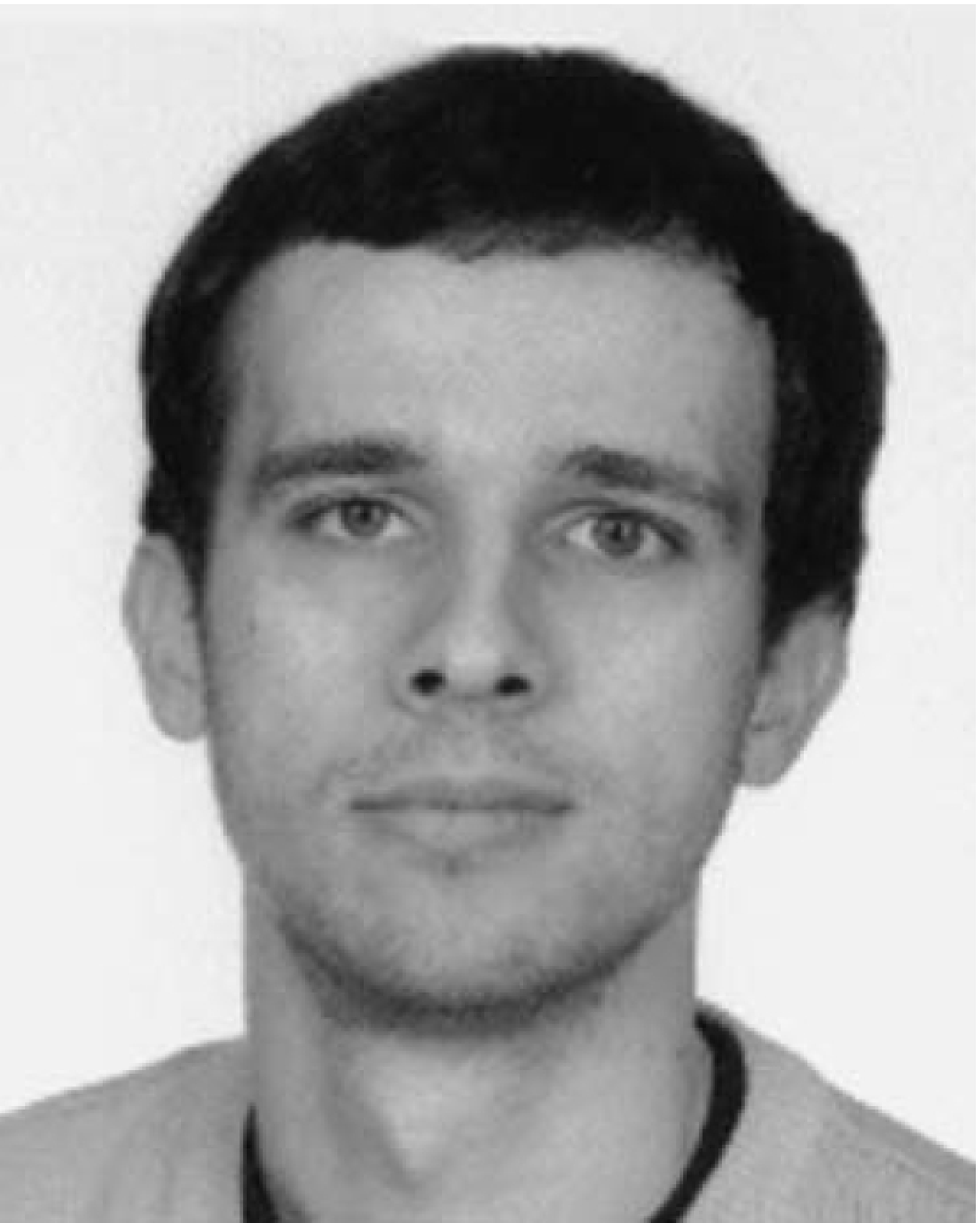}}]{Mate Boban}
Mate Boban is a Senior Researcher at Huawei European Research Center, Munich.
He received the Ph.D. degree in electrical and computer engineering from Carnegie Mellon University and the Diploma in Informatics from University of Zagreb. Before joining Huawei, he worked for NEC Labs Europe and Carnegie Mellon University and interned at Apple. He is an alumnus of the Fulbright Scholar Program. His current research is in the areas of cooperative intelligent transportation systems, wireless communications, and networking. He received the Best Paper Award at the IEEE VTC Spring 2014 and at IEEE VNC 2014. More information can be found on his website \url{http://mateboban.net}.
\end{IEEEbiography}

\begin{IEEEbiography}[{\includegraphics[width=1in,height=1.25in,clip,keepaspectratio]{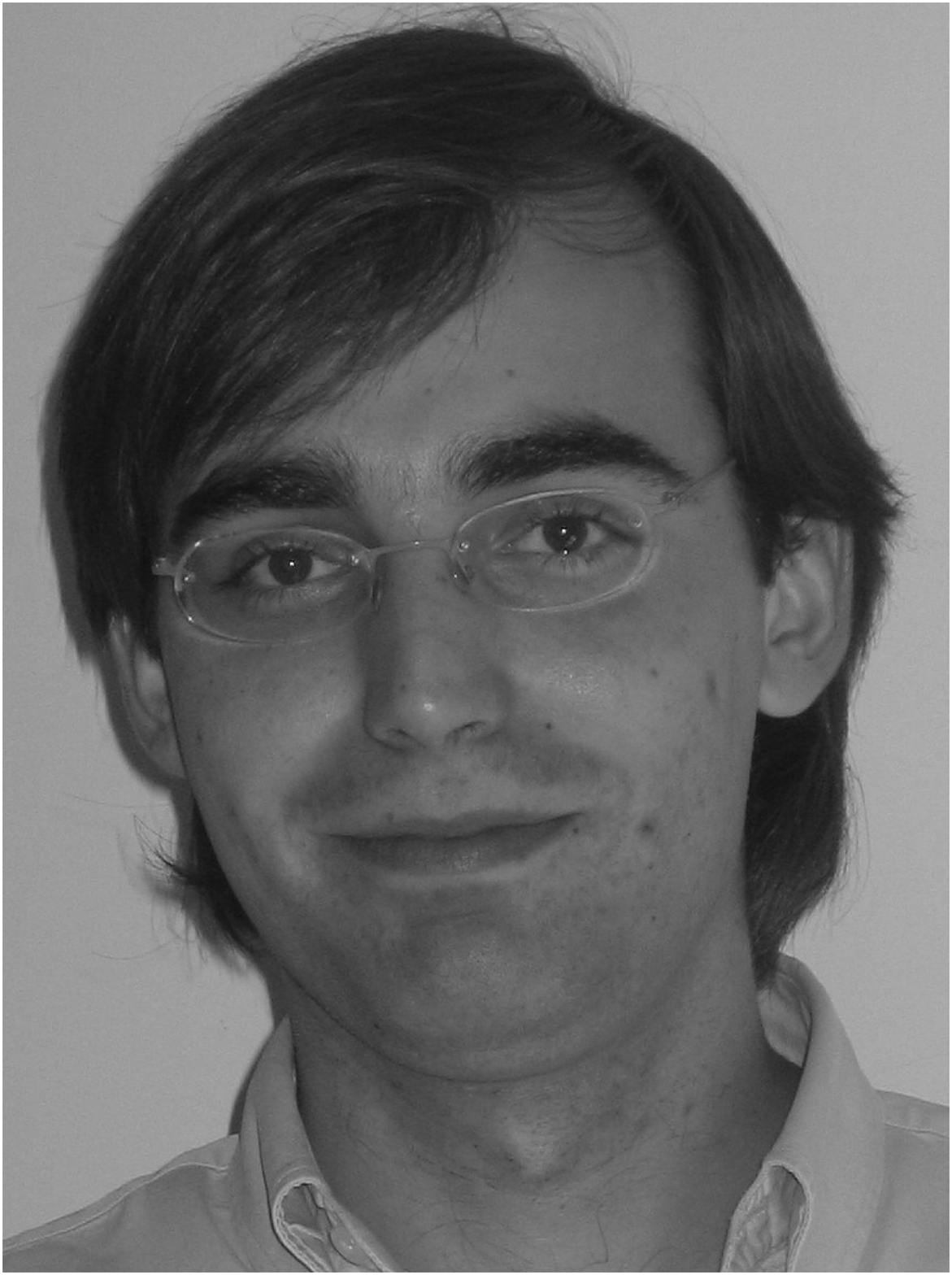}}]{Pedro M. d'Orey}
 received the \textit{Licenciatura} degree in electrical and computer engineering from 
the University of Porto, Portugal, in 2004,  the MSc. degree in Telecommunications from Queen Mary, University of London,
 UK, in 2008 and the Ph.D. degree in Telecommunications from the University of Porto, Portugal, in 2014.
He is currently a Researcher at the Instituto de Telecomunica\c{c}\~{o}es, University of Porto, Portugal.
 Previously, he was a Research Scientist at NEC Laboratories Europe and a R\&D Engineer in the area of Mobile Communications.
His main research interest are in the area of Intelligent Transportation Systems, Vehicular Networks and Autonomous Vehicles.
He received the Best Paper Award at the IEEE VTC Spring 2014 and at IEEE VNC 2014.
\end{IEEEbiography}
\end{document}

%% file: acronyms.tex
\begin{acronym}
	\acro{BTP}{Basis Transport Protocol}
	\acro{NARR}{Neighbors Above Range Ratio}
	\acro{CACC}{Cooperative Adaptive Cruise Control}
	\acro{CAM}{Cooperative Awareness Message}
	\acro{DENM}{Decentralized Environmental Notification Message}
	\acro{DSRC}{Dedicated Short Range Communication}
	\acro{FOT}{Field Operational Test}
	\acro{GPS}{Global Positioning System}
	\acro{GEMV2}[$GEMV^2$]{Geometry-based Efficient propagation Model for V2V communication}
	\acro{ITS}{Intelligent Transportation Systems}
	\acro{I2V}{Infrastructure to Vehicle}
	\acro{IRT}{Inter-Reception Time}
	\acro{LAN}{Local Area Network}
	\acro{LOS}{Line of Sight}
	\acro{MAC}{Medium Access Control}
	\acro{NAR}{Neighborhood Awareness Ratio}
	\acro{NIR}{Neighborhood Interference Ratio}	
	\acro{NLOS}{Non-LOS}
	\acro{NLOSv}{Non-LOS due to vehicles}
	\acro{NLOSb}{Non-LOS due to buildings/foliage}
	\acro{PIR}{Packet Inter-Reception time}
	\acro{PDR}{Packet Delivery Ratio}
	\acro{RNARI}{Ratio of Neighbors Above Region of Interest}
	\acro{RNAR}{Ratio of Neighbors Above Range}
	\acro{RSU}{Road Side Unit}
	\acro{SUMO}{Simulation of Urban MObility}
	\acro{VANET}{Vehicular Ad Hoc Network}
	\acro{V2V}{Vehicle to Vehicle}
	\acro{V2I}{Vehicle to Infrastructure}
	\acro{VTL}{Virtual Traffic Lights}
	\acro{ETSI}{European Telecommunications Standards Institute}
	\acro{ROI}{Region of Interest}
	\acro{EU}{European Union}
	\acro{U.S.}{United States}
	\acro{BSM}{Basic Safety Message}
\end{acronym}

%% file: intro.tex
Cooperative awareness is the ability to provide information on presence, position, direction, as well as basic status of communicating vehicles to neighboring vehicles (those located within a single hop distance)~\cite{etsi14}. Enabled by periodic message exchange, cooperative awareness is the basis for a large number of \ac{ITS} applications proposed by standardization bodies~\cite{etsi09}. Using the information provided by cooperative messaging, vehicles and \acp{RSU} are able to create a map of their surroundings, which is then used as input for safety applications that detect potentially hazardous situations. To enable cooperative awareness, standardization bodies have proposed specific messages for that purpose: in the EU, \acp{CAM} have been specified as part of the standard~\cite{etsi14}, whereas in the U.S., the same functionality is enabled by the \ac{BSM}~\cite{saeBsm}. These messages are exchanged periodically and contain location, speed, and direction of the vehicle, among other information.

\begin{figure}[!t]
  \begin{center}
    \includegraphics[width=0.3\textwidth]{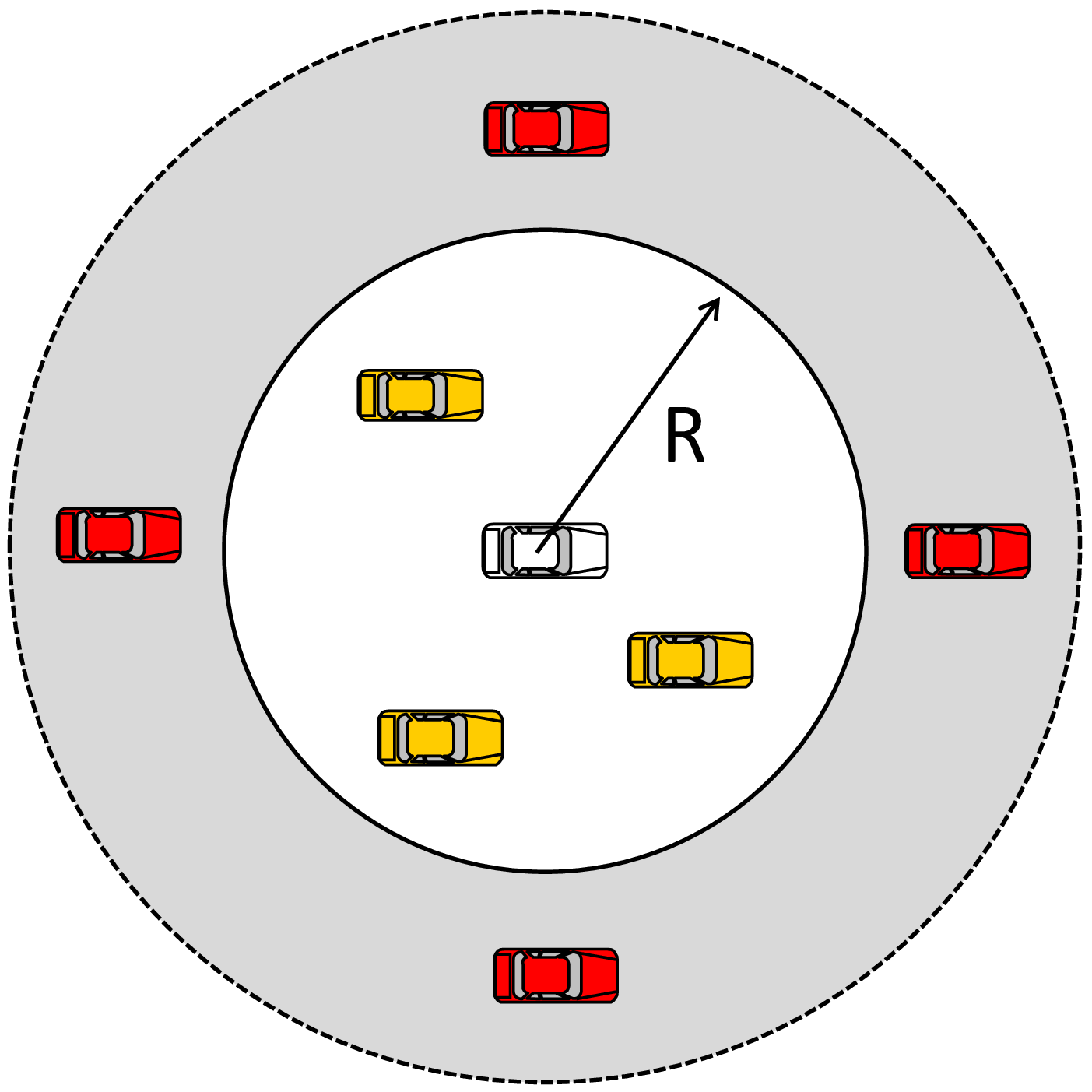}
     \caption{Neighborhood awareness is based on the broadcast of periodic messages and allows gathering relevant information on the  evolving neighborhood within a given range (e.g., the white zone with radius R). However, increasing awareness results in increased interference from distant vehicles (vehicles outside the designated awareness zone, i.e., grey unbounded zone), whose information might be less relevant than that of nearby vehicles.}
      \label{fig:metrics}
   \end{center}
\end{figure}

The IEEE 802.11p and ETSI ITS G5 standards were proposed in the US and the EU, respectively, 
as the underlying communication technology for \ac{V2V} and \ac{V2I} communication. 
To understand the capability of these technologies to support awareness-based \ac{ITS} applications, 
several research works have been conducted mostly resorting to analytical models or simulations.
These studies demonstrate the capability of the vehicular communication system to support cooperative awareness in restricted conditions (e.g., low channel load) 
and that improved accuracy on awareness information comes at the cost of higher channel load and increased interference (e.g., \cite{rateControl}).
Other studies (e.g., \cite{abbas12},\cite{boban14TVT}) have shown that shadowing caused by vehicles, buildings or other objects,
 can impair the perform of single-hop broadcasting schemes.   Adaptive power  (e.g., \cite{kloiber2012update}), rate (e.g., \cite{rateControl}) or joint power/rate (e.g., \cite{Tielert2013JPC}) beaconing schemes  can be used to improve network conditions, especially in high interference scenarios.
However, previous research did not realistically study the performance limits of cooperative awareness in large-scale. 
 Regarding empirical works studying cooperative awareness in \acp{VANET}, previous studies were limited to understanding the performance of communication system (e.g., using metrics such as packet delivery ratio \cite{Bai2010TUC}), and have neglected investigating the level of service provided to awareness-based \ac{ITS} applications.

Our previous studies on the topic (\cite{dorey14vtc,boban14vnc}) focused on measurement-based evaluation of cooperative awareness 
 using the following three metrics: 1) \acf{PDR}, a well established metric in the evaluation of communication systems, and two metrics to measure the efficacy of cooperative awareness; 2) \acf{NAR}; and 3)~\acf{RNAR}. For completeness, we formally define all three metrics in Section~\ref{subsec:Metrics}. 
In this paper, we extend those studies by: 
\begin{itemize}
\item Developing and validating a simple model for calculating \ac{NAR}, which requires as input only the information on \ac{PDR} statistics of the desired environment; for a given environment and parameter settings (transmit power, rate), the model can provide the achievable awareness level as a function of distance; 
\item Analyzing how the duration of \ac{NAR} measurement period $t$ and the number of messages sent per period impact the performance of cooperative awareness. 
\item Performing large scale simulations -- validated against measurements -- to determine the achievable performance of cooperative awareness; simulations can determine required values for cooperative message transmit power and rate to achieve the target awareness at the target distance; 
\end{itemize}

The rest of the paper is organized as follows. 
Section~\ref{sec:RelWork} discusses the related work.
Section~\ref{sec:Evaluation} describes the measurement-based aspect of the study, including the DRIVE C2X communications platform used to perform measurements, locations where measurements were performed, and the results of the measurements. 
Section~\ref{sec:Model} details the model for calculating \ac{NAR} using \ac{PDR}, including the results of the comparison between measurements and model.
Section~\ref{sec:Simulation} describes the results obtained through large-scale simulations, exploring the limits of cooperative awareness in terms of transmit power, rate, and target distance. Section~\ref{sec:Conclusions} concludes the paper.

%% file: relatedWork.tex
Extensive research has been conducted to study cooperative awareness in \acp{VANET},  with most studies resorting to analytical models or simulations. While previous work has mainly focused on the assessment of communication performance,  fewer studies looked at the cooperative awareness level provided to applications. In addition, the vast majority of previous studies have focused solely on the evaluation  of \acf{V2V} performance of periodic beaconing. 

With respect to assessment of communication performance in Vehicular Networks using analytical studies or simulations,
 Mittag et al.~\cite{Mittag2009CSM} compared single and multi-hop broadcast performance. 
 They concluded that limited benefit is achieved when using multi-hop communication instead of single-hop for cooperative awareness. 
 Van Eenennaam et al.~\cite{CACCbeacon} verified analytically that the three main dimensions that make the solution space of beaconing in \acp{VANET} are transmission power, generation rate and message duration,  and showed how different beaconing configurations support \acf{CACC}. Noori et al.~\cite{noori2013novel} performed simulations to study the probability of beacon delivery in an urban scenario and showed how packet delivery is impacted by increasing vehicle density and different road types. 
In \cite{rateControl}, Schmidt et al. study the trade-off between information accuracy and channel load derived from the dependable frequency of periodic beaconing. The authors also propose a scheme to control channel overloading by dynamically adjusting the beacon frequency to the current traffic situation (e.g., current vehicle density), while ensuring appropriate information accuracy.
  Kloiber et al.~\cite{kloiber2011performance} analyzed the ability of cooperative message exchange to inform the vehicles about hazardous situations
under challenging \ac{MAC} conditions. 
Several studies (e.g., \cite{kloiber2012update, Tielert2013JPC}) proposed improving awareness levels or reducing the channel load in \acp{VANET} by adaptive modification of beacon transmission power or generation rate.
Yin et al. \cite{multiChannel}  consider  multi-channel operation in their analytical studies
 and study the relevant impact of channel switching on the performance and reliability of safety message broadcasting on the Control Channel.

Regarding empirical evaluation of communication and application performance in \acp{VANET}, 
Martelli et al.~\cite{martelli2012measurement} analyzed the \ac{PIR}. Their results showed that PIR follows a power-law distribution (i.e., long-lasting outages occur with certain periodicity). Furthermore, \ac{PIR} is strongly affected by \ac{LOS} conditions, with up to five-fold performance drop in case of \ac{LOS} obstruction by vehicles. Bai et al.~\cite{Bai2010TUC} performed an extensive study on the impact of controllable parameters (transmit power, modulation scheme) and uncontrollable factors (distance, environment, velocity) on the performance of IEEE 802.11p radios in terms of \ac{PDR}. In a similar study, Santa et al.~\cite{Santa2014} analyzed the influence of several parameters on the performance of \acp{CAM} using an experimental testbed 
and showed that the \ac{LOS} conditions, equipment installation point and hardware capabilities are key variables in the network performance. 

Apart from analyzing the conventional communication performance (e.g., throughput, delay), several studies proposed
using information-centric metrics  (e.g., awareness quality~\cite{Mittag2009CSM, An2011}, update delay~\cite{kloiber2012update}, and PIR~\cite{Tielert2013JPC}). For instance, Kloiber et al.~\cite{kloiber2012update} proposed the \textit{Update Delay} metric, which is defined for a pair of vehicles  as the time interval between the expected \ac{CAM} reception and the actual message reception. These metrics allow for a better understanding of the impact of the underlying vehicular communication system on application-level performance.

%% file: experimental.tex
This section describes the metrics we use in our evaluation, DRIVE C2X experimental platform, the measurement test sites environments and the results of the measurement data analysis in terms of delivery rate (\ac{PDR}), awareness (\ac{NAR}) and interference (\ac{RNAR}). 

\begin{table}[tb!]
\centering
\caption{Test Site Italy RSU locations} 
\label{tab:RSU}
\footnotesize{
\begin{tabular}{l c c c}
\hline
\textbf{Id} & \textbf{Position (lat, lon)} &  \textbf{Height} & \textbf{Installed} \\ \hline
251 & 45.909728, 11.03248  	& 9~m  	& Pole\\
252 & 45.905776, 11.02953 	& 11~m 	& Overhead Gantry\\
253 & 45.86776,  11.005438 	& 9~m  	& Pole \\
254 & 45.871724, 11.007555 	& 9~m 	& Pole \\
255 & 45.8569,   11.000692  & 9~m 	& Overhead Gantry \\ \hline
\end{tabular}
}
\end{table}

\begin{table*}[t!]
\centering
\caption{Description of Measurement Test Sites and Parameters} 
\label{tab:params} 
\tiny{

\begin{tabular}{l c c c c}
\hline
\textbf{Location} & \textbf{Gothenburg}, Sweden & \textbf{Helmond}, the Netherlands & \textbf{Tampere}, Finland & \textbf{Trento}, Italy \\
Scenarios & Suburban  	& Suburban  			& Suburban 									& \\
 & (57.710316,11.94238) & (51.472803, 5.622418) & (lon $<$ 23.847835, lat $<$ 61.45894) 	&\\ \cline{2-5}
 & Highway 				&  Highway 				& Highway 									& Highway\\
 & (57.718424,11.918331)& (51.477243,5.620085) 	& (lat $>$ 61.45894 and lat $<$ 61.491023) 	& (45.934435, 11.087010)\\
 & 						& 						& (lon $>$ 23.790289 and lon $<$ 23.843118) & \\ \cline{2-5}
 & 						& 						& Urban 									&\\
 & 						& 						& Otherwise 								&\\ \hline
Route Length (Max.)  & 11~km	& 5.5~km  		& 22~km 									& 60~km\\ \hline
Time & June 2013 		& September 2012  		& April and May 2013  						& July to October 2013 \\
& (9~a.m. to 5~p.m) 	& (9~a.m. to 5~p.m.)	& (7~a.m. to 1~p.m.) 						& (7~a.m. to 2~p.m.) \\ \hline
Number of Vehicles 	& 6 & 9				& 3 										& 3/4\\ \hline
Vehicle Type & Personal & Personal 				& Personal 									& Personal \\ \hline
Antenna Type & Omni-directional & Omni-directional & Omni-directional 						& Omni-directional\\ \hline
Antenna Location & Rooftop & Rooftop & Rooftop & Rooftop\\ \hline
Antenna Height & approx.~1.55~m & approx.~1.44 - 1.66~m & approx.~1.5~m & approx.~1.49~m \\ \hline
Number of RSUs & 0 & 0  &0 & 5\\ \hline
RSU Antenna & N/A & N/A & N/A & Two Corner Reflector \\ \hline
\end{tabular}
}
\end{table*}

\subsection{Performance evaluation metrics}\label{subsec:Metrics}

\begin{enumerate}
 \item \textit{\acf{PDR}}: the ratio of the number of correctly received packets to the number of transmitted packets.  

Formally, for a transmitting vehicle,the combined \ac{PDR} to all receiving vehicles within a certain distance range denoted by $r$ (e.g., between 25 and 50~meters from receiving vehicle) is given by $PDR_{i,r} = \frac{PR_{i,r}}{PT_{i,r}}$, where $PT_{i,r}$ is the total number of messages sent by $i$ to vehicles within $r$ from $i$, whereas $PR_{i,r}$ is the subset of $PT_{i,r}$ packets that was correctly received.  We measure \ac{PDR} during the entire experiment duration, i.e., the time interval $t$ over which \ac{PDR} is measured equals the experiment duration. This metric provides the indication of the link quality and effective and maximum communication range. Effective communication range is defined as the maximum distance below which the \ac{PDR} is above a given threshold (e.g., 0.9), whereas maximum communication range is the distance above which the \ac{PDR} is equal to 0.
 \item \textit{\acf{NAR}}: the proportion of vehicles in a specific range from which a message was received 
 in a defined time interval. Formally, for vehicle $i$, range $r$, and time interval $t$, $NAR_{i,r,t} = \frac{ND_{i,r,t}}{NT_{i,r,t}}$, where $ND_{i,r,t}$ is the number of vehicles within $r$ around $i$ from which $i$ received a message in $t$ and $NT_{i,r,t}$ is the total number of vehicles within $r$ around $i$ in $t$ (we use $t$=1 second). 
Referring to Fig.~\ref{fig:metrics}, for the  white vehicle in the center, \ac{NAR} is the proportion of nodes in the inner (white) circle (which encompasses the distance range from 0 to R) from which the observed vehicle received a message. This metric measures the efficacy of cooperative awareness messaging.
 \item \textit{\acf{RNAR}}: for a vehicle $i$, distance $R$, and time interval $t$, the ratio of neighbors that are above a certain distance from the observed vehicle is defined as $RNAR_{i,R,t} = \frac{NA_{i,R,t}}{N_{i,t}}$, where $NA_{i,R,t}$ is the number of vehicles above $R$ from which $i$ received a message in $t$ 
and $N_{i,t}$ is the total number of vehicles from which $i$ received a message in $t$ (irrespective of distance from $i$).
Referring to Fig.~\ref{fig:metrics}, for the white vehicle in the center, \ac{RNAR} is the proportion of vehicles outside the inner (white) circle from which at least one message was received within $t$  to the total number of vehicles from which a message was received.
This metric gives an indication of potentially unnecessary traffic overheard from distant neighbors.
\end{enumerate}

Note that, when representing \ac{PDR} and \ac{NAR}, we consider a set of uniformly spaced distance bins (i.e., an annulus -- region between two concentric circles around the vehicle, represented by $r$). On the other hand, for \ac{RNAR} we consider the region outside a certain radius $R$. Furthermore, note that $t$ and $R$ can assume different values, as these are application specific (e.g., $t=$ 1 second might be sufficient for basic awareness service, whereas more stringent applications such as platooning might require $t\leq$ 100~ms). Since the purpose of cooperative message exchange is timely notification of vehicles and infrastructure about existence of other vehicles, the \ac{NAR} metric measures the proportion of vehicles in a given \ac{ROI} that receive \emph{at least one} message from the transmitting vehicle in time interval $t$ and are thus aware of the transmitting vehicle. Conversely, the more distant the transmitting vehicle, the less relevant the messages from that vehicle are for majority of safety applications. To that end, the \ac{RNAR} metric measures the proportion of vehicles outside the \ac{ROI} $R$, from which the messages are received. In future scenarios, where a high percentage of vehicles will be equipped with the communication equipment, high \ac{RNAR} would imply high interference, and thus low overall system throughput. Therefore, in terms of the communications performance, a well-functioning transmit system would aim to increase \ac{NAR}, while at the same time keeping \ac{RNAR} reasonably low.

\subsection{Experimental platform}\label{subsec:ExpSetup}
DRIVE-C2X project designed and evaluated a set of applications enabled by  \ac{V2V} 
and \ac{V2I} communication  in test sites throughout Europe.   
The DRIVE-C2X system uses ITS-G5 compliant radios that operate in the 5.9 GHz frequency band.
 The default value for transmit power was set to 21~dBm. On vehicles, whose heights ranged from 1.4~meters to 1.7~meters, omni-directional antennas were placed on the roof. 
Across test sites, vehicles had different communication system setup, including different radios, cable losses, antenna gains and placements, etc. All of these parameters resulted in significant variations of the effective transmit power output at each vehicle -- this is in line with what is expected in the production-grade systems once the communication devices are installed in the cars due to different system designs across manufacturers.
 The radios transmit \acp{CAM} that are in line with the ETSI standard~\cite{etsi14}. 
 \acp{CAM} contain node information (e.g., position, speed, and sensor information) and are broadcast to one-hop neighbors over the control channel. Positioning information was provided by \acs{GPS} receivers on the vehicles.
In the analyzed datasets, \acp{CAM} were sent at 10~Hz frequency and had the size of 100 Bytes.

\subsection{Measurement test sites}\label{subsec:Scenarios}

The empirical evaluation of cooperative awareness in \acp{VANET} presented in this paper is based on analysis of logging information. All nodes (vehicles and \acp{RSU}) record all received and transmitted messages during the several test runs. In all test sites, vehicles were driven in normal traffic conditions  with the presence of other vehicle types and respecting traffic rules. Traffic conditions (e.g., vehicle density) varied between test sites and trials but detailed information was not reported during field tests.

In test sites in Sweden and Finland, combined with antenna gains and cable losses, the effective vehicle transmit power ranged between 10 and 20~dBm. 
In Trento, Italy, 
there were 5 \acp{RSU} with the antenna placed at heights between 9 and 11~m  at the positions and locations indicated in Table~\ref{tab:RSU}.
One \ac{RSU} is installed on a highway on an overhead gantry 11~m above the road surface. It is equipped with two corner reflector antennas each having 14 dBi nominal gain, beam width 30 degrees in azimuth and 60 degrees in elevation. Remaining \acp{RSU} are installed next to the highway at the height of 9~m. Both vehicles and \acp{RSU} have a nominal output power of 21~dBm. Combined with antenna gains and cable and insertion losses, this yields 27~dBm transmit power on the vehicles, and 32~dBm on \acp{RSU}. These power settings are markedly higher than in remaining test sites, where the transmit power on vehicles was between 10 and 20~dBm. In the Netherlands, the vehicles used for testing were a combination of vehicles used in the other test sites.

In the test sites Sweden, the Netherlands and Finland, no logging data was available for \acp{RSU}, 
therefore the results contain \ac{V2V} communication tests only.  In Italy, on the other hand, logging data was available for both \ac{V2V} and \ac{V2I} communication. More details on the experimental setup are given in Table~\ref{tab:params}.

%% file: results.tex
\begin{figure*}[!t]
        \centering
		\subfigure[\scriptsize Sweden -- Highway.]{\label{fig:PDRV2VSH}\includegraphics[width=0.24\textwidth]{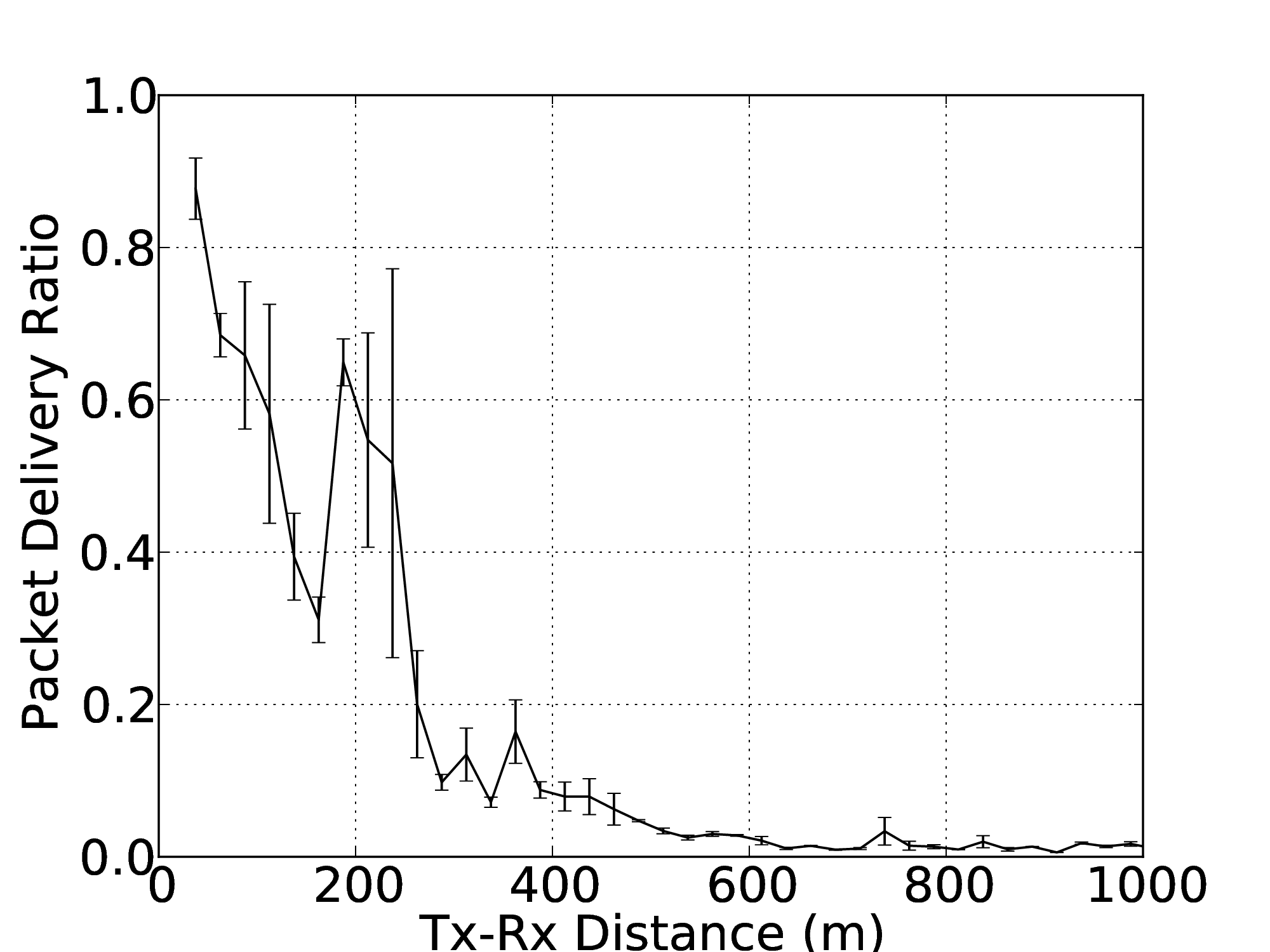}}
		\subfigure[\scriptsize Sweden -- Suburban.]{\label{fig:PDRV2SS}\includegraphics[width=0.24\textwidth]{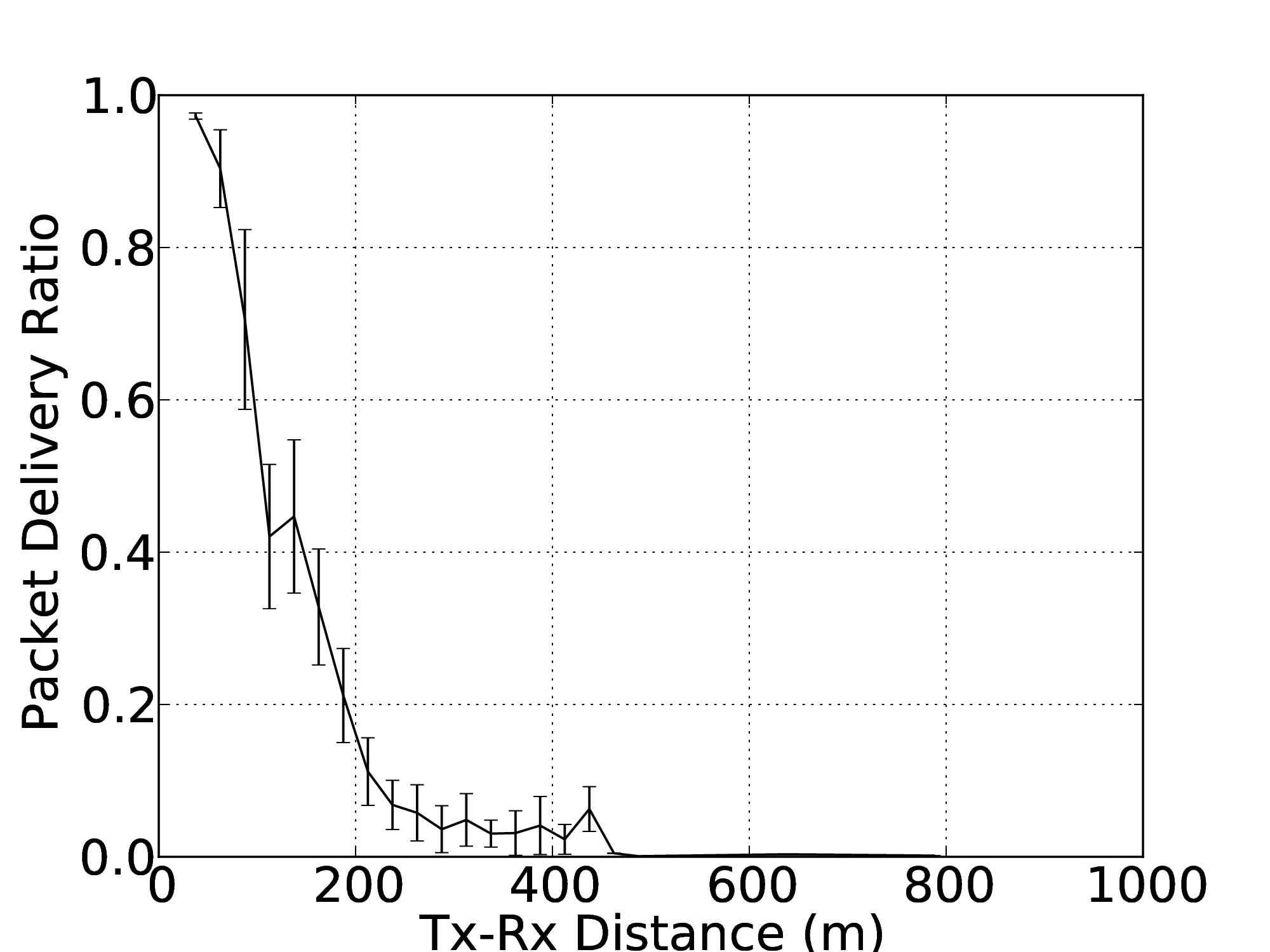}}      
		\subfigure[\scriptsize The Netherlands -- Highway.]{\label{fig:PDRV2VNH}\includegraphics[width=0.24\textwidth]{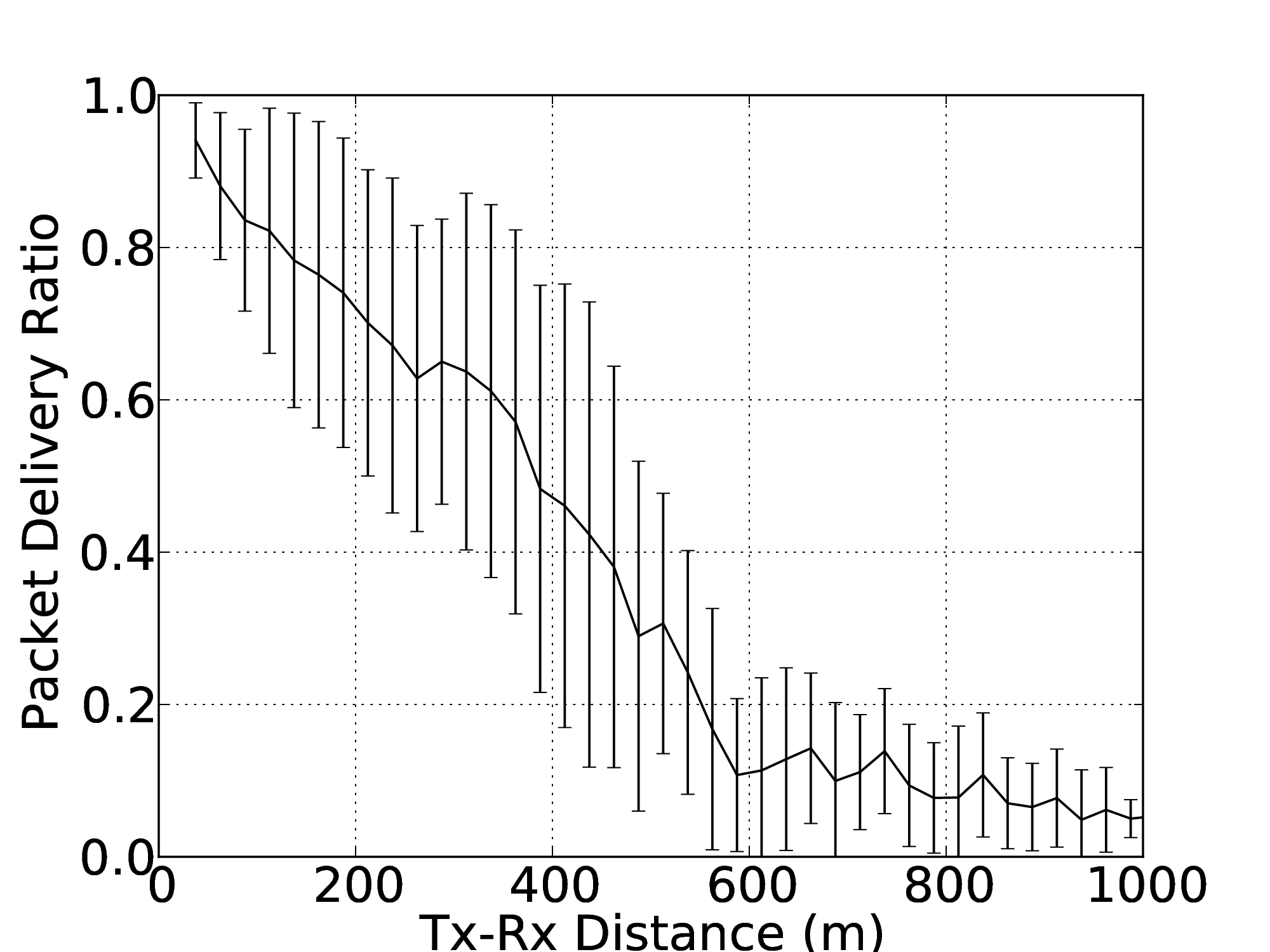}}
		\subfigure[\scriptsize The Netherlands -- Suburban.]{\label{fig:PDRV2VNS}\includegraphics[width=0.24\textwidth]{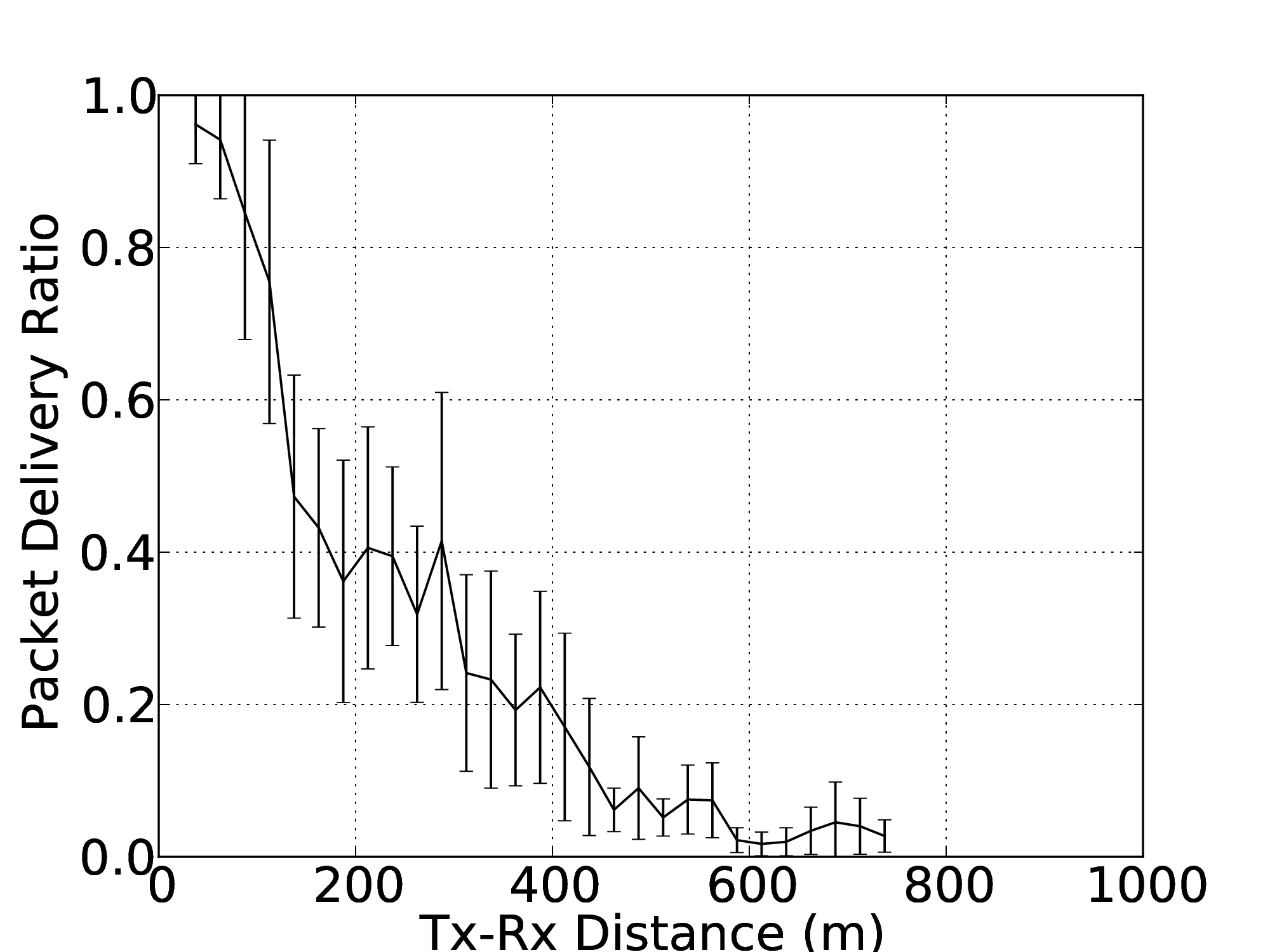}}                      
        \subfigure[\scriptsize Italy -- Highway.]{\label{fig:PDRV2VI}\includegraphics[width=0.24\textwidth]{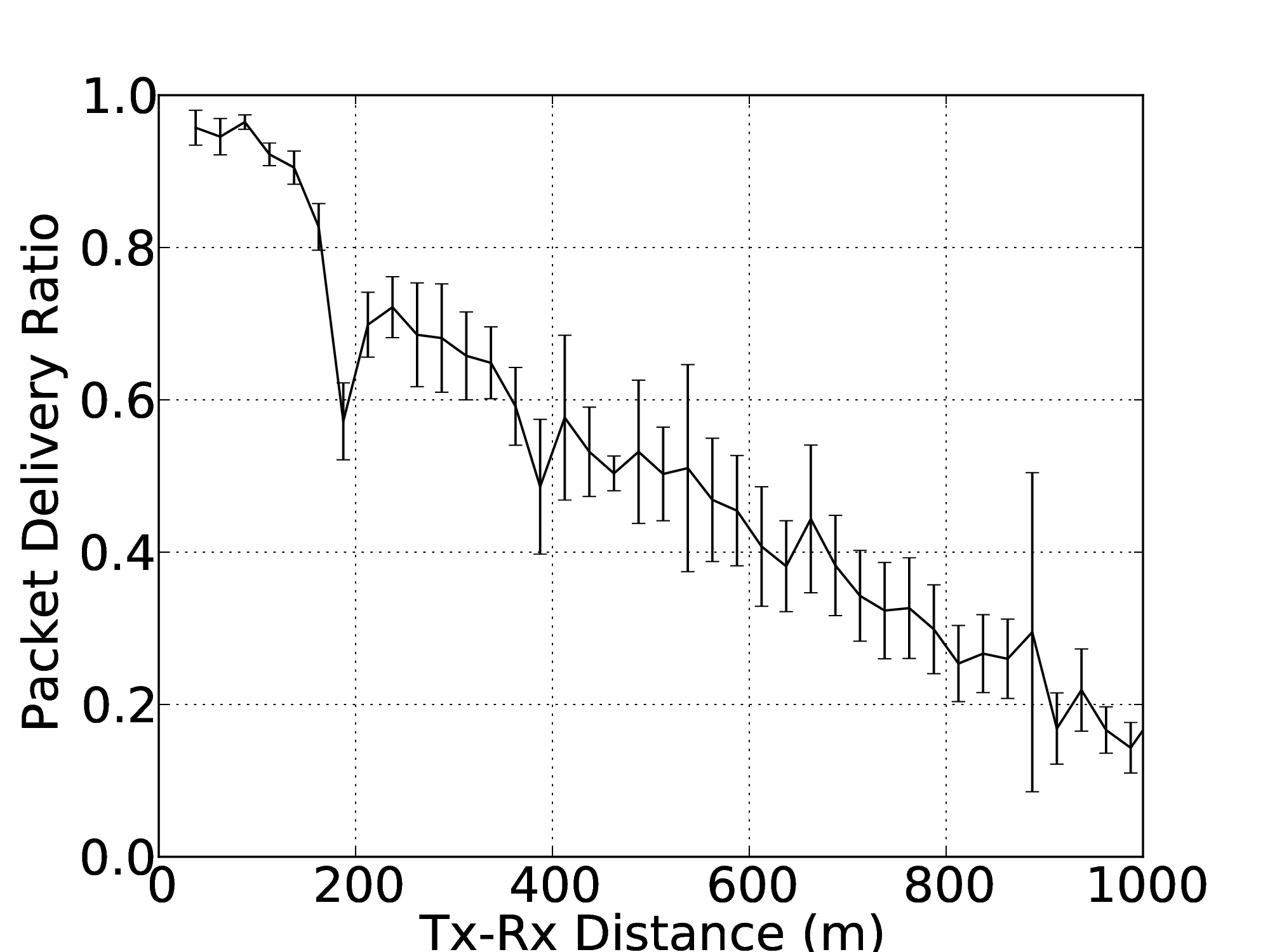}}  
		\subfigure[\scriptsize Finland -- Highway]{\label{fig:PDRV2VFH}\includegraphics[width=0.24\textwidth]{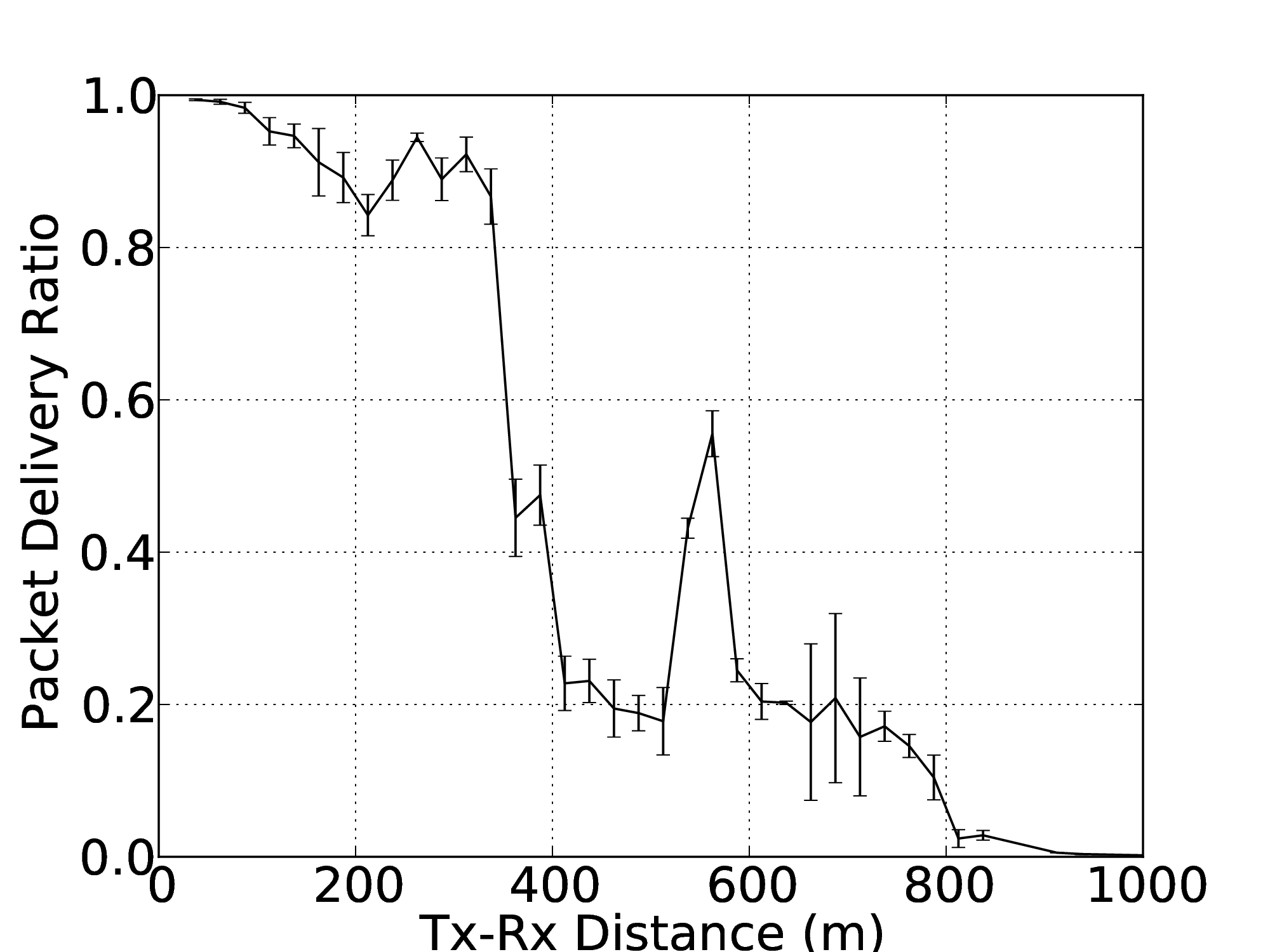}}   
		\subfigure[\scriptsize Finland -- Urban]{\label{fig:PDRV2VFU}\includegraphics[width=0.24\textwidth]{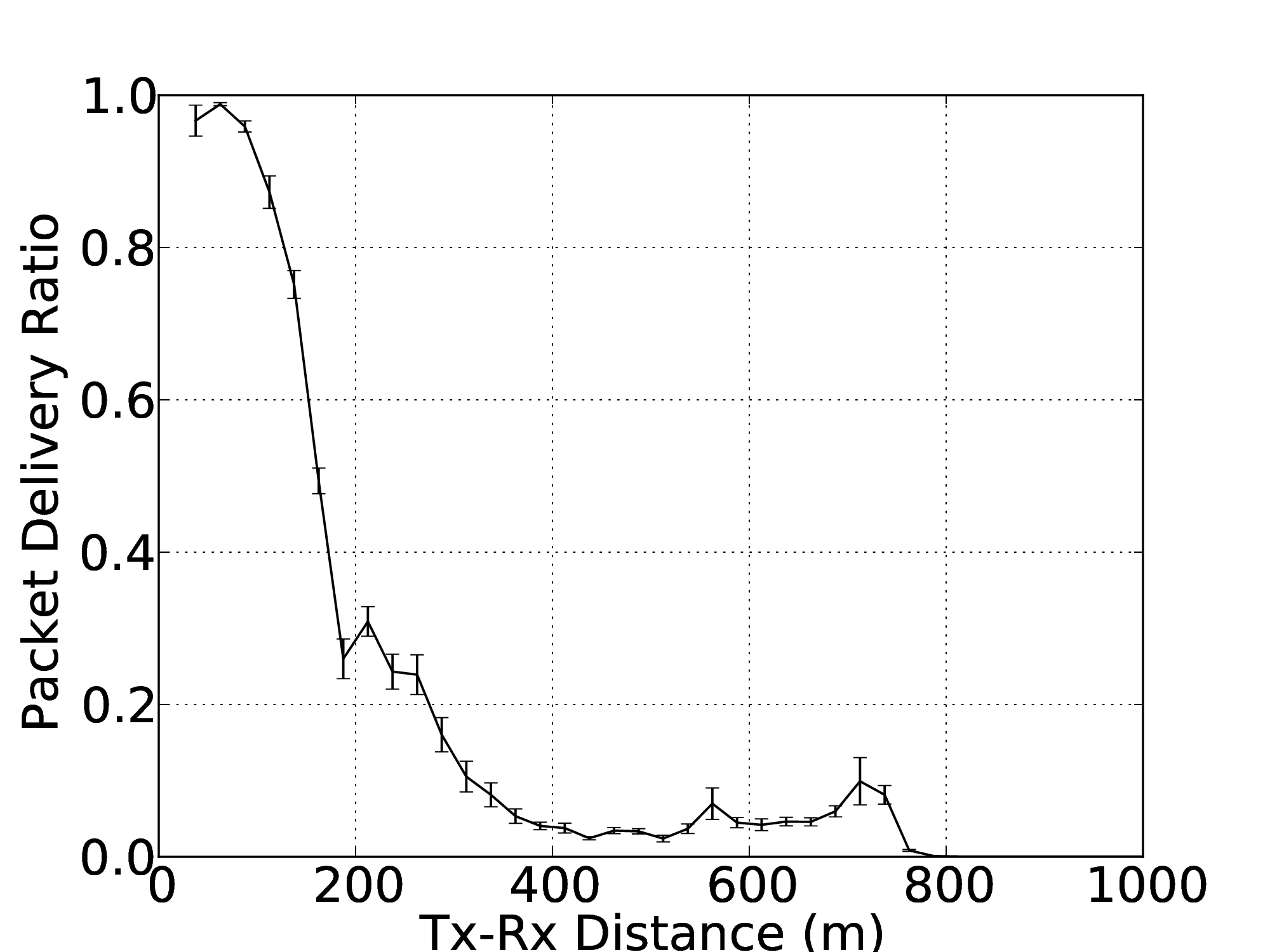}}   
		\subfigure[\scriptsize Finland -- Suburban]{\label{fig:PDRV2VFSU}\includegraphics[width=0.24\textwidth]{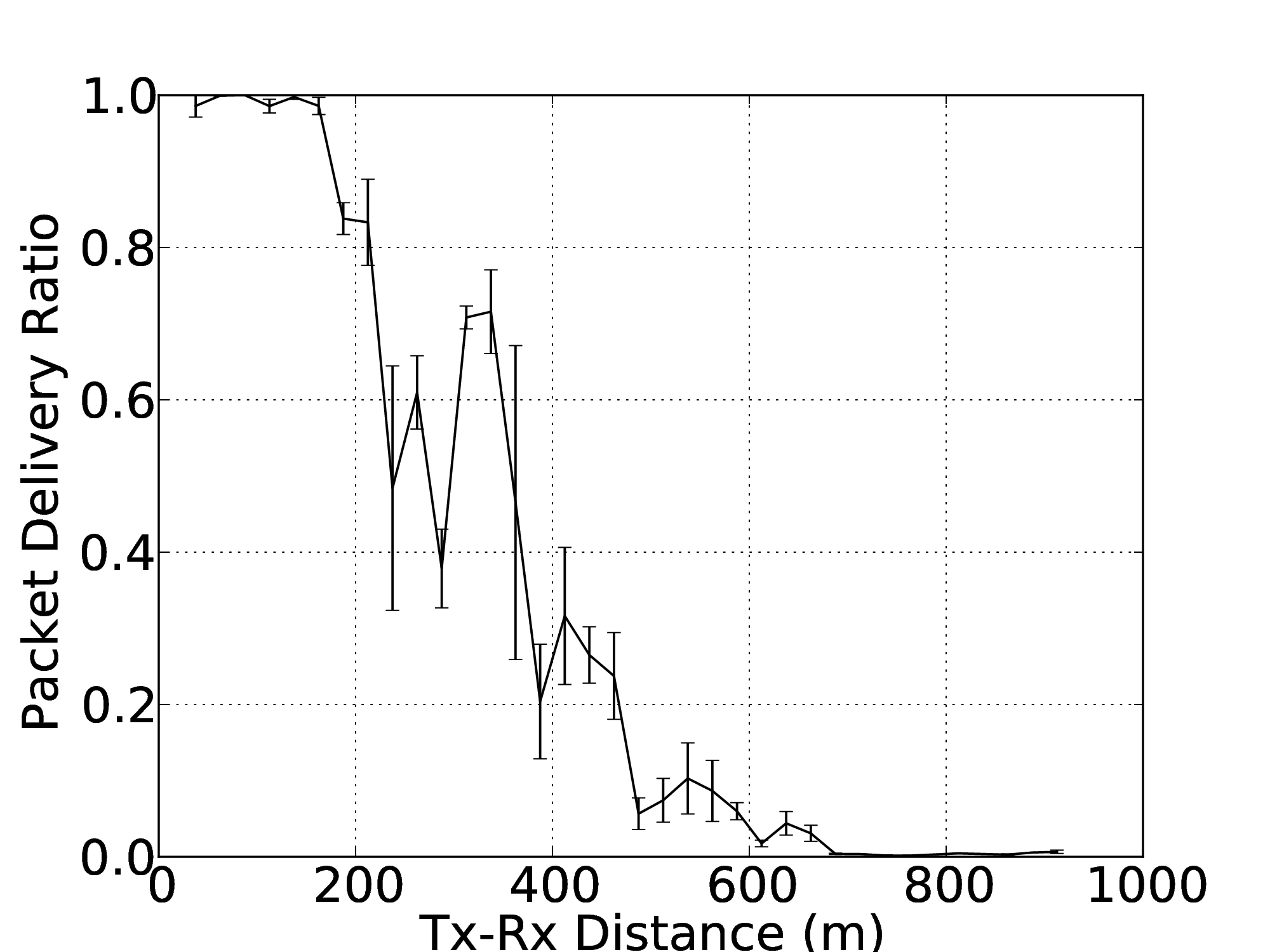}}        
        \caption{Overall V2V Packet Delivery Ratio (PDR) for  Test Site Sweden, the Netherlands, Italy and Finland.} \label{fig:PDRV2VOverall}      
\end{figure*} 

Below, we present and discuss the results of collected during the DRIVE-C2X measurement campaign in terms of \ac{PDR}, \ac{NAR}, and \ac{RNAR}.
The results are aggregated per vehicle (over messages transmission) and per test site (over different vehicles)  for different environments (urban, suburban, and highway). Each distance bin is 25 meters for \ac{PDR} and 50 meters for \ac{NAR} and \ac{RNAR}, with the plotted data point centered in the middle of the distance bin. Error bars represent one standard deviation around the mean of the measured variable for each vehicle. For statistical relevance, we consider solely bins with at least 40 data points. With respect to \ac{NAR} and \ac{RNAR}, for all results and plots shown in the following, one second window ($t=1$~s) was used for determining the reception of messages from direct neighbors.

\subsubsection{Packet Delivery Rate}\label{subsec:PDR}

\begin{figure}
        \centering    
        \subfigure[\scriptsize  The Netherlands -- Highway.]{\label{fig:PDRV2VNHI}\includegraphics[width=0.49\linewidth]{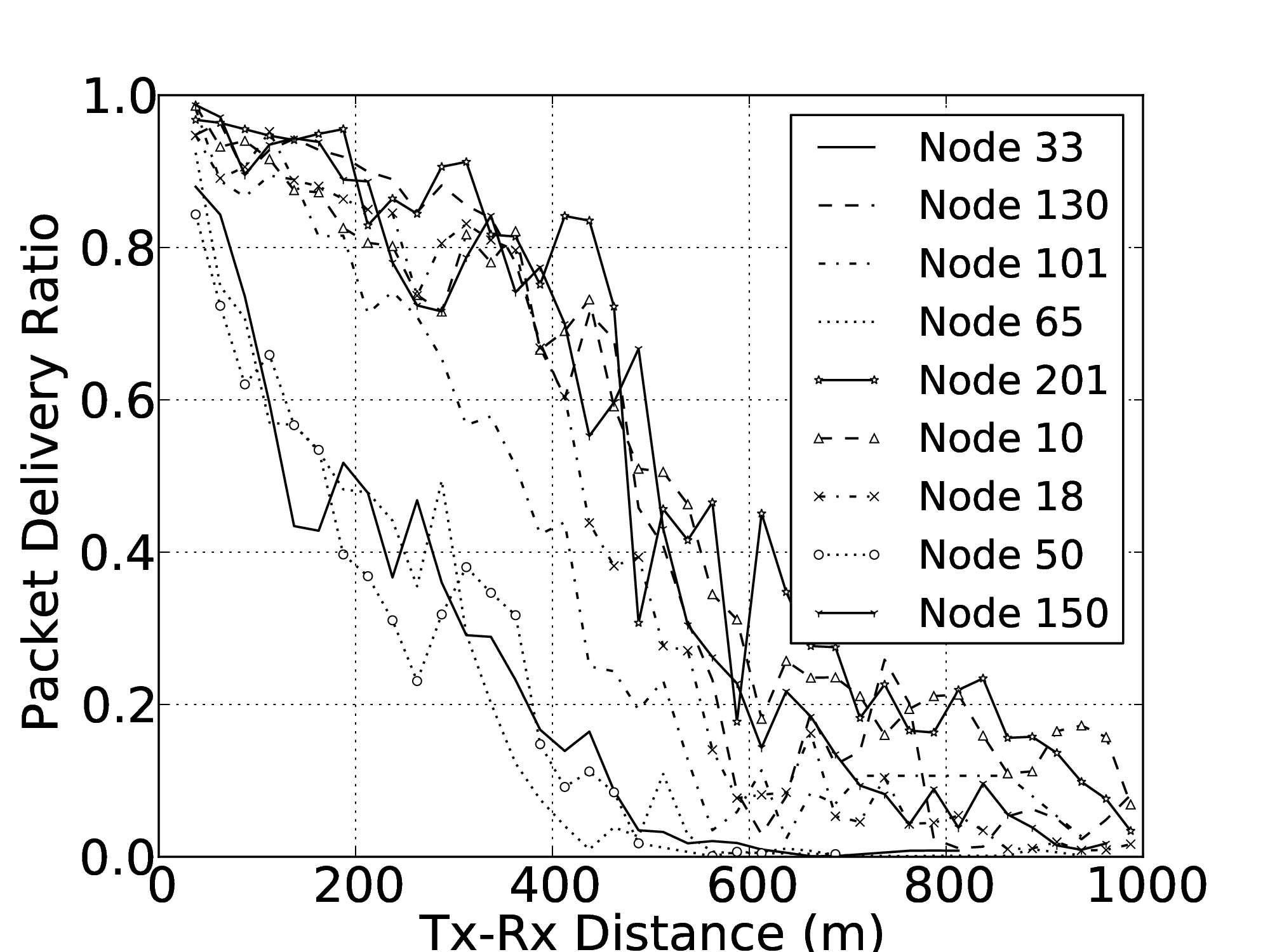}}                  
		\subfigure[\scriptsize  Italy -- Highway]{\label{fig:PDRV2VII}\includegraphics[width=0.49\linewidth]{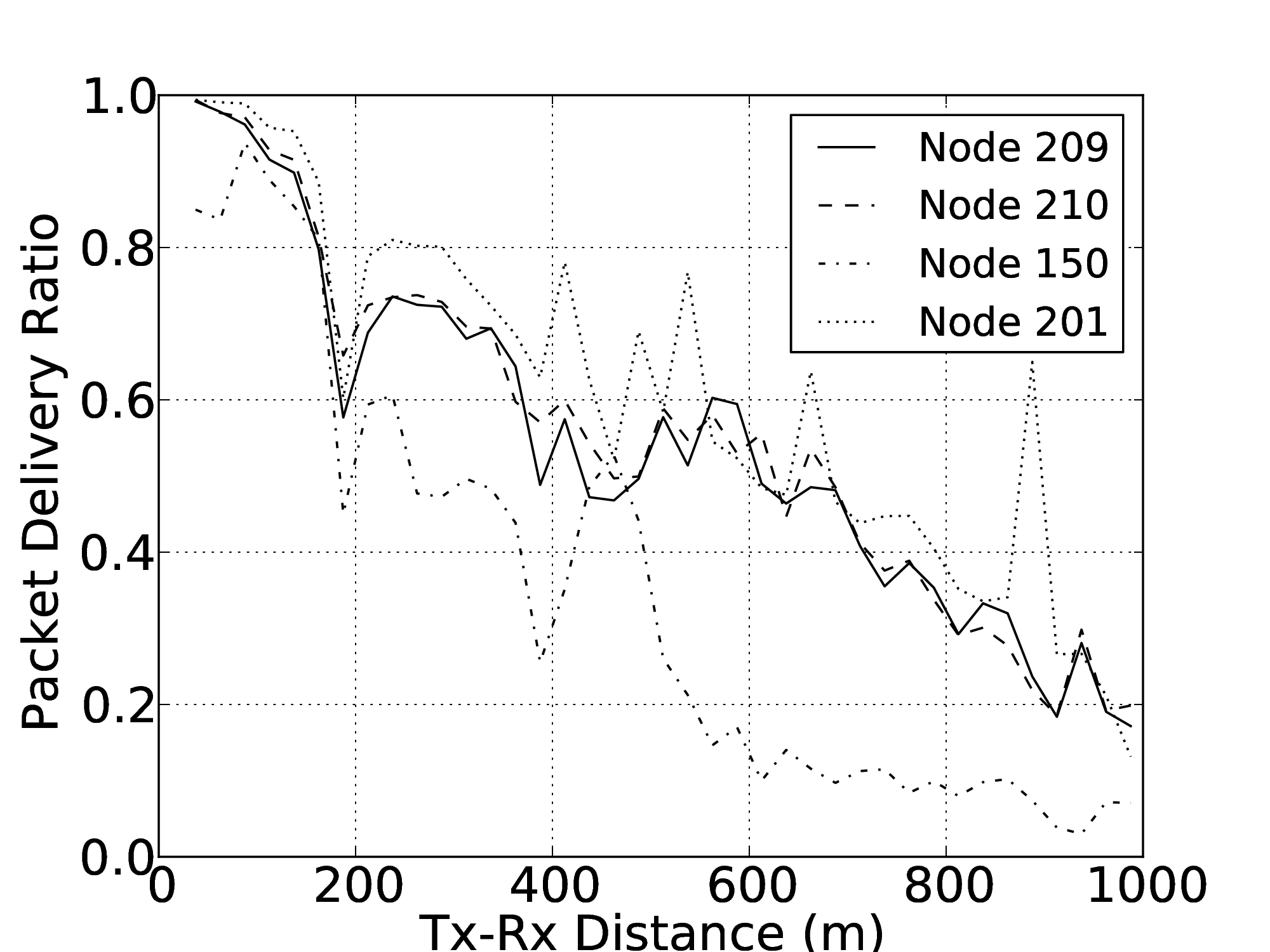}}    
        \caption{Per-vehicle V2V Packet Delivery Ratio (PDR) for Test Site the Netherlands and Italy.} \label{fig:PDRV2VPerVeh}     
\end{figure}

\begin{figure}
        \centering            
       	\subfigure[\scriptsize Overall V2I results]{\label{fig:PDRV2II}\includegraphics[width=0.49\linewidth]{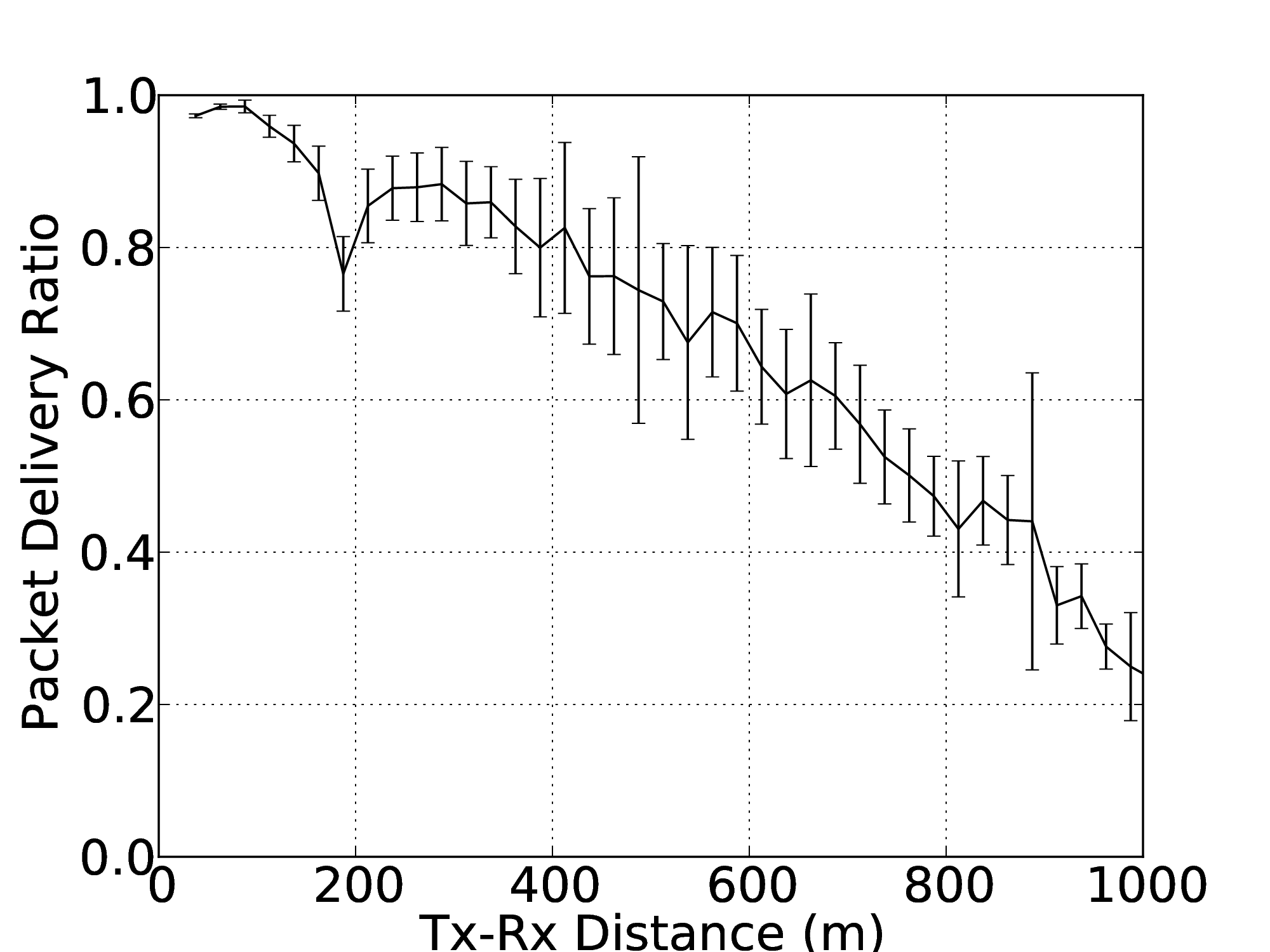}}   
		\subfigure[\scriptsize Per-node V2I results (3 vehicles (201, 209, 210) / 5~RSUs 25x)]{\label{fig:PDRV2III}\includegraphics[width=0.49\linewidth]{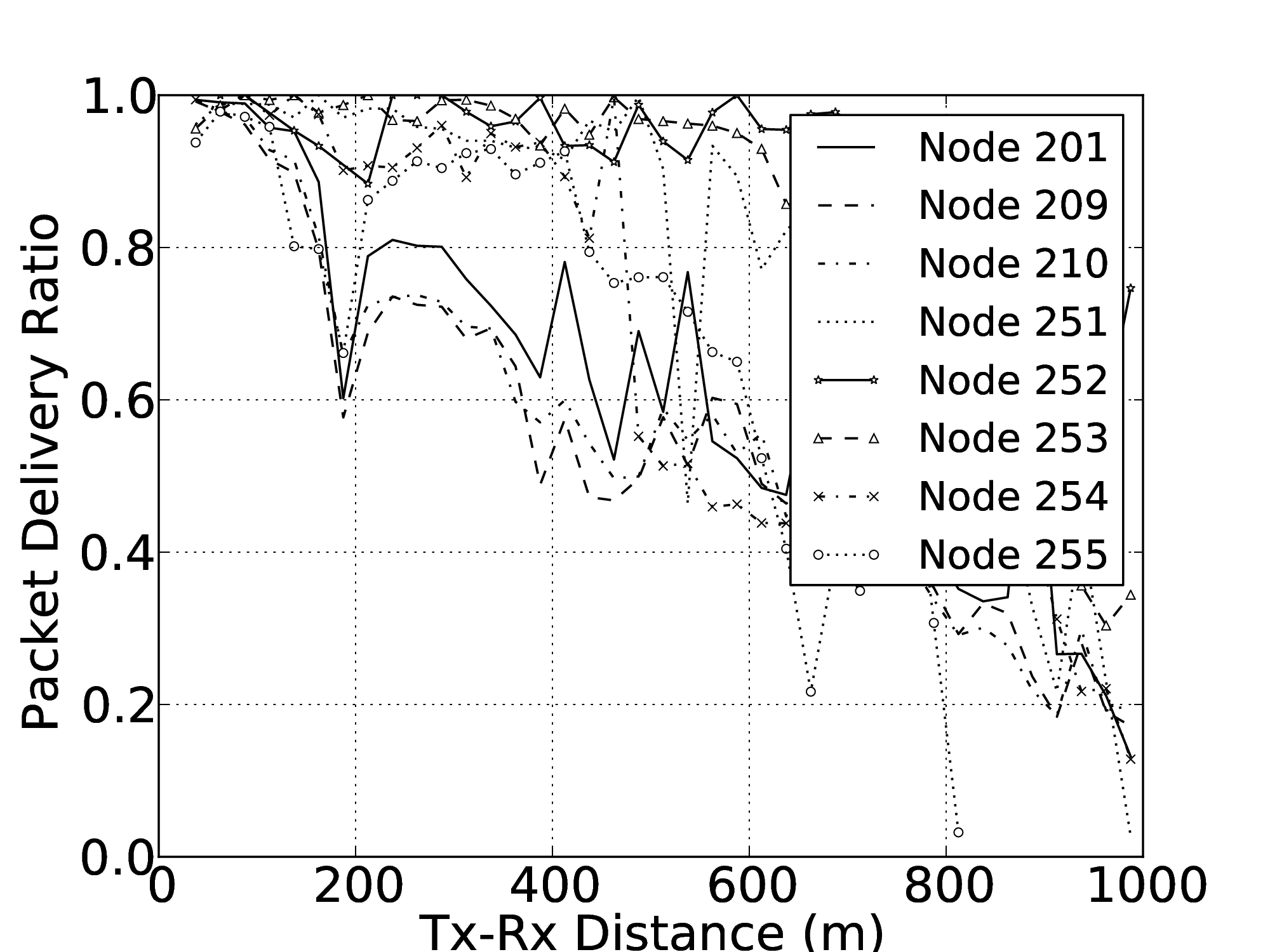}}             
        \caption{V2I Packet Delivery Ratio (PDR) for  Test Site Italy.} \label{fig:PDRV2I}      
\end{figure}

Figs.~\ref{fig:PDRV2VOverall} ,~\ref{fig:PDRV2VPerVeh}  and ~\ref{fig:PDRV2I} show the \acf{PDR} as a function of distance  for \ac{V2V} and \ac{V2I} communications
for different measurement locations.

\textbf{\ac{V2V}} -- As expected, for all test sites,  the \ac{PDR} decreases, albeit non-monotonically, as the node separation increases. 
The non-monotonic behavior of \ac{PDR} over distance is mainly due to: i) in case of \ac{LOS} communication, the dominating two-ray ground reflection model~\cite{Boban2013MVL}; and ii) in case of non-\ac{LOS} communication, variations in \ac{LOS} obstruction level.
 Our results in terms of \ac{PDR} are in line with the analytic results obtained by An et al.~\cite{An2011} and the empirical results by Visintainer et al.~\cite{crf1} for the highway scenario.  
 
 The \ac{PDR}  varies greatly between test sites and between qualitatively classified propagation environments.
 When considering the environment type, the communication ranges are increasing in the following order for a given test site: urban, suburban, and highway (e.g., see Fig.~\ref{fig:PDRV2VNH} and Fig.~\ref{fig:PDRV2VNS} for difference between highway and suburban \ac{PDR}). 
  The harsher propagation environment present in (sub)urban scenarios, including frequent non-\ac{LOS} conditions due to surrounding objects (e.g., other vehicles, buildings, and trees), affects considerably the link quality and consequently the successful packet delivery. This is in line with previous measurements studies (e.g.,~\cite{abbas12}).    However, the results for the same environment may vary substantially from one test site to another. 
  This is most evident for the highway scenario where the maximum communication range varies from approximately 600~m in Sweden (Fig.~\ref{fig:PDRV2VSH}) and Finland (Fig.~\ref{fig:PDRV2VFH}) to more than 1000~m in Italy (Fig.~\ref{fig:PDRV2VI}). This is the result of different propagation environments in different test sites (even for qualitatively same type of environment) as well as differences in installations in test vehicles (including antenna placement and gain).
For example, one of the main reasons for the improved performance in the Italian test site (Fig.~\ref{fig:PDRV2VI}) is the higher effective transmit power in test vehicles.

\textbf{\ac{V2I}} -- The \ac{V2I} \ac{PDR} results are presented in Fig.~\ref{fig:PDRV2II} and Fig.~\ref{fig:PDRV2III} for the Italian test site (highway scenario). These results are in line with the study by Paier et al.~\cite{paier10}, with the increased \ac{PDR} in case of our measurements due to higher transmit powers (32~dBm EIRP on \acp{RSU} and 27~dBm on vehicles, compared to 15.5~dBm in Paier et al.). 
Compared to \ac{V2V} results in the same location (Fig.~\ref{fig:PDRV2VI}), \ac{V2I} \ac{PDR} is significantly higher due to two main reasons:
1) advantageous position of \acp{RSU} (9-11~m above ground), giving the \acp{RSU} unobstructed \ac{LOS} at larger distances; and 2) the 
increased effective transmission power of \acp{RSU}.

\subsubsection{Neighborhood Awareness Ratio}\label{subsec:NeighborAwareness}

\begin{figure*}
\centering
\subfigure[\scriptsize Sweden -- Highway.]{\label{fig:NARSH}\includegraphics[width=0.24\textwidth]{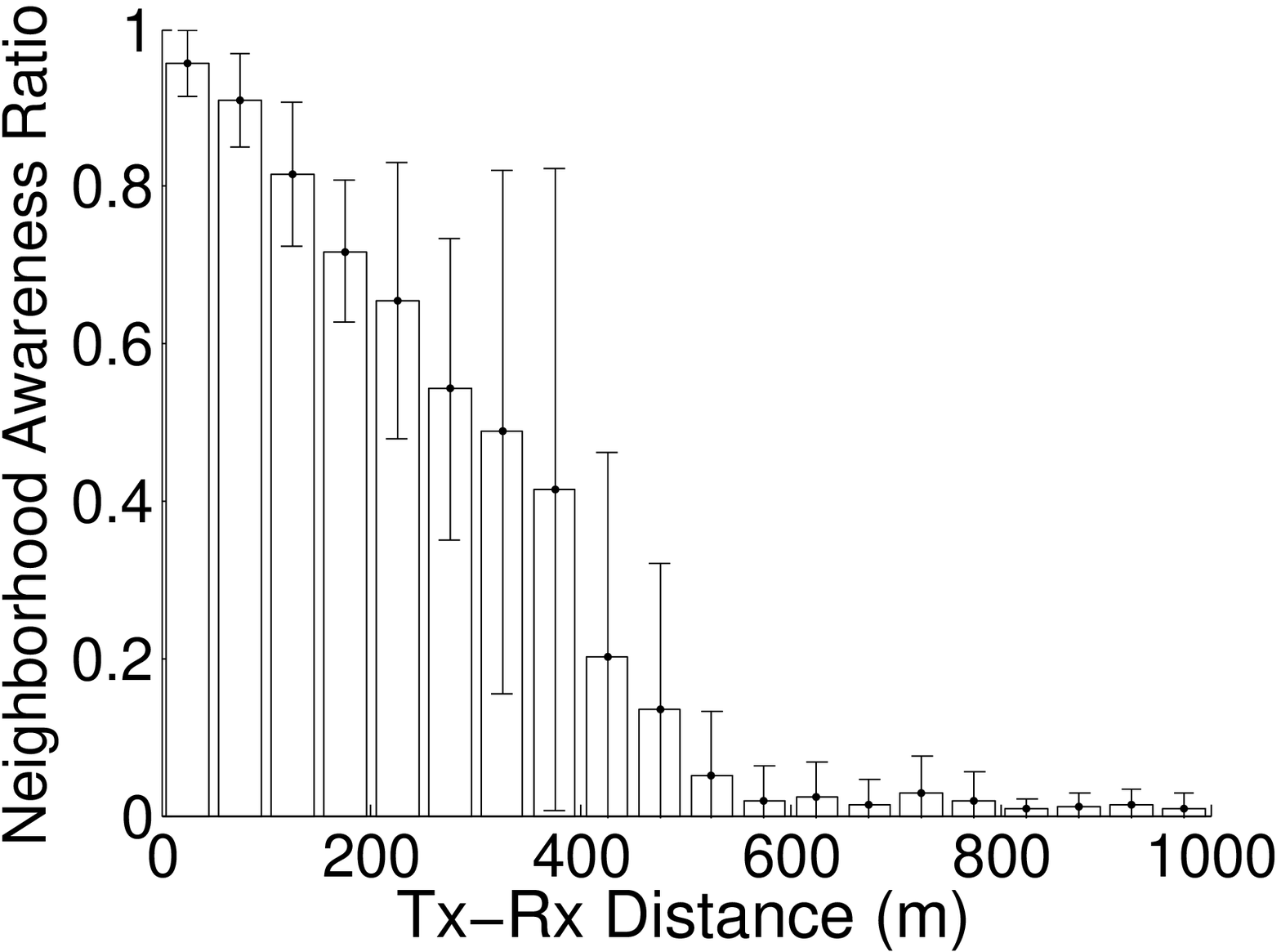}}
\subfigure[\scriptsize Sweden -- Suburban.]{\label{fig:NARSS}\includegraphics[width=0.24\textwidth]{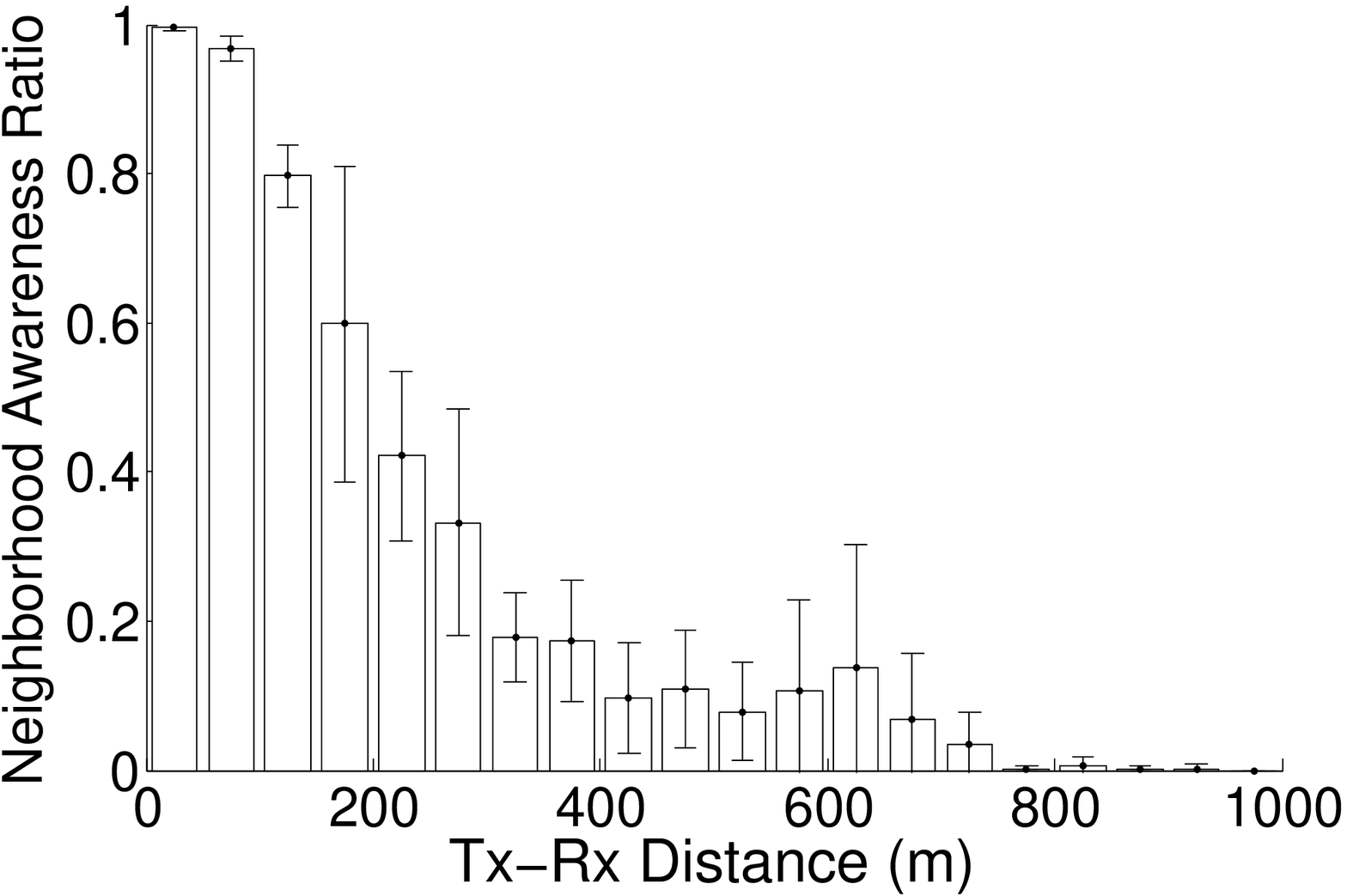}}
\subfigure[\scriptsize The Netherlands -- Highway.]{\label{fig:NARNH}\includegraphics[width=0.24\textwidth]{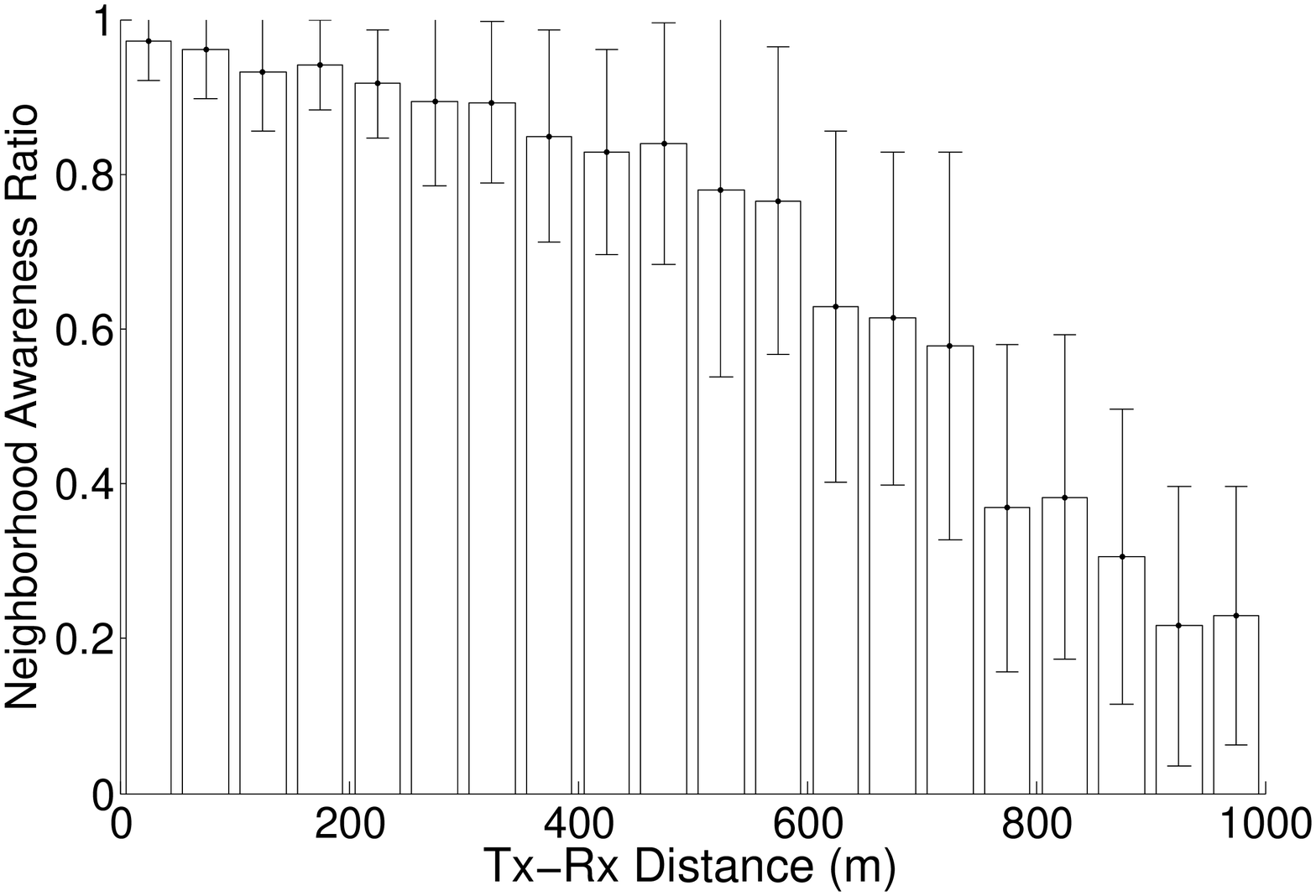}}
\subfigure[\scriptsize The Netherlands -- Suburban.]{\label{fig:NARNS}\includegraphics[width=0.24\textwidth]{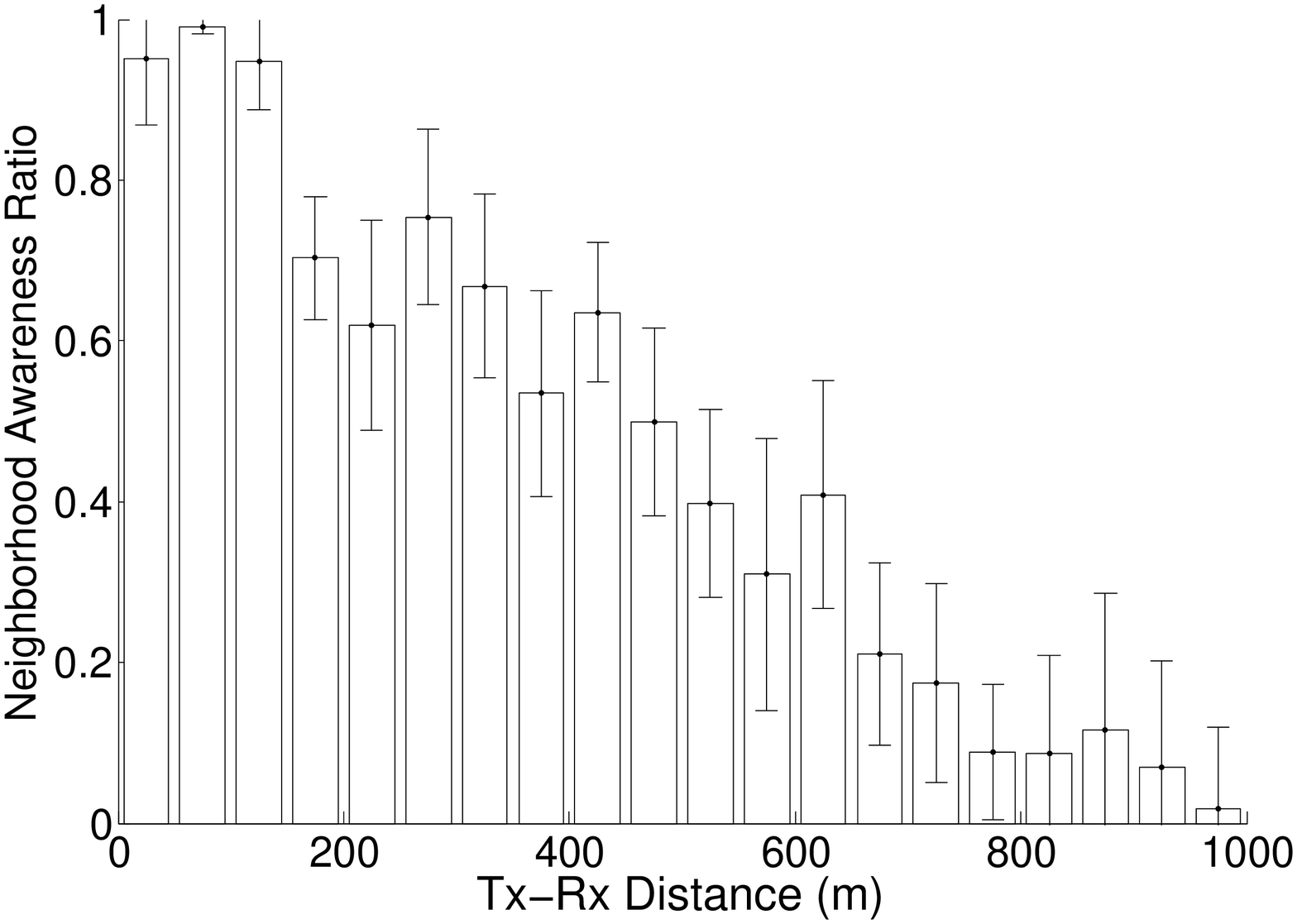}}
\subfigure[\scriptsize Italy -- Highway.]{\label{fig:NARIV2V}\includegraphics[width=0.24\textwidth]{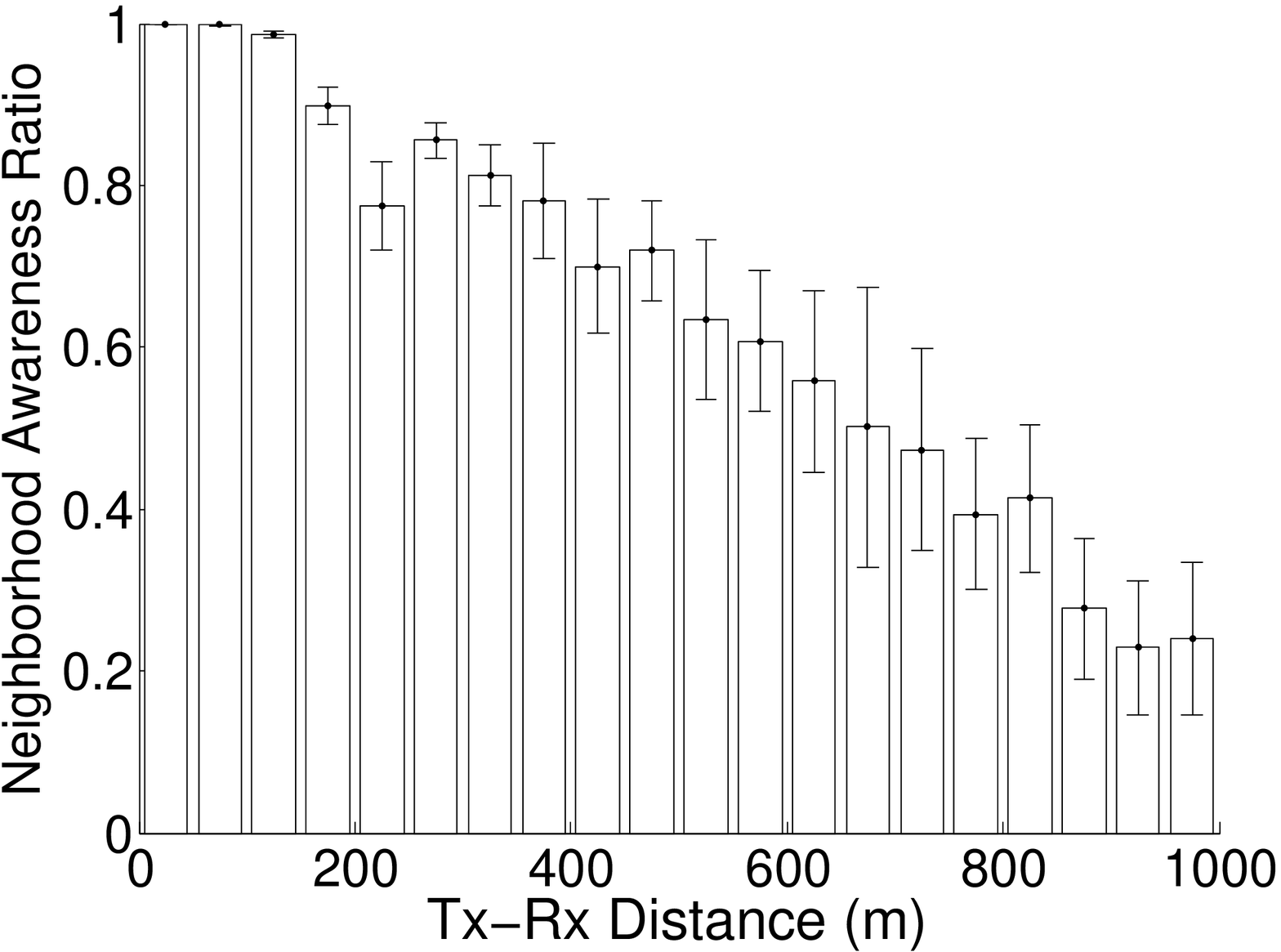}}
\subfigure[\scriptsize Finland -- Highway.]{\label{fig:NARFH}\includegraphics[width=0.24\textwidth]{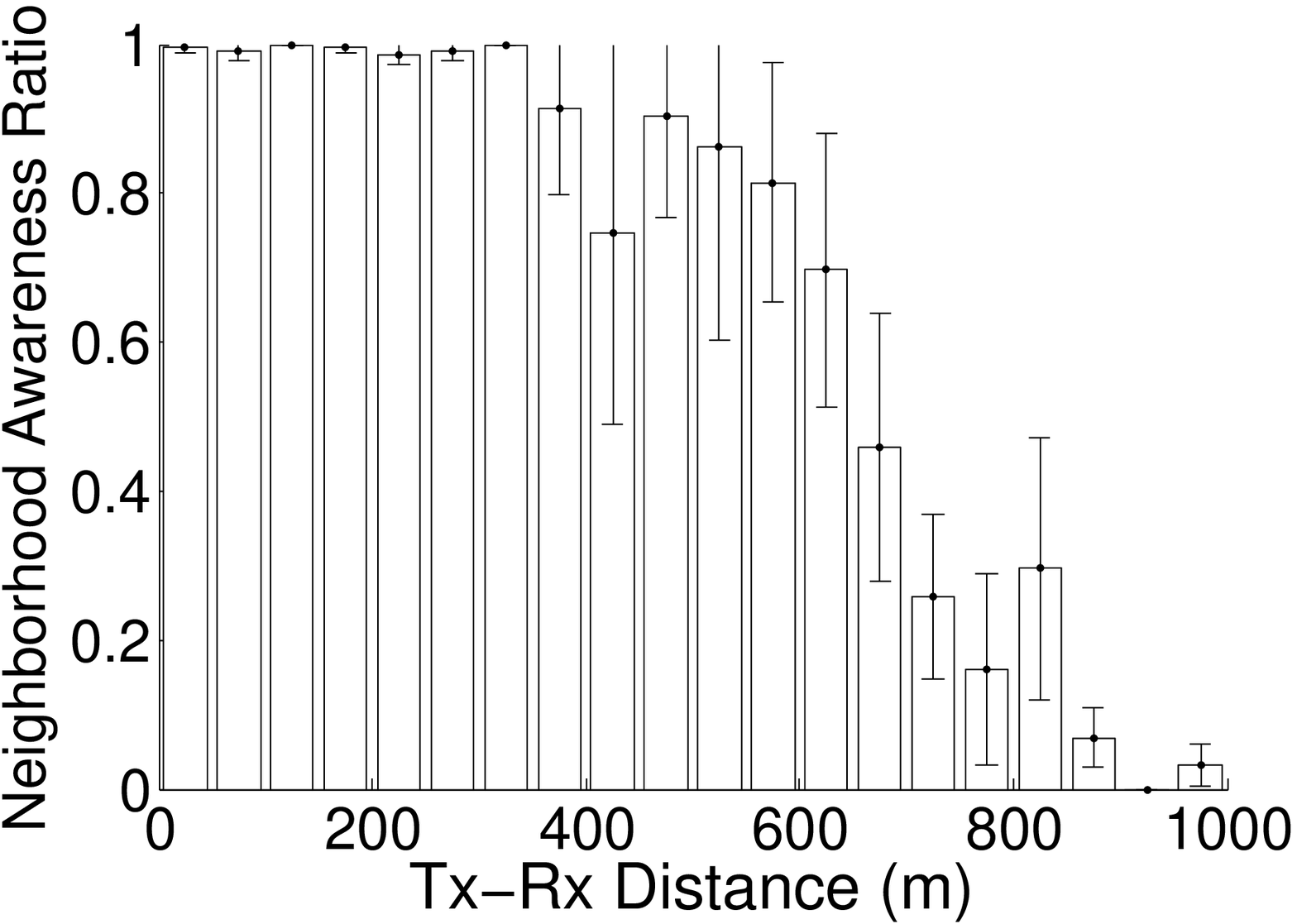}}
\subfigure[\scriptsize Finland -- Urban.]{\label{fig:NARFU}\includegraphics[width=0.24\textwidth]{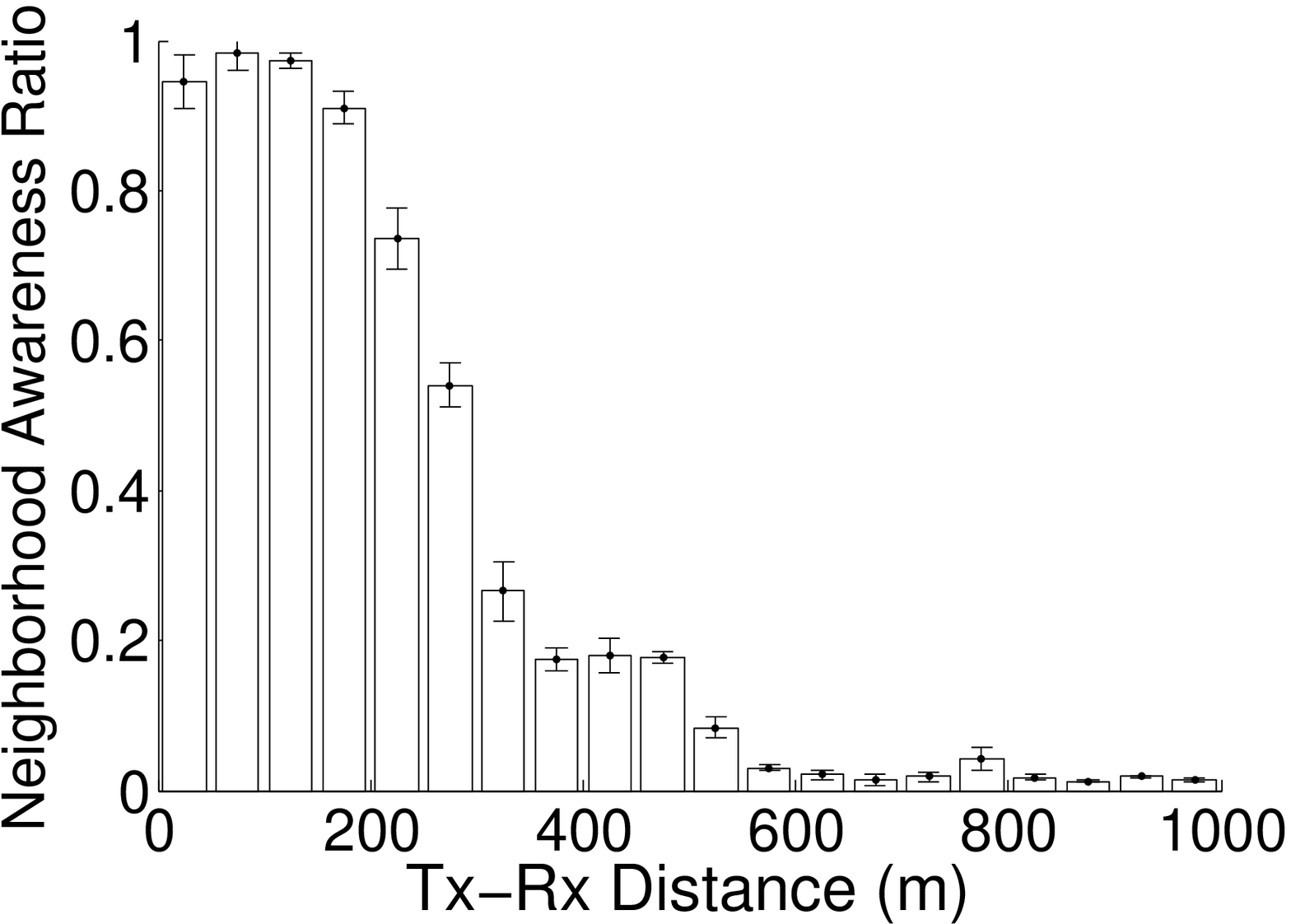}}
\subfigure[\scriptsize Finland -- Suburban.]{\label{fig:NARFS}\includegraphics[width=0.24\textwidth]{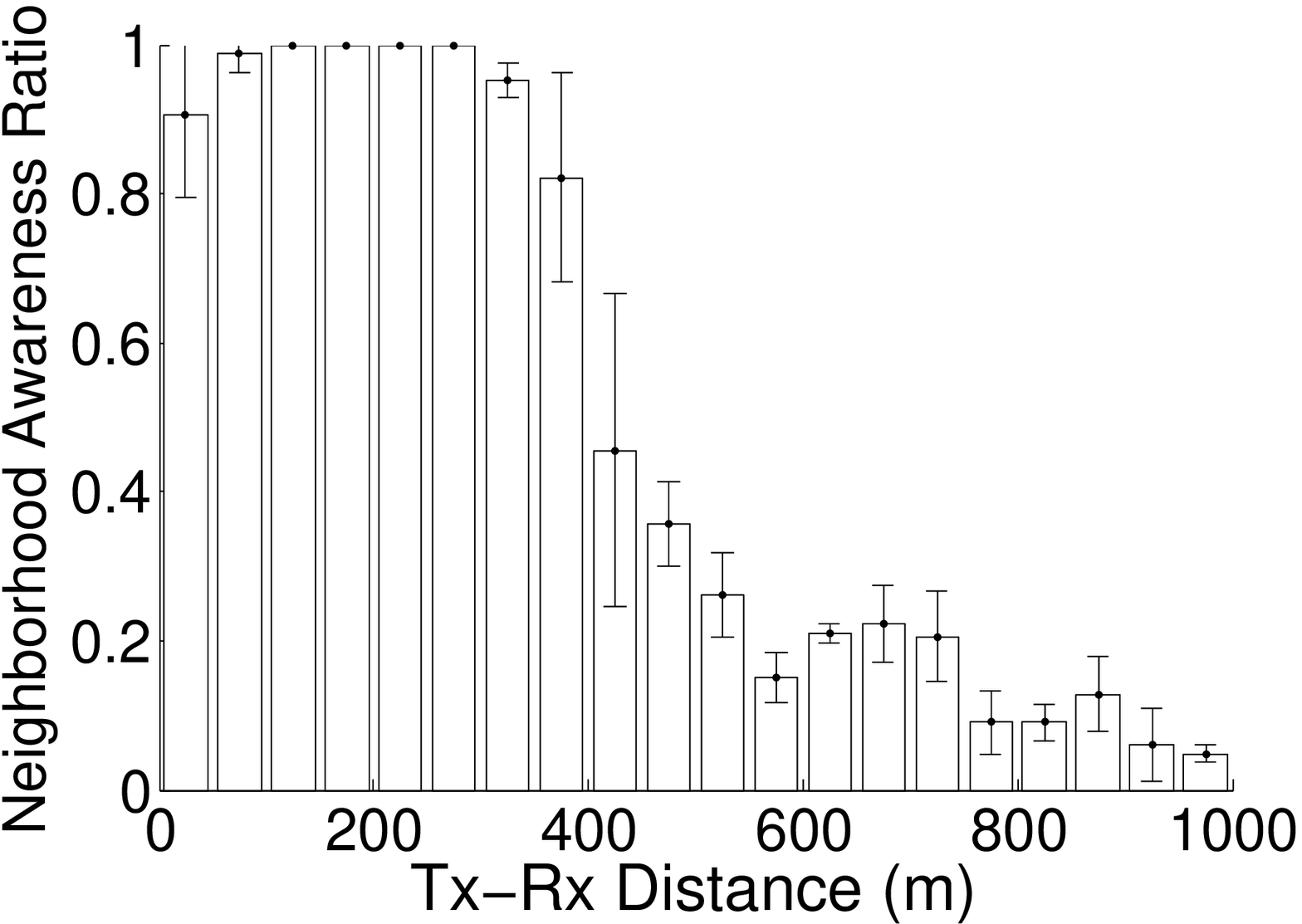}}
\caption{Overall V2V \acf{NAR} for  Test Site Sweden, the Netherlands, Italy and Finland.}\label{fig:NARV2VOverall} 
\end{figure*}

\begin{figure}
\centering
\subfigure[\scriptsize The Netherlands -- Highway.]{\label{fig:NARNHI}\includegraphics[width=0.49\linewidth]{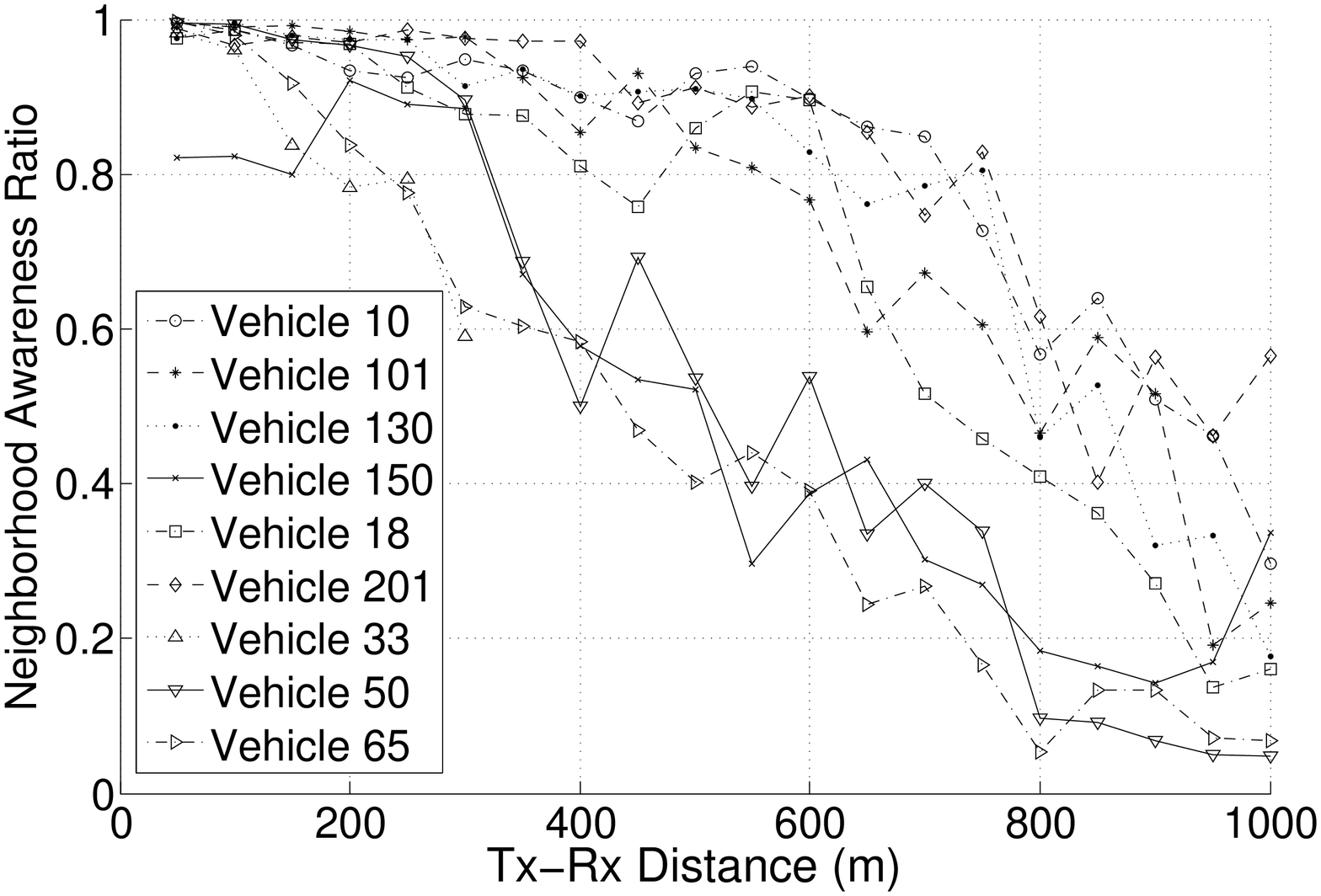}}
\subfigure[\scriptsize Italy -- Highway.]{\label{fig:NARIV2VI}\includegraphics[width=0.49\linewidth]{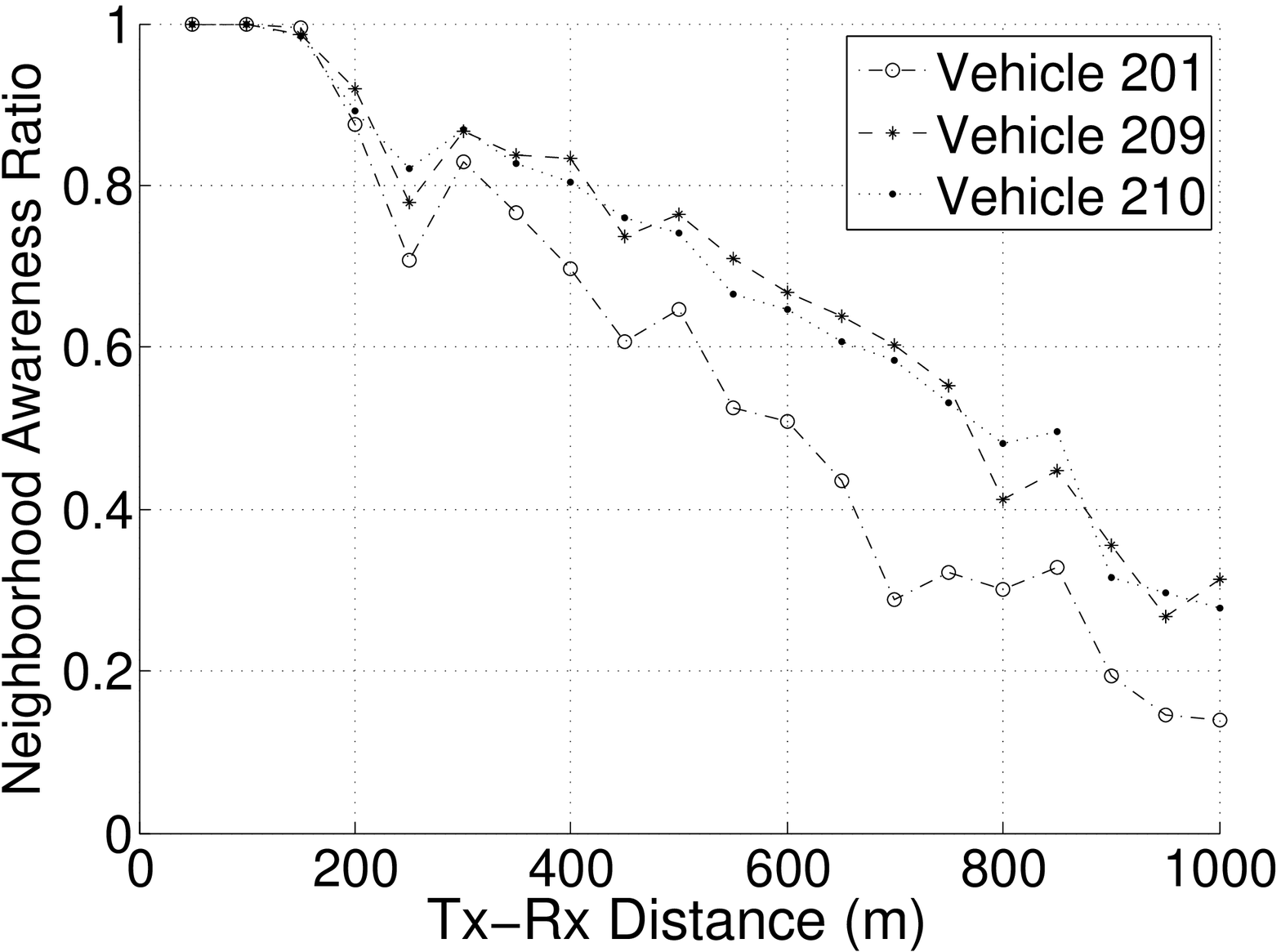}}
\caption{Per-vehicle V2V \acf{NAR} for Test Site the Netherlands and Italy.}\label{fig:NARV2VPerVeh} 
\end{figure}

\begin{figure}
\centering
\subfigure[\scriptsize Overall V2I results.]{\label{fig:NARIV2IO}\includegraphics[width=0.49\linewidth]{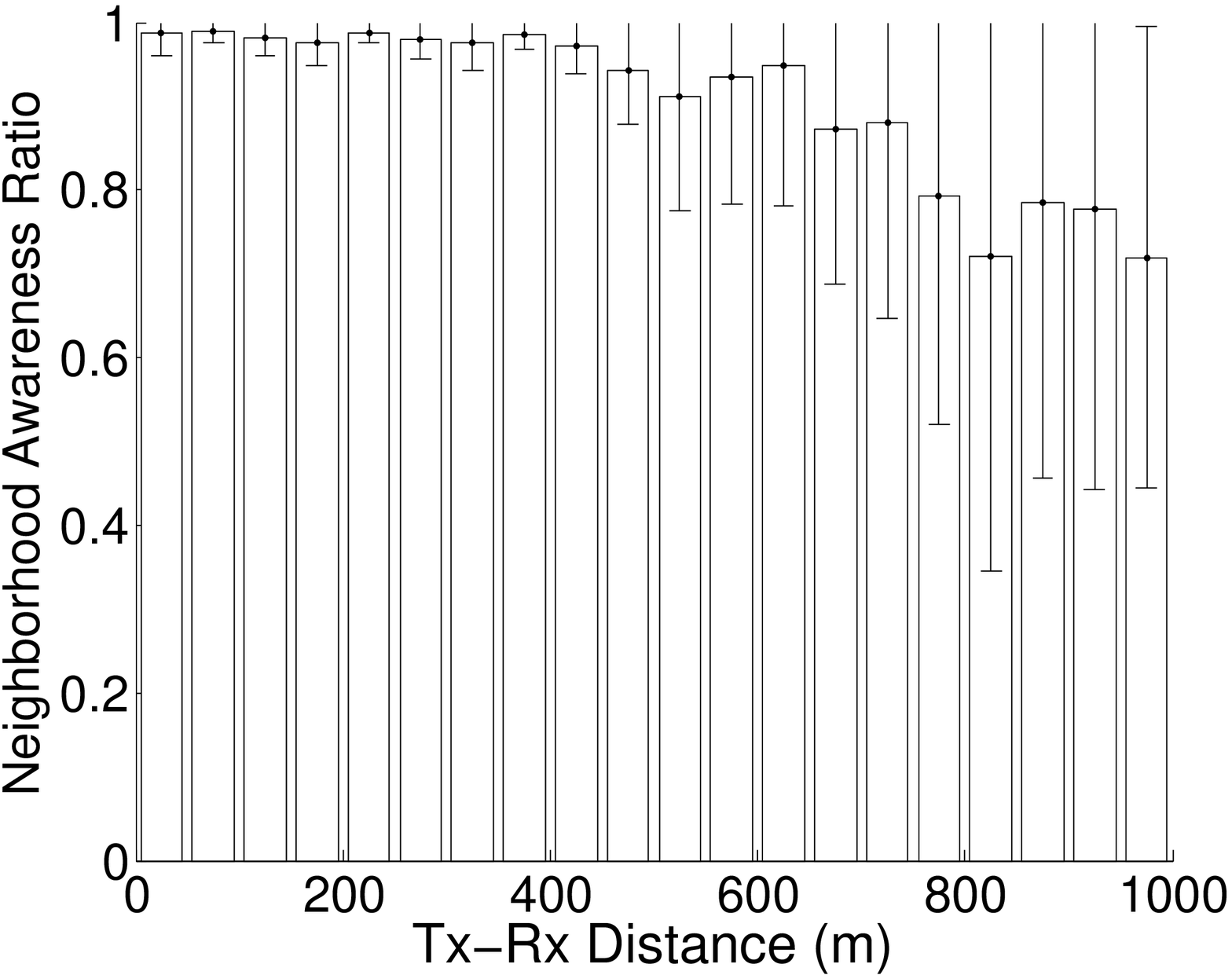}}
\subfigure[\scriptsize Per-vehicle V2I results.]{\label{fig:NARIV2II}\includegraphics[width=0.49\linewidth]{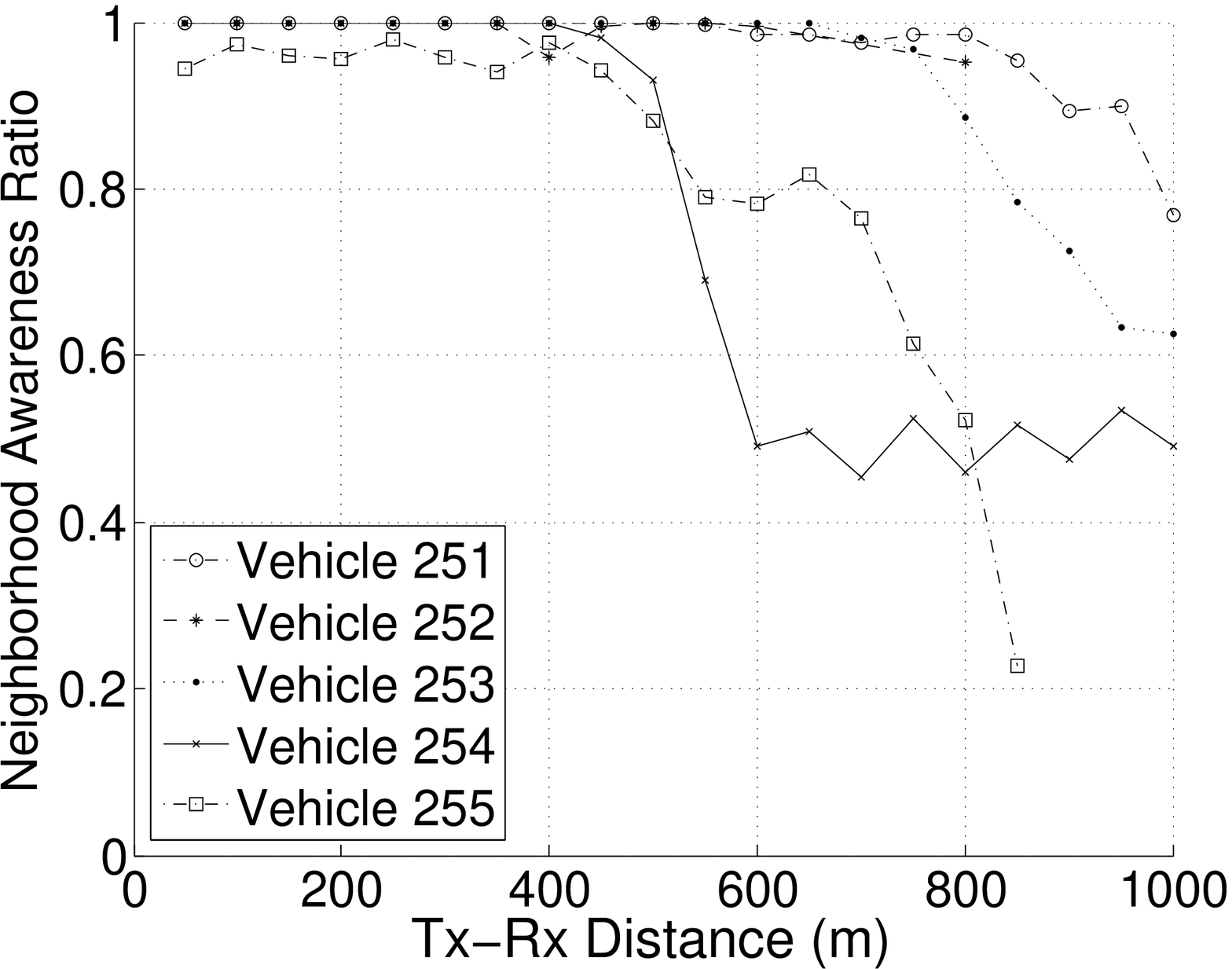}}
\caption{V2I \acf{NAR} for  Test Site Italy.}\label{fig:NARV2I} 
\end{figure}

Figs.~\ref{fig:NARV2VOverall},~\ref{fig:NARV2VPerVeh} and ~\ref{fig:NARV2I} present the \ac{NAR} results  for \ac{V2V} and \ac{V2I} communications
in different locations. As evidenced in our previous work~\cite{dorey14vtc}, there is a clear relation between \ac{PDR} and neighborhood awareness. 

\textbf{\ac{V2V}} --
Across test sites, the relationship between different environments and \ac{NAR} is quite clear: the more complex the environment, the lower the \ac{NAR} at a given distance. The most clear comparison can be seen on test site Finland (Table~\ref{tab:90perAwareness} and Figs.~\ref{fig:NARFH}, ~\ref{fig:NARFU} and ~\ref{fig:NARFS}): in urban environment, 90\% \ac{NAR} can be achieved at a maximum of 200~m, compared to 350~m and 400~m in suburban and highway environments, respectively. Furthermore, looking more deeply at Fig.~\ref{fig:NARV2VOverall}
 we can see that qualitative separation of environments into urban, suburban, and highway cannot be generalized across test sites, 
 which is in line with the \ac{PDR} results discussed in Section~\ref{subsec:PDR}.
 Therefore, a protocol that is able to dynamically adjust to the current environment would be useful for adapting the power of transmitted \acp{CAM}.

When analyzing the per-vehicle neighborhood results (Fig.~\ref{fig:NARV2VPerVeh}), we can observe that, for a given distance bin, the  performance fluctuations between different  vehicles is pronounced in all scenarios. This is the result of both the environment changes over small distance as well as different system setup on vehicles (e.g., antenna placement, cable loss).

\textbf{\ac{V2I}} --
Results for \ac{V2I} communications (Fig.~\ref{fig:NARV2I}) prove that the advantageous antenna positions and higher gain of \acp{RSU} antennas create a better propagation environment, which results in \ac{NAR} that is above 90\% up to 700+~m (Fig.~\ref{fig:NARIV2VI}).

\begin{table}[t!]
\centering
\caption{Distance Above Which \acf{NAR} Falls Below 90\% \label{tab:90perAwareness}}
\footnotesize{
\begin{tabular}{c c c c c}
\hline
\textbf{Environment} 	& \textbf{Sweden} 	& \textbf{The Netherlands} & \textbf{Finland} & \textbf{Italy} \\ 
\hline
Highway V2V 			& 100~m  			& 250~m  					& 400~m  	& 200~m\\
Suburban V2V 			& 100~m 			& 150~m						& 350~m 	& N/A \\
Urban V2V 				& N/A  				& N/A						& 200~m  	& N/A\\
Highway V2I 			& N/A 				& N/A						& N/A 		& 650~m \\ \hline
\end{tabular}
}
\end{table}

\subsubsection{\acf{RNAR}}\label{subsec:NIR} 

\begin{figure*}
\centering
\subfigure[\scriptsize Sweden -- Highway.]{\label{fig:NIRSH}\includegraphics[width=0.24\textwidth]{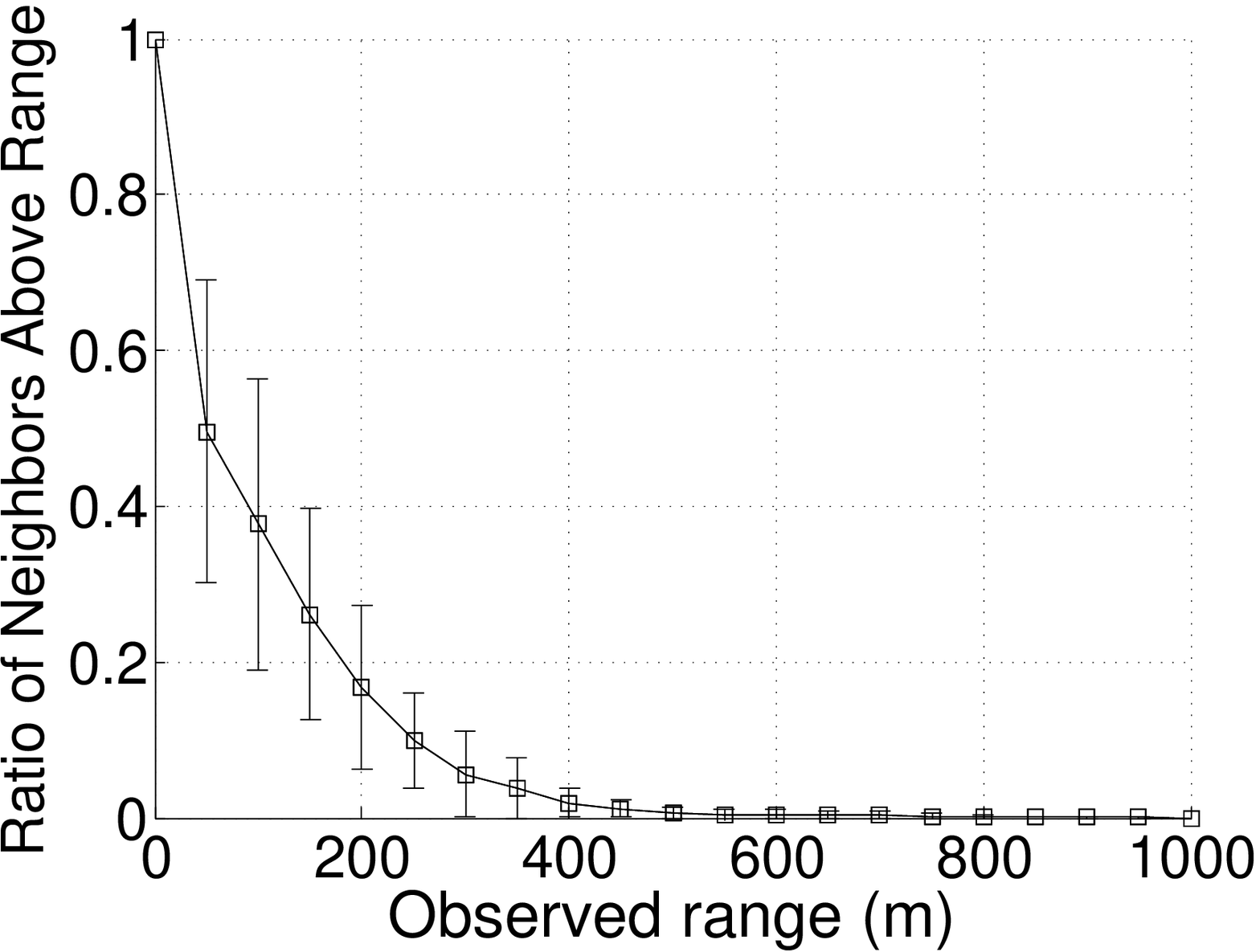}}
\subfigure[\scriptsize Sweden -- Suburban.]{\label{fig:NIRSS}\includegraphics[width=0.24\textwidth]{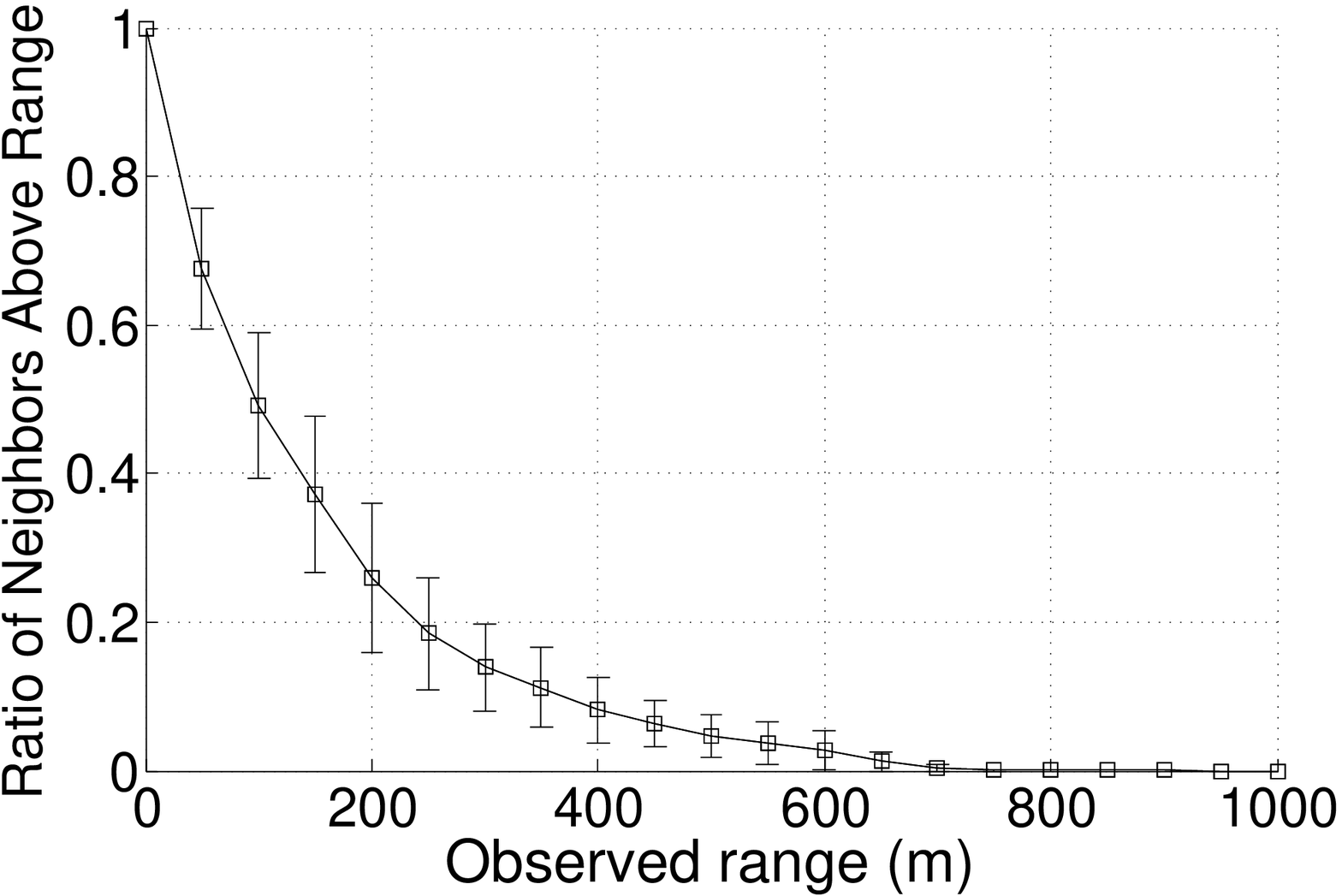}}
\subfigure[\scriptsize The Netherlands -- Highway.]{\label{fig:NIRNH}\includegraphics[width=0.24\textwidth]{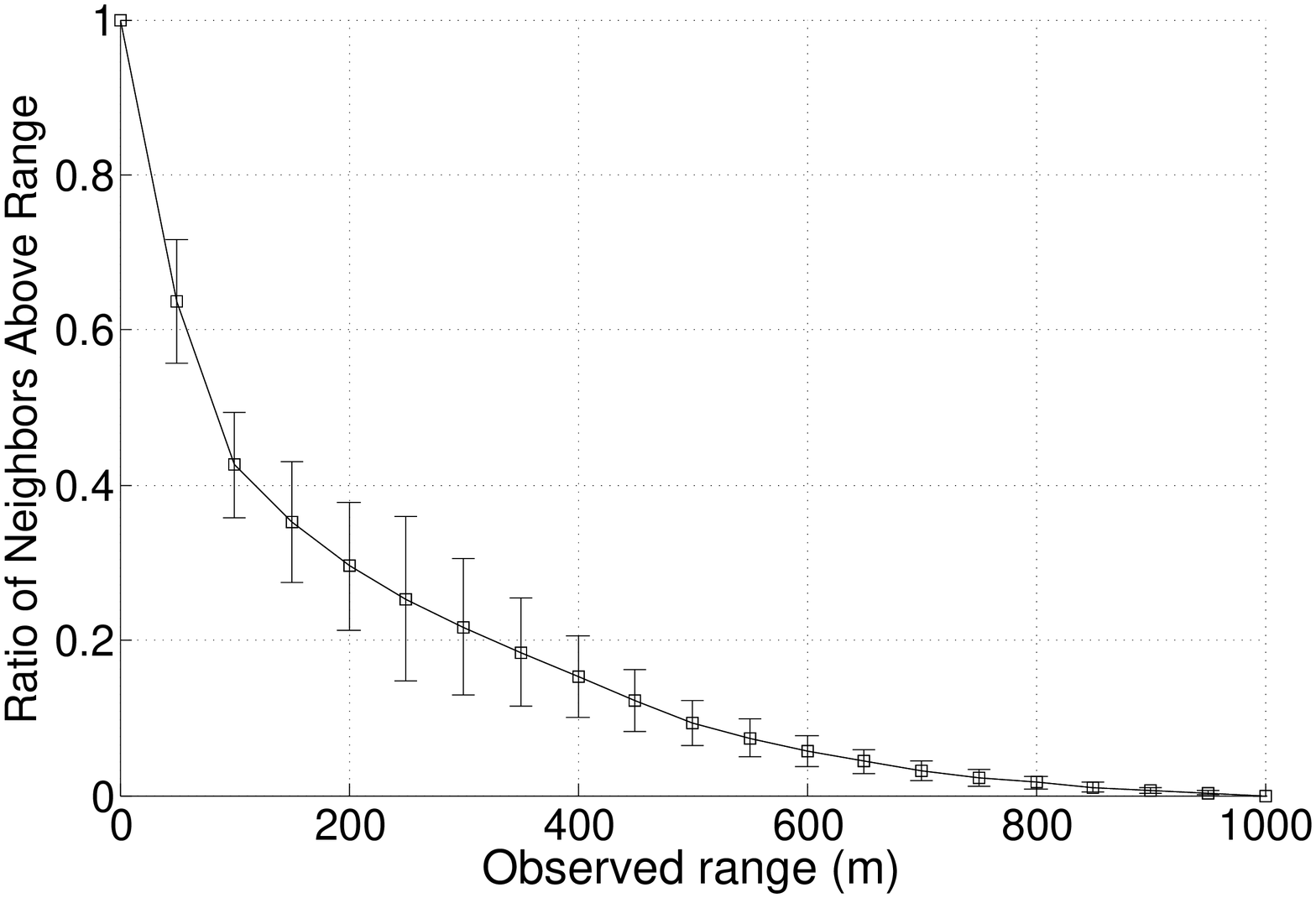}}
\subfigure[\scriptsize The Netherlands -- Suburban.]{\label{fig:NIRNS}\includegraphics[width=0.24\textwidth]{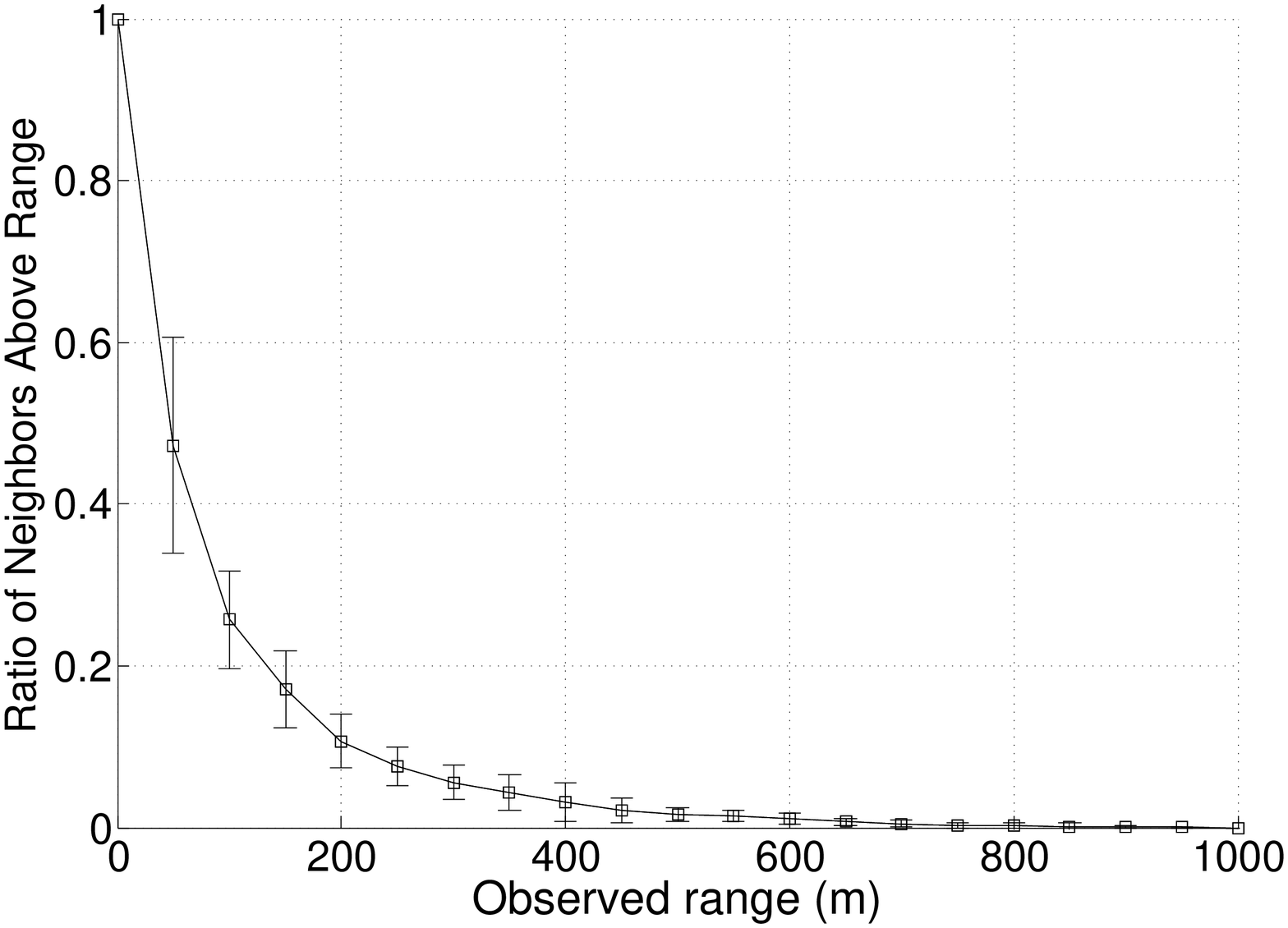}}
\subfigure[\scriptsize Italy -- Highway.]{\label{fig:NIRV2VIH}\includegraphics[width=0.24\textwidth]{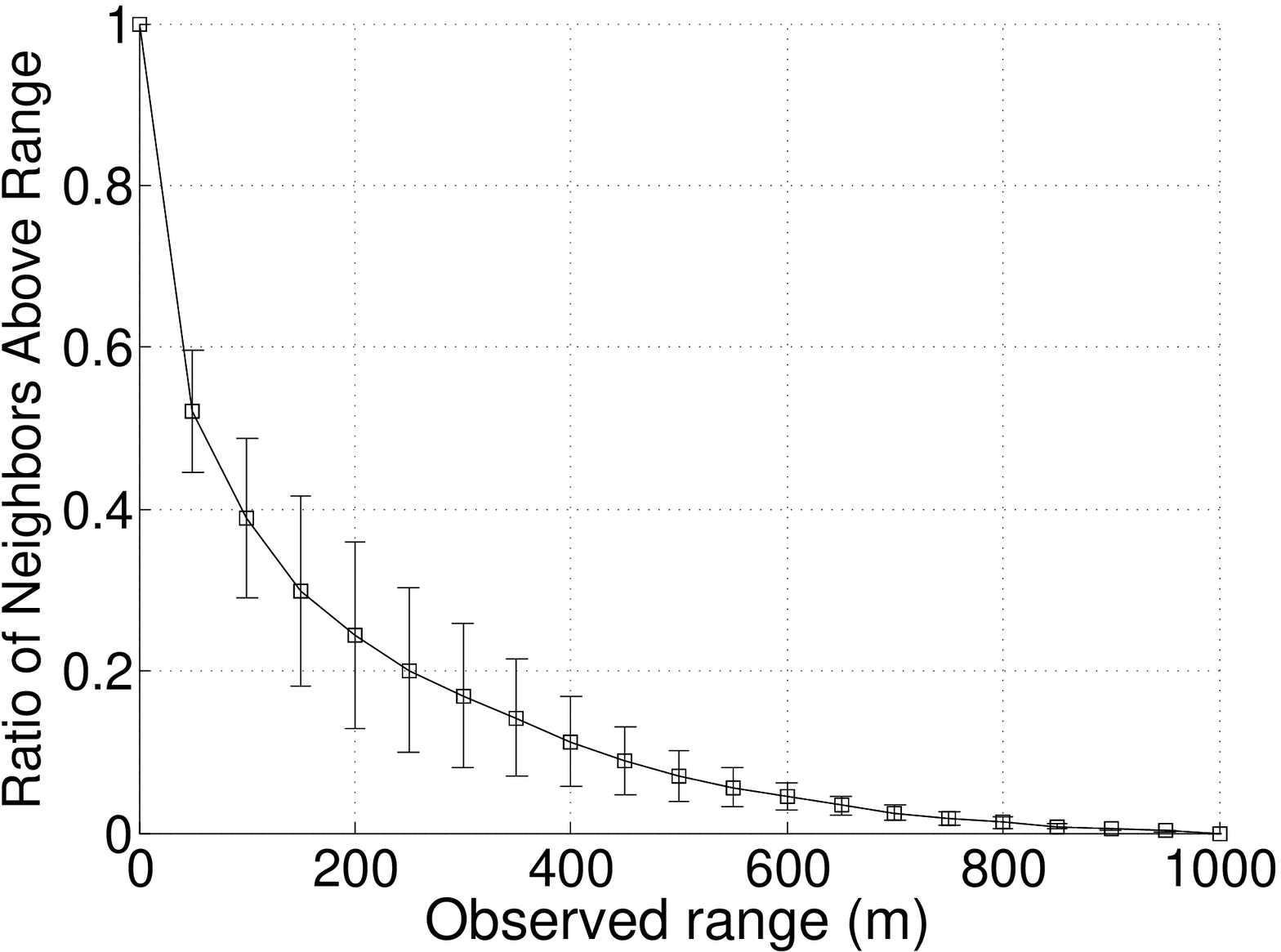}}
\subfigure[\scriptsize Finland -- Highway.]{\label{fig:NIRFH}\includegraphics[width=0.24\textwidth]{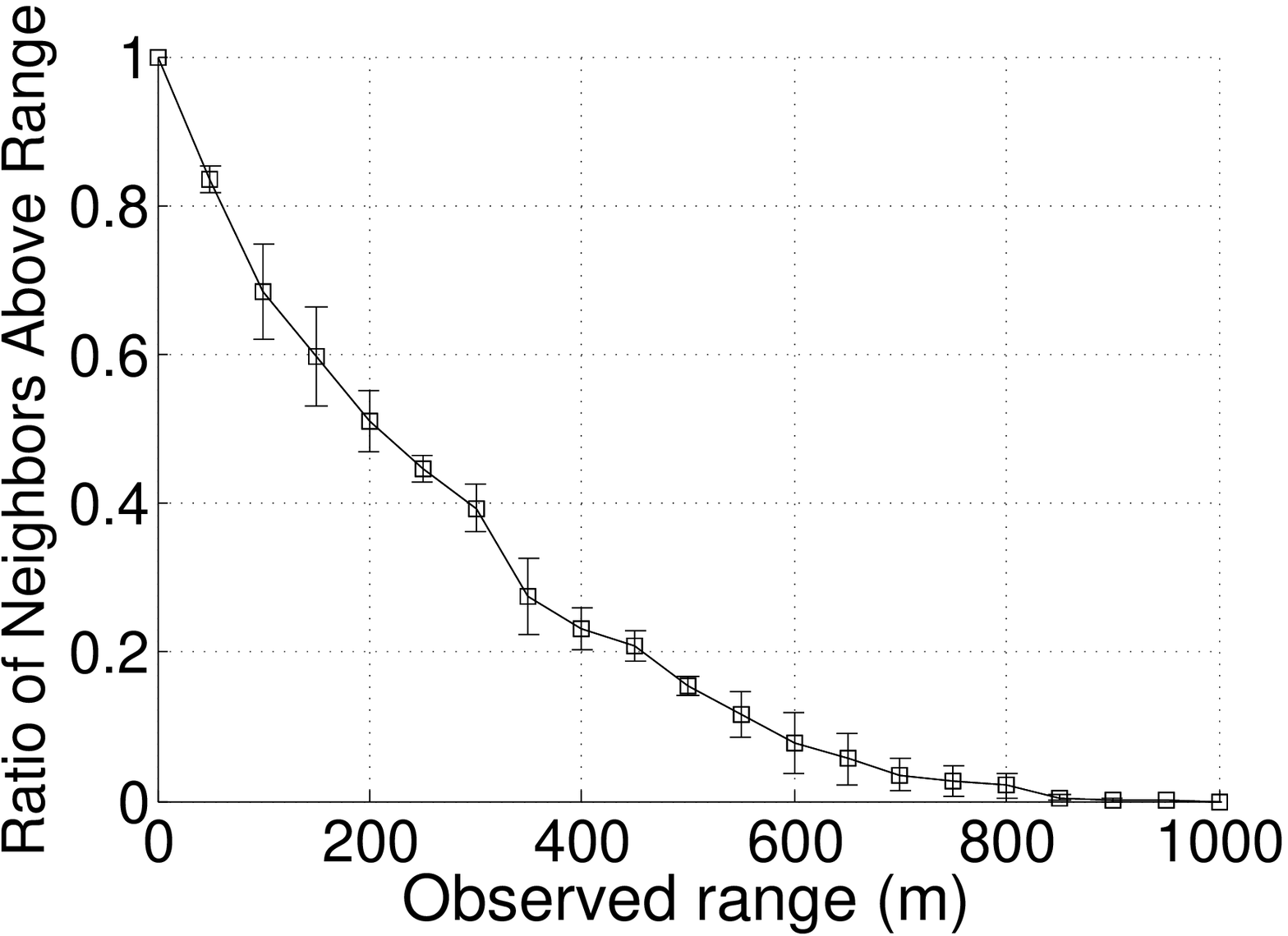}}
\subfigure[\scriptsize Finland -- Urban.]{\label{fig:NIRFU}\includegraphics[width=0.24\textwidth]{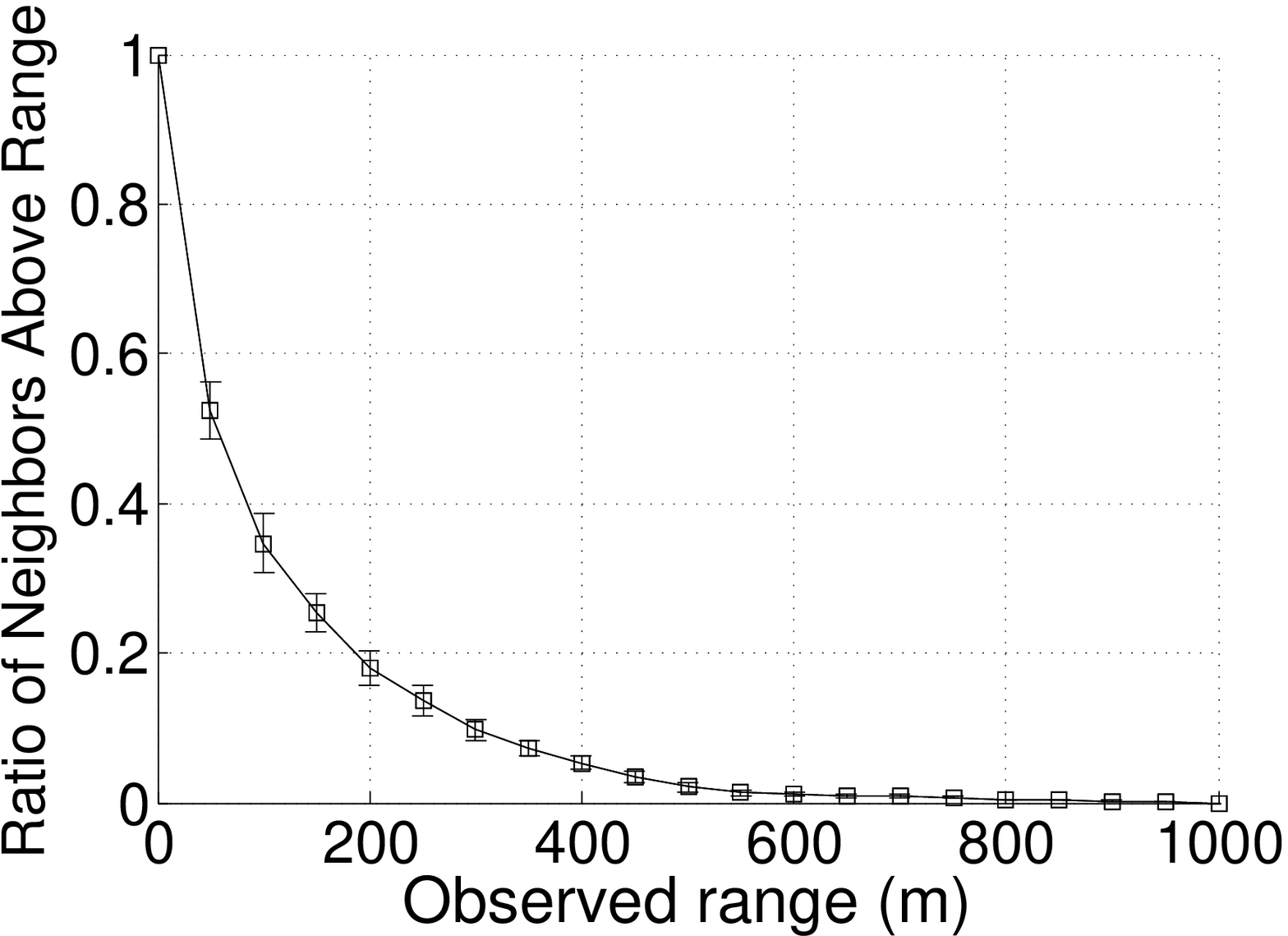}}
\subfigure[\scriptsize Finland -- Suburban.]{\label{fig:NIRFS}\includegraphics[width=0.24\textwidth]{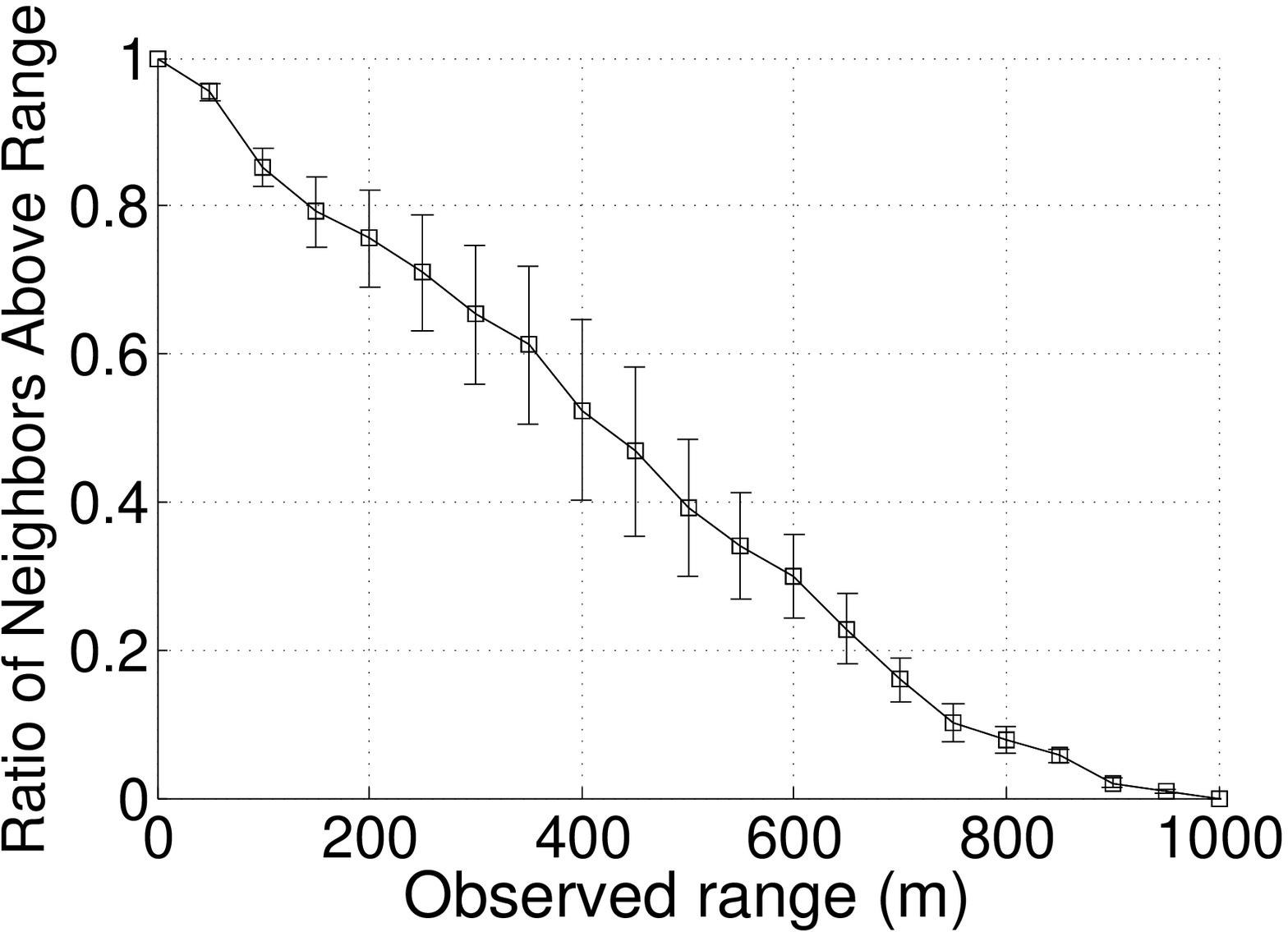}}
\caption{Overall V2V \acf{RNAR} for  Test Site Sweden, the Netherlands, Italy and Finland. \label{fig:NIRV2VOverall}} 
\end{figure*}
\begin{figure}
\centering
\subfigure[\scriptsize The Netherlands -- Highway.]{\label{fig:NIRNHI}\includegraphics[width=0.49\linewidth]{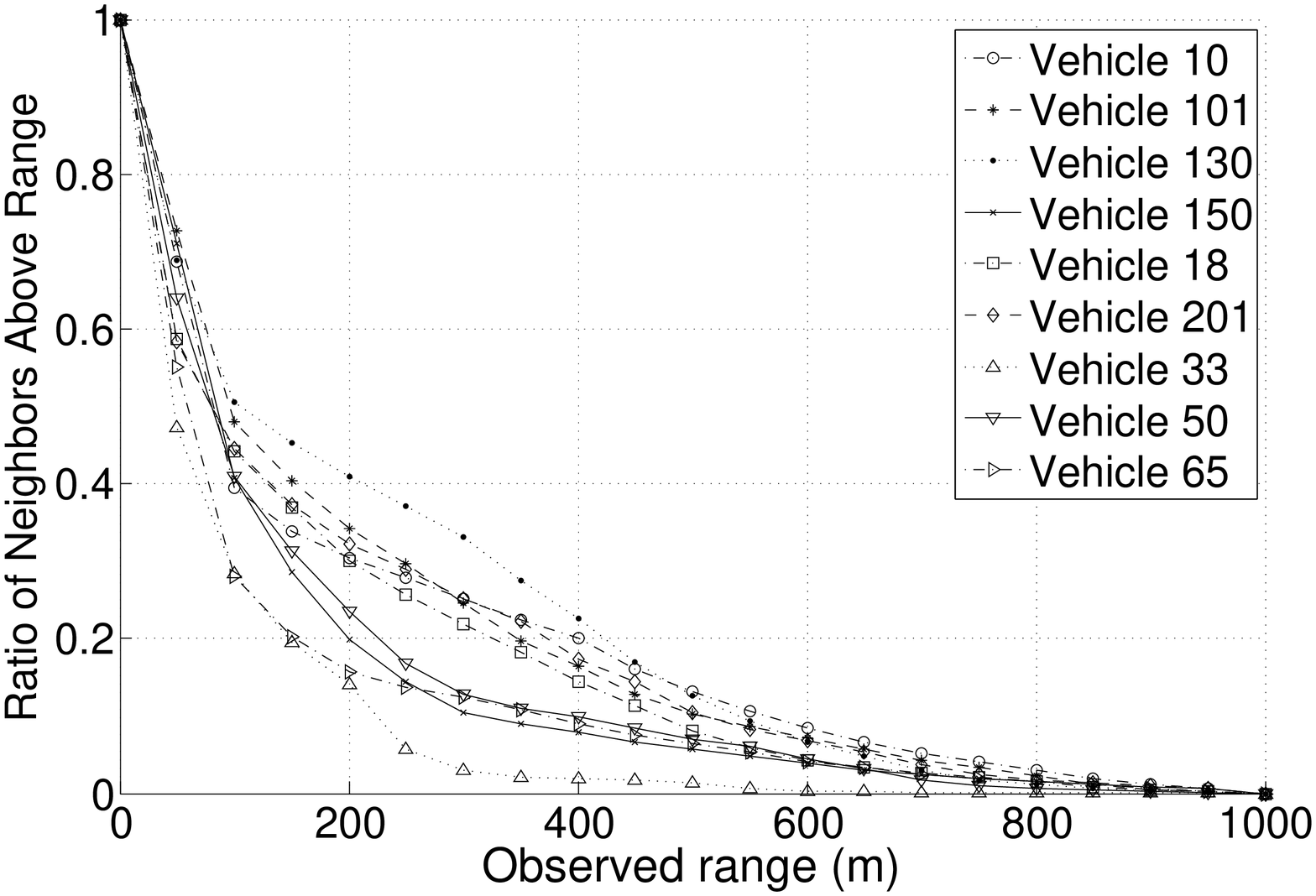}}
\subfigure[\scriptsize Italy -- Highway.]{\label{fig:NIRV2VIHI}\includegraphics[width=0.49\linewidth]{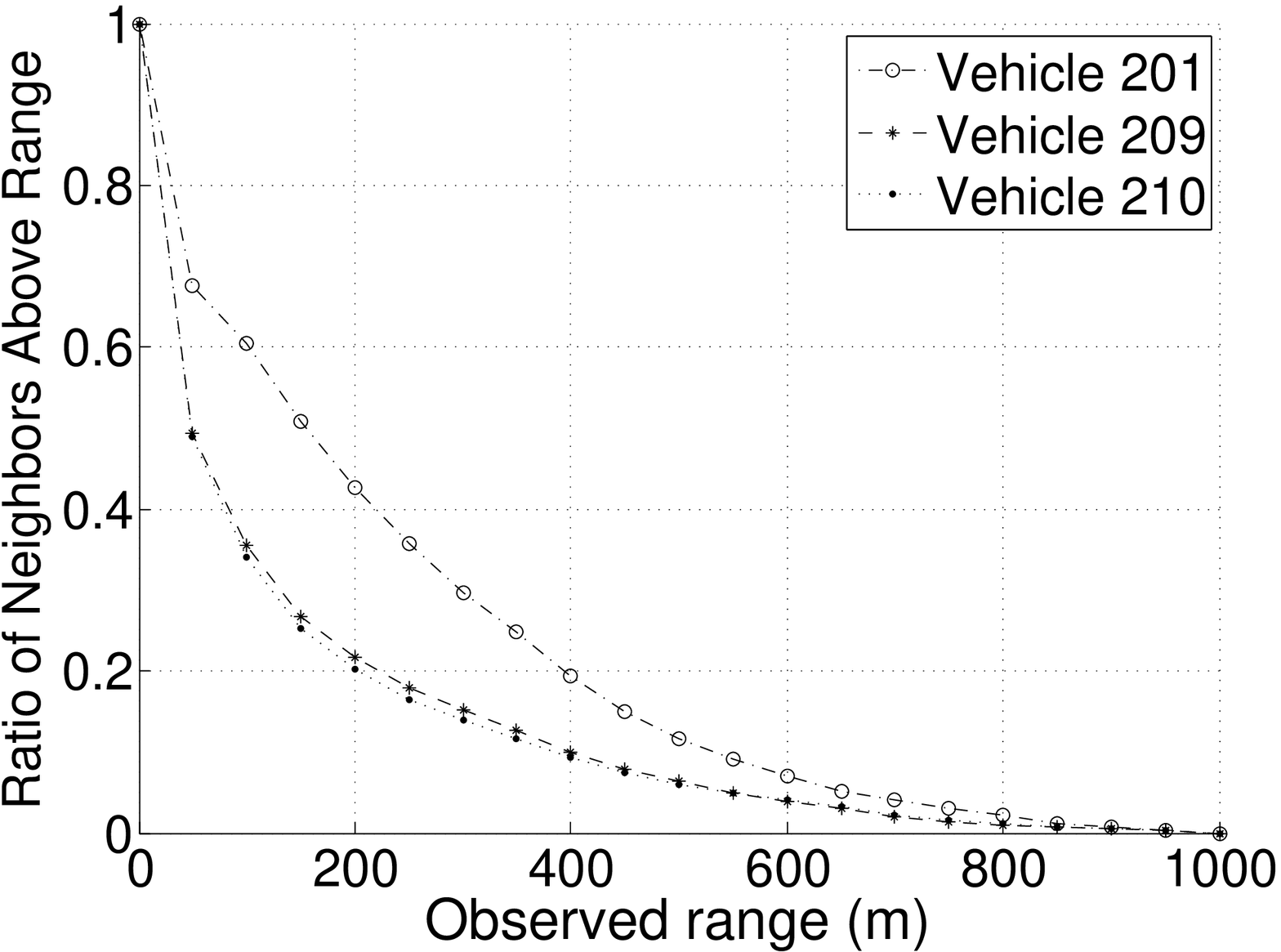}}
\caption{Per Vehicle \acf{RNAR} for Test Site The Netherlands and Italy. \label{fig:NIRV2VPerVeh}} 
\end{figure}
\begin{figure}
\centering
\subfigure[\scriptsize Overall V2I results.]{\label{fig:NIRV2IO}\includegraphics[width=0.49\linewidth]{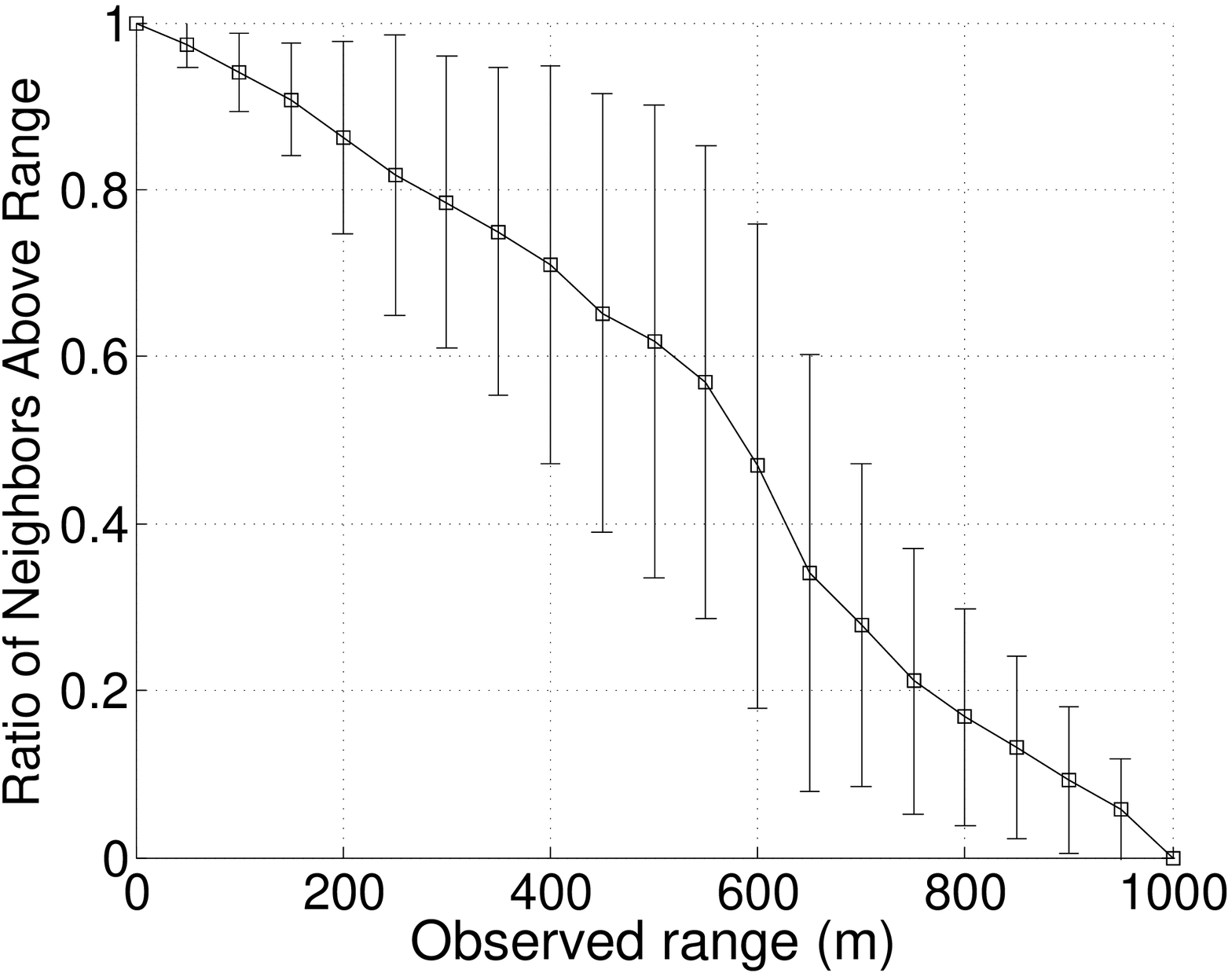}}
\subfigure[\scriptsize Per-vehicle V2I results.]{\label{fig:NIRV2IP}\includegraphics[width=0.49\linewidth]{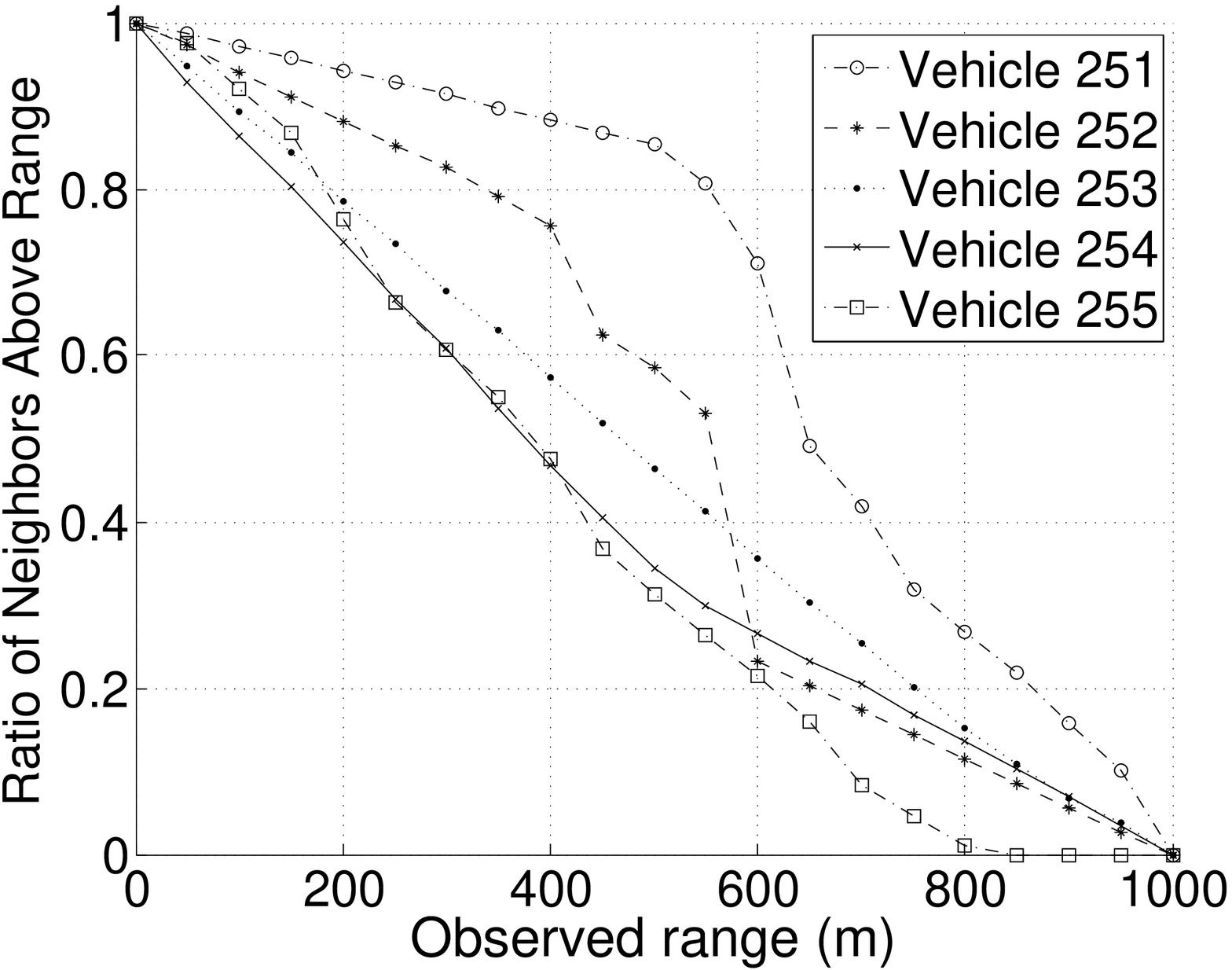}}
\caption{V2I \acf{RNAR} for Test Site Italy. \label{fig:NIRV2I}} 
\end{figure}

Figs.~\ref{fig:NIRV2VOverall},~\ref{fig:NIRV2VPerVeh} and ~\ref{fig:NIRV2I} show the \acf{RNAR} for different test sites.

\textbf{\ac{V2V}} -- \ac{RNAR} exhibits an exponentially decreasing behavior, with progressively fewer vehicles detected at higher distances (e.g., proportion of vehicles above 400~meters mostly contained within 10\%). For safety applications requiring information from immediate neighborhood, such behavior is beneficial, since it implies that most periodic messages that a vehicle receives are useful. While the trend of \ac{RNAR} is similar across the environments, 
different surroundings and effective transmit powers lead to
significantly different \ac{RNAR} values. For instance, for a highway scenario and a reference distance of 200~m, 
\ac{RNAR} is 20\% in Sweden (Fig.~\ref{fig:NIRSH}) and 50\% Finland (Fig.~\ref{fig:NIRFH}.

\textbf{\ac{V2I}} -- Whereas in \ac{V2V} scenarios, the \ac{RNAR} tapers off after at most 500~m, the large effective range of \acp{RSU} results in a large number of detected far-away vehicles (e.g., more than half of detected vehicles were farther than 500~m away in Fig.~\ref{fig:NIRV2IO}). 
As explained previously, the large \ac{RSU} range arises from their advantageous positions on tall gantries and higher-gain antennas. 

\subsubsection{Discussion}
Measurement results  show that \ac{V2V} links with low effective transmit power can suffer from low neighborhood awareness, particularly in built-up urban areas; at the same time, \ac{V2I} links can exhibit high awareness rates even above 1~km. On one hand, it is questionable if the neighborhood awareness information is relevant at distances above those required by safety-critical applications. High awareness is closely related to the potentially high interference, which reduces the frequency reuse and negatively impacts the throughput of future vehicular networks.
On the other hand, within distances relevant for safety applications, there is a need for as high awareness as possible. 

%% file: modeling.tex
In this section, we design a model that can provide insight into the effectiveness of future cooperative awareness. By using the information on \ac{PDR} only, the model is able predict awareness in terms of \ac{NAR}. In addition to giving system designers a quick insight
into the effectiveness of cooperative awareness message exchange for a given environment, the model allows incorporating awareness into 
mathematical models for optimizing cooperative message sending.

\begin{figure*}[t]
\centering
\subfigure[\scriptsize Test Site Finland -- Highway.]{\label{fig:NARMFinHw}\includegraphics[width=0.3\textwidth]{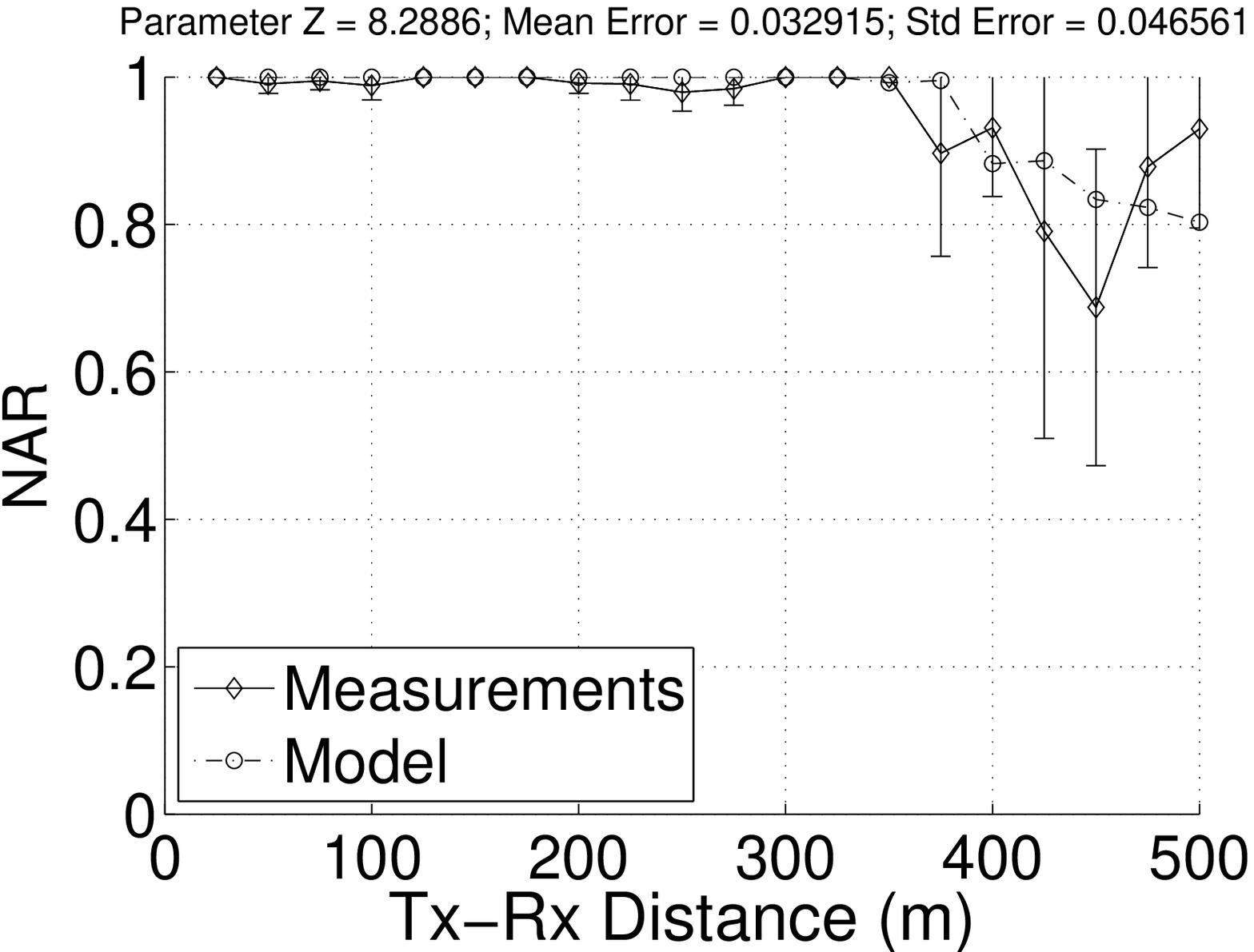}}
\subfigure[\scriptsize Test Site Finland -- Suburban.]{\label{fig:NARMFinSub}\includegraphics[width=0.3\textwidth]{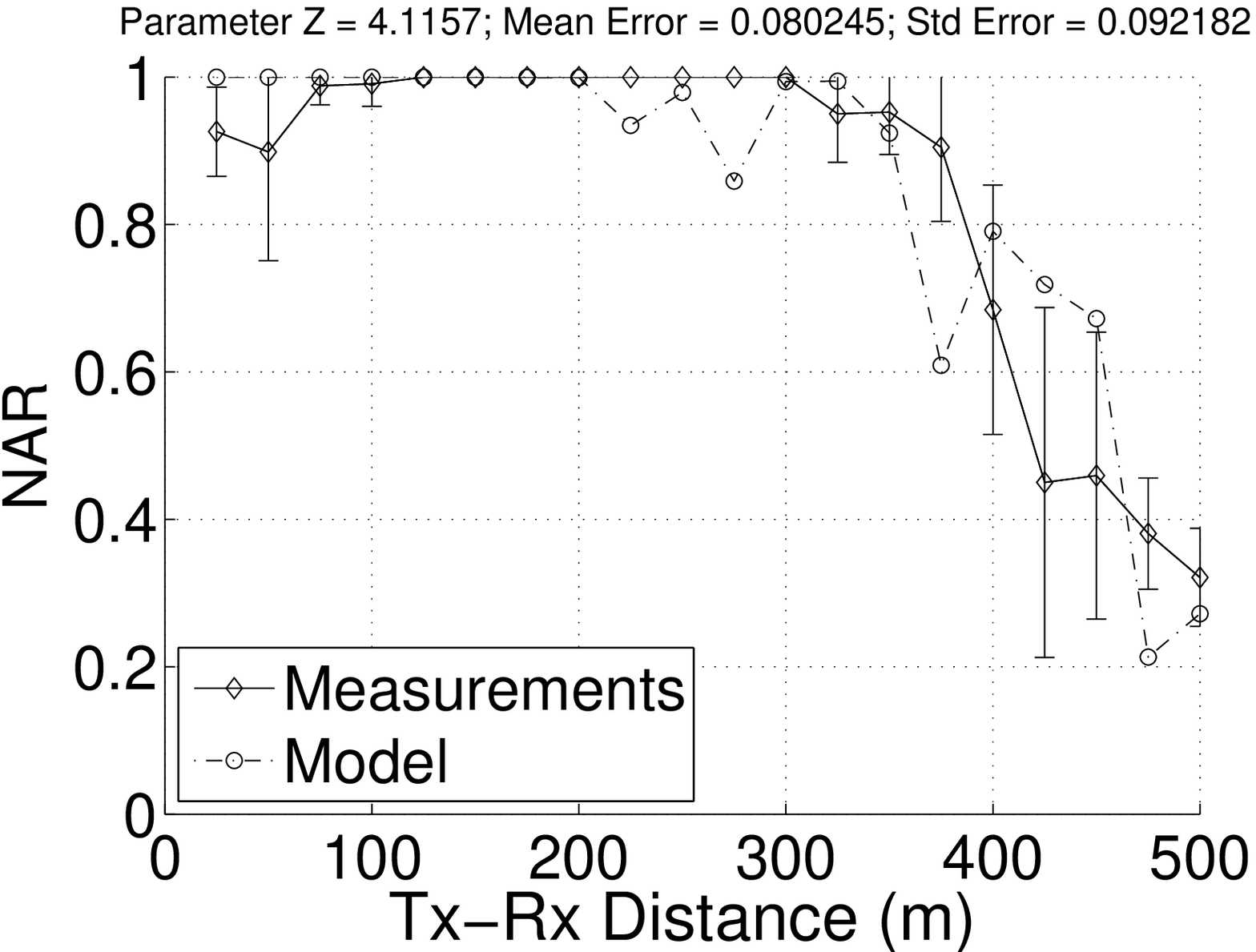}}
\subfigure[\scriptsize Test Site Finland -- Urban.]{\label{fig:NARMFinUrb}\includegraphics[width=0.3\textwidth]{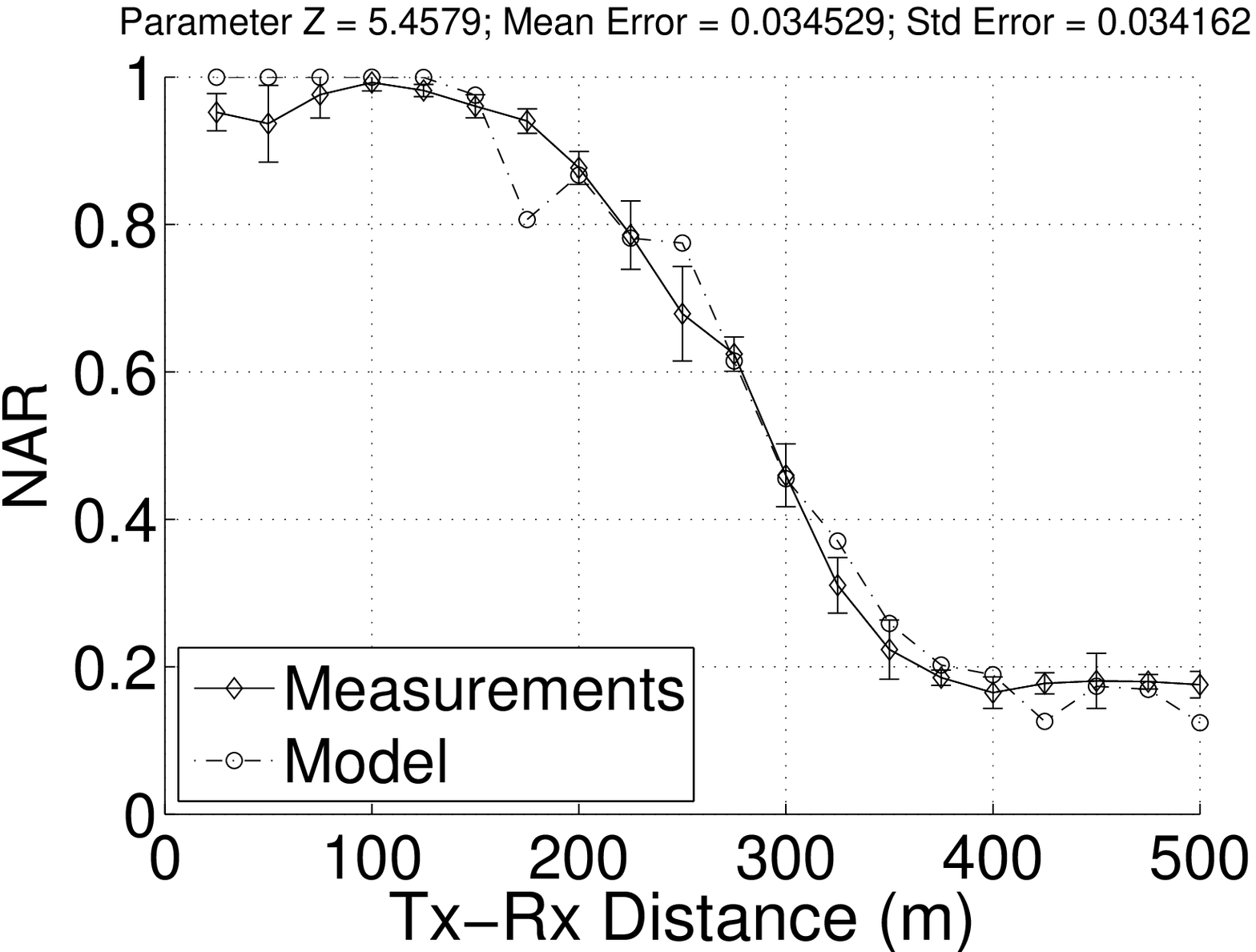}}
\subfigure[\scriptsize Test Site Italy -- Highway.]{\label{fig:NARMI}\includegraphics[width=0.3\textwidth]{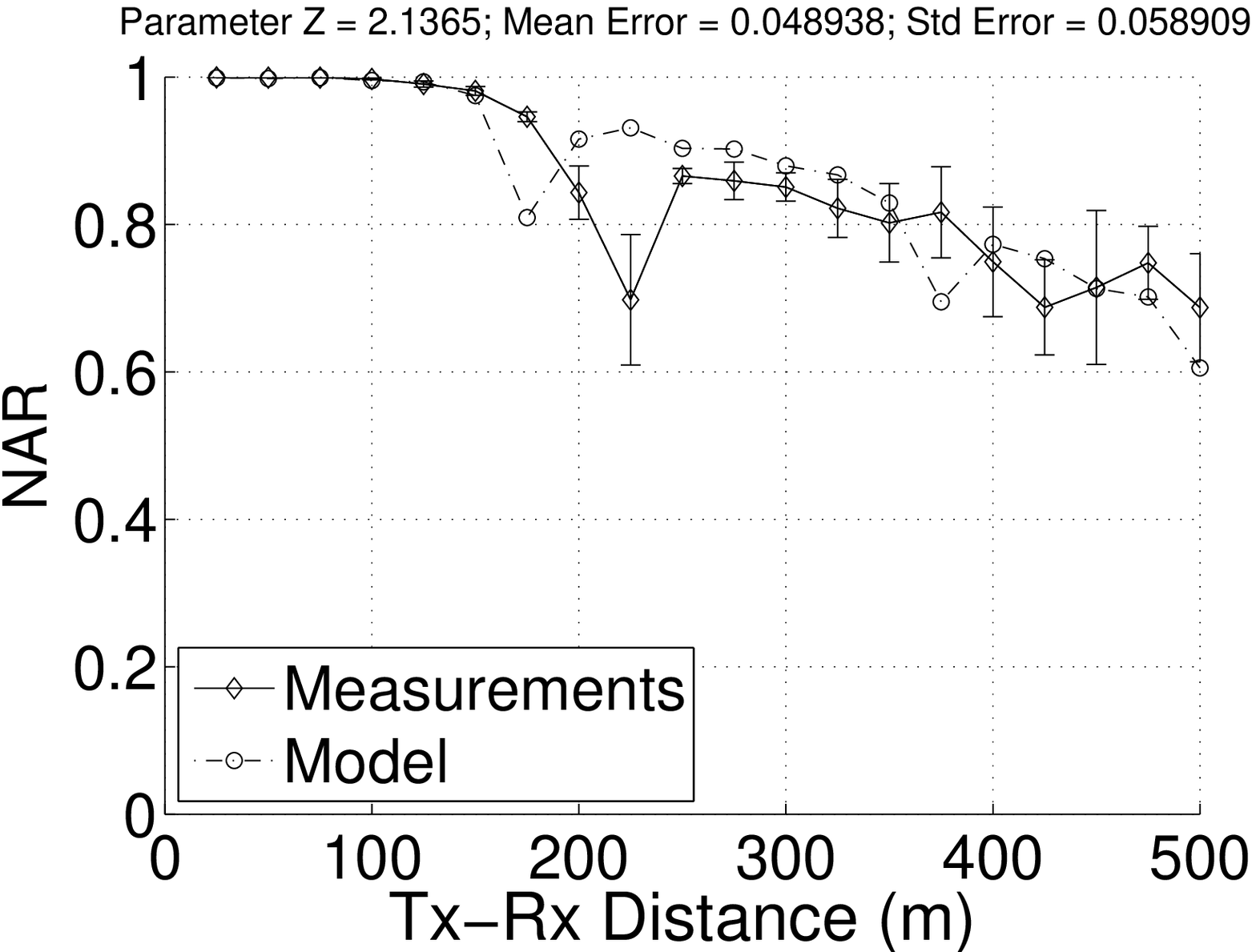}}
\subfigure[\scriptsize Test Site Sweden -- Overall.]{\label{fig:NARMS}\includegraphics[width=0.3\textwidth]{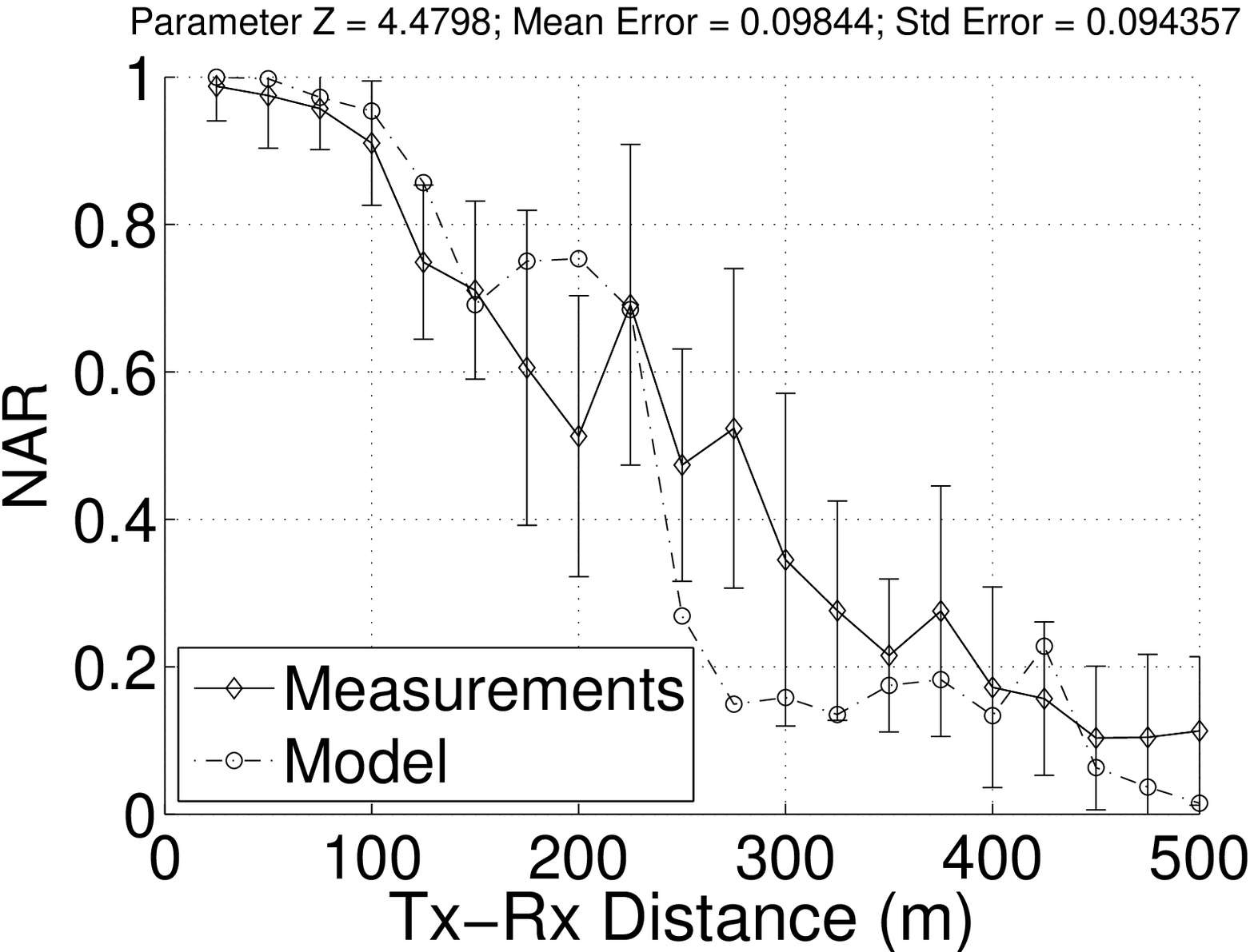}}
\subfigure[\scriptsize Test Site The Netherlands -- Overall.]{\label{fig:NARMN}\includegraphics[width=0.3\textwidth]{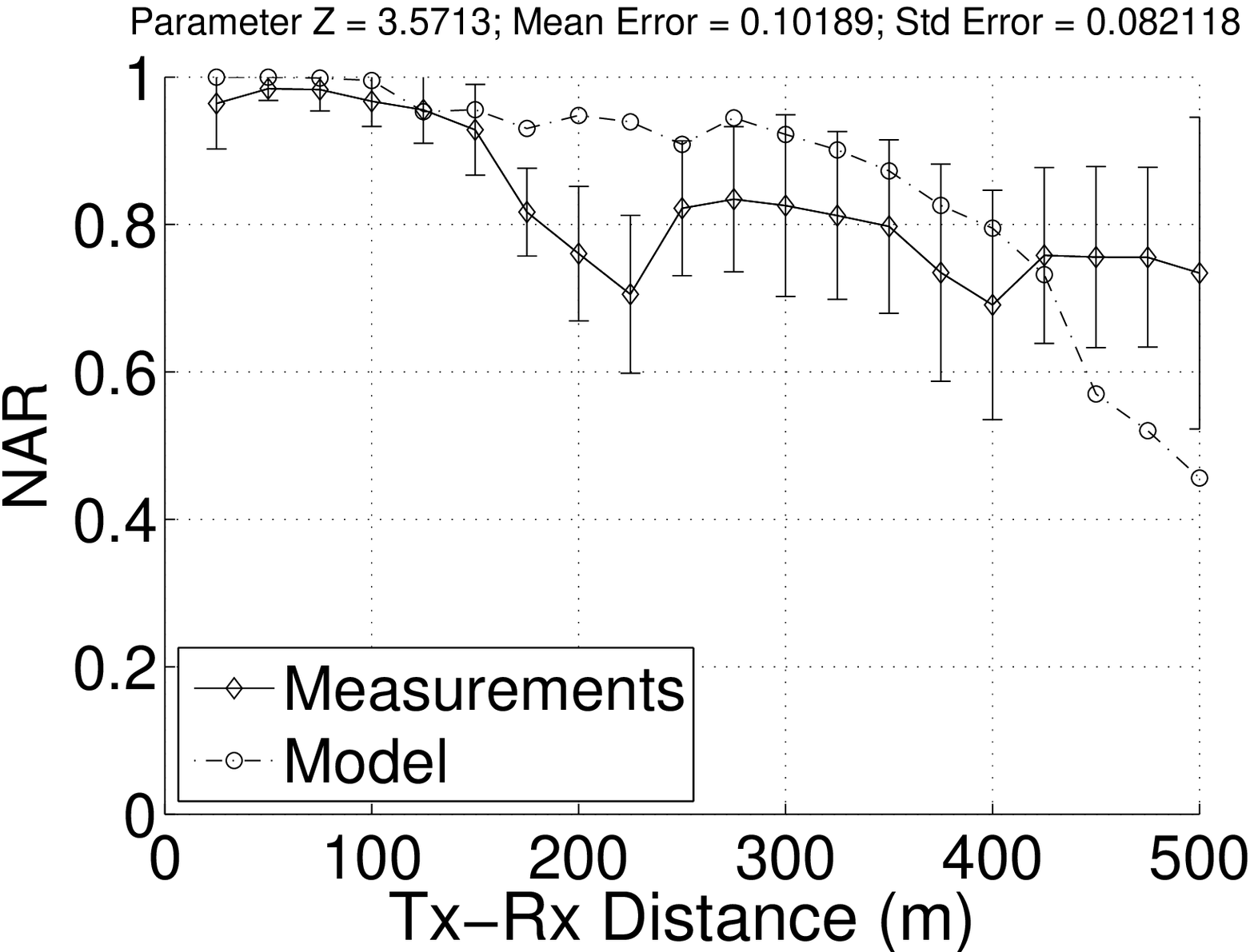}}
\caption{\acf{NAR} results: measurements vs. model. Mean and standard error is expressed in terms of absolute difference in \ac{NAR} between the measurements and the model. The error bars for the measurement data represent one standard deviation around the mean.} \label{fig:NARModelAll}
\end{figure*}

We start by observing that the inter-reception time (IRT)
 of cooperative messages (i.e., the time interval between successful packet receptions) depends on the number of effectively 
lost packets. If the probability of packet reception $p$ (which is tantamount to \ac{PDR}) is assumed to be constant, IRT follows a geometric
distribution~\cite{Tielert2013JPC}:

\begin{align}
P(IRT=k) &= (1-p)^{k-1}p
\end{align}

\ac{NAR} is defined as probability of receiving \emph{at least} one message from a vehicle in time $t$ (see Section~\ref{sec:Introduction}). In other words,  
\emph{at least} one out of $N$ sent messages in time $t$ needs to be received to make the receiving vehicle aware of the sending vehicle. Therefore, we model \ac{NAR} using the cumulative geometric distribution over the number of CAM transmissions, where the probability of success for each CAM transmission is equal to the \ac{PDR} for distance $r$, $PDR_r$:
\begin{align}
NAR_{r,t} &= \sum_{k=1}^N(1-PDR_{r})^{k-1} \times PDR_{r}.
\end{align}
The above expression can also be written in terms of the probability that all messages in time $t$ from the transmitting vehicle fail to reach the designated recipient:
\begin{align}
NAR_{r,t} &= 1-(1-PDR_{r})^N.
\end{align}

\begin{figure}
  \begin{center}
      \subfigure[\scriptsize Test Site Sweden -- Overall. The value of $Z$ for the fit: 4.4798.]{\label{fig:NARPDRrelationshipSweden}\includegraphics[width=0.49\linewidth]{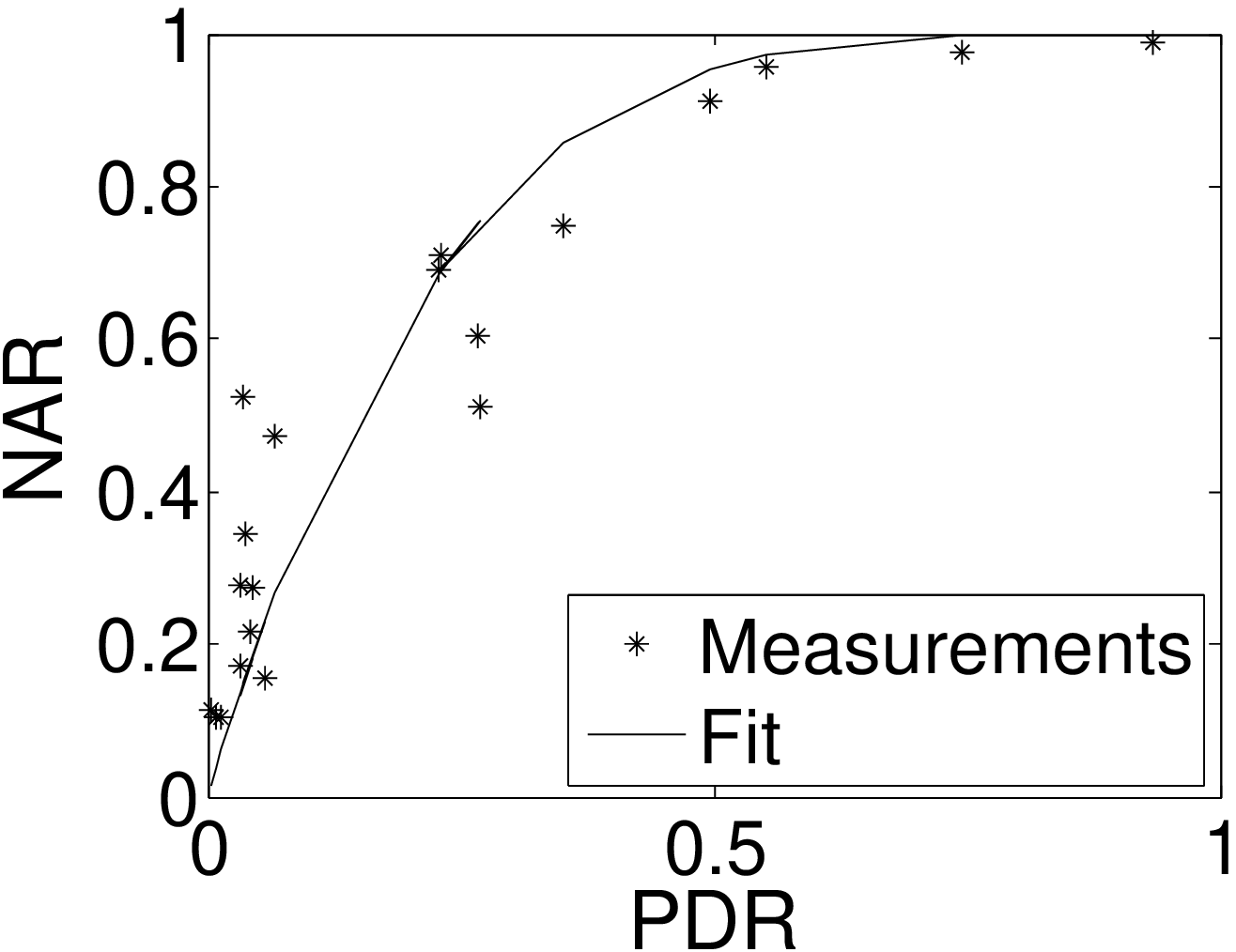}}
      \subfigure[\scriptsize Test Site Finland -- Overall. The value of $Z$ for the fit: 6.4821.]{\label{fig:NARPDRrelationshipFinland}\includegraphics[width=0.49\linewidth]{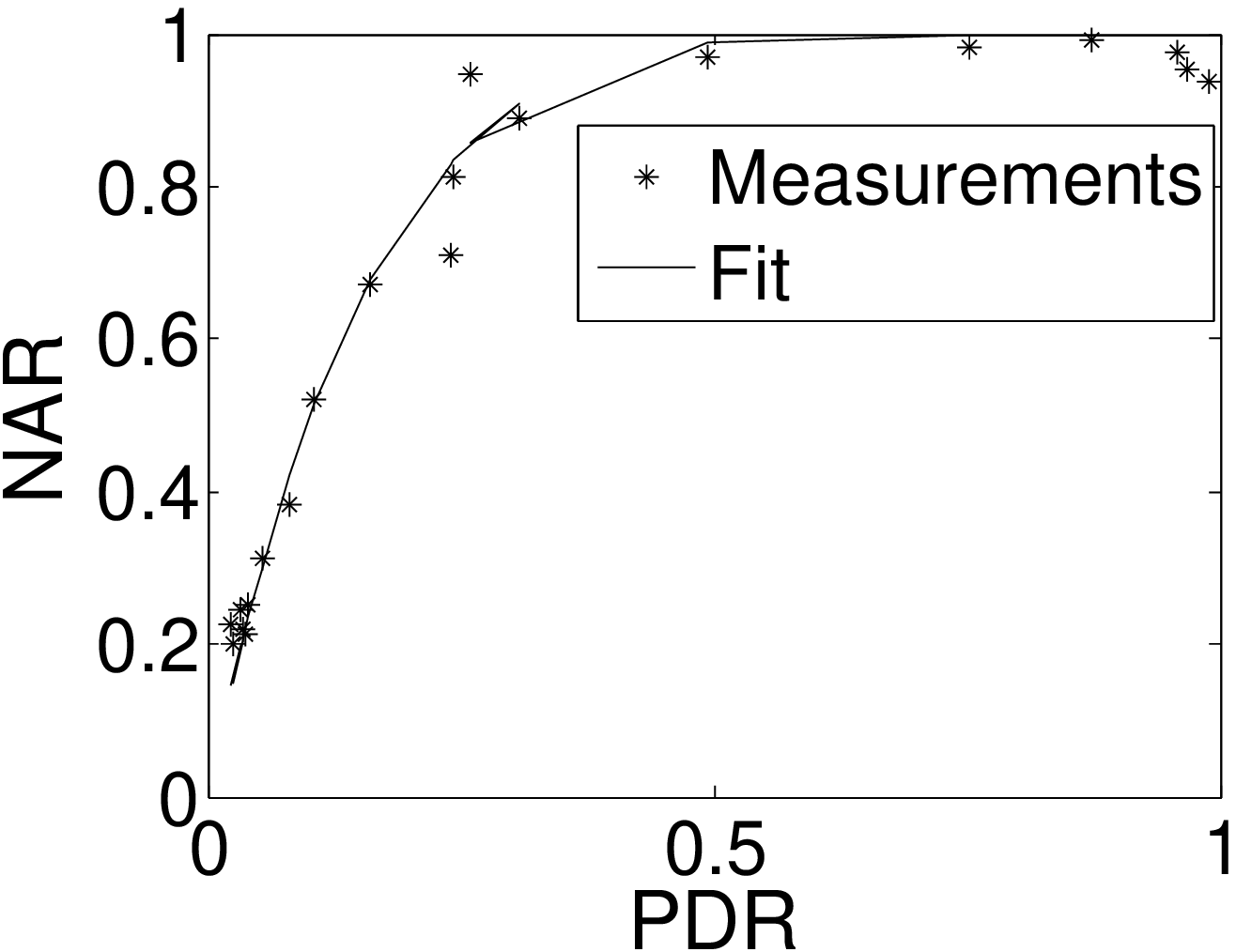}}
     \caption{Relationship between PDR and NAR. }
      \label{fig:NARPDRrelationship}
   \end{center}
\end{figure}

However, geometric distribution assumes independent trials (i.e., independent CAM transmissions). In reality, measurements have shown that there is a correlation between subsequent CAM transmissions, provided that the time between successive transmission is sufficiently small (e.g., below one second).
In other words, probability of success in time $t$ increases if the transmission in time $t-1$ was successful: 
$P(CAM_t|CAM_{t-1}) > P(CAM_t)$. Analogously, the probability of success of CAM reception decreases if the previous CAM transmission failed: 
$P(CAM_t|\overline{CAM_{t-1}})<P(CAM_t)$.
This observation was confirmed by previous measurement studies reported by Martelli et al.~\cite{martelli2012measurement} and Bai et al.~\cite{bai2006reliability}. 
This dependency affects the cooperative awareness and consequently the \ac{NAR} calculations.   
Since \ac{NAR} does not benefit from bursts of received messages (e.g., receiving one message in $t$ is equal to receiving 10 messages in $t$ in terms of \ac{NAR}), we need to implement a ``discount'' function to the cumulative geometric distribution to account for the negative effect (loss bursts) in the calculation of \ac{NAR}. This can be seen as reducing the effective number of transmissions to achieve awareness. Thus, we consider
\begin{align}
NAR_{r,t} = 1-(1-PDR_{r})^Z, 
\label{eq:Z}
\end{align}
where $Z \leq N$.

\begin{figure}
\centering
\subfigure[\scriptsize Test Site The Netherlands.]{\label{fig:NEDVAR}\includegraphics[width=0.49\linewidth]{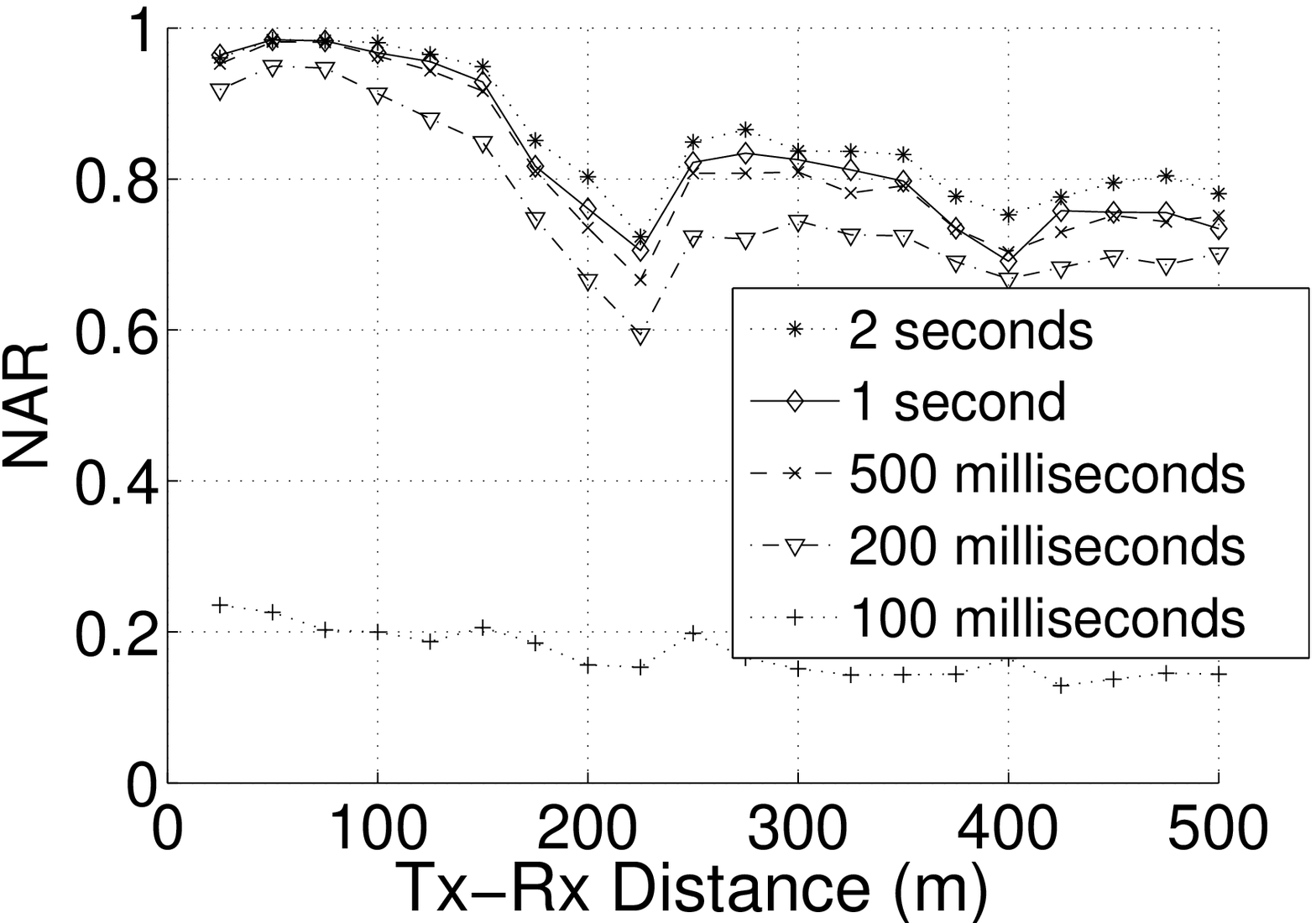}}
\subfigure[\scriptsize Test Site Italy.]{\label{fig:ITVAR}\includegraphics[width=0.49\linewidth]{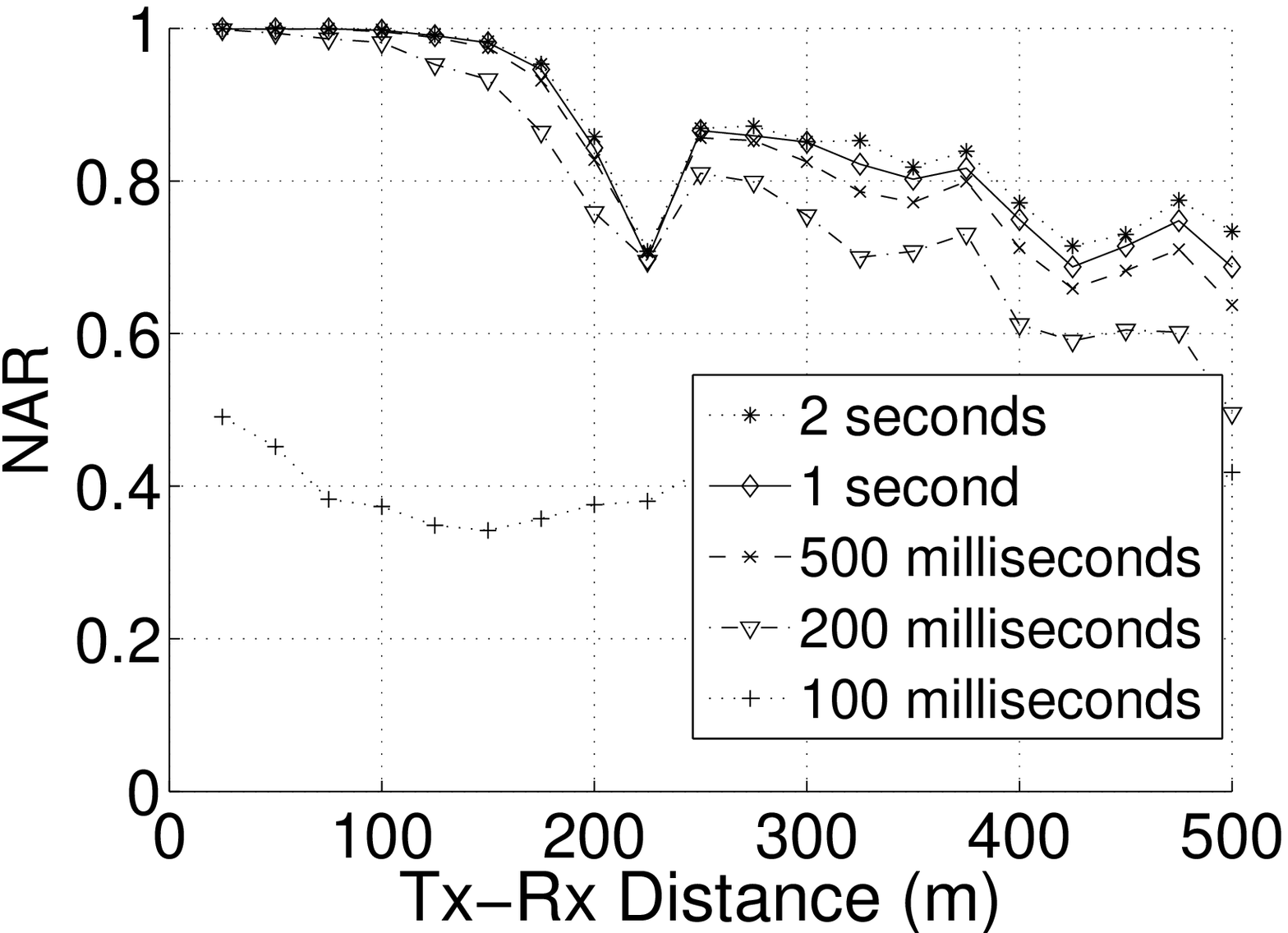}}
\caption{Behavior of \acf{NAR} with varying time period $t$. Since CAM messages were sent with 10~Hz frequency, the time periods $t$ of 100~ms, 200~ms, 500~ms, 1~s, and 2~s contain 1, 2, 5, 10, and 20 CAM transmissions, respectively.\label{fig:NARDifferentT}}
\end{figure}

To estimate $Z$ for different environments, we compared \ac{PDR} and \ac{NAR} results from measurements described in section~\ref{sec:Evaluation}. 
Using a non-linear MMSE estimator, we fit the value of $Z$ in eq.~\eqref{eq:Z} for each of the datasets. Fig.~\ref{fig:NARPDRrelationship} shows the relationship between \ac{PDR} and \ac{NAR} for two measurement test sites, whereas
Fig.~\ref{fig:NARModelAll} shows the resulting \ac{NAR} estimation using eq.~\eqref{eq:Z} and fitted $Z$ parameter, compared with the measured \ac{NAR}.  
The relatively large range of $Z$ values (between approx. 2 and 8) can be explained by analyzing Fig.~\ref{fig:NARDifferentT}, which 
gives some indication of this relationship: with the fixed CAM transmission rate of 10~Hz, the figure shows that, by increasing the number of sent messages above a certain threshold (in this case, 200~ms period equivalent to 2 messages per time period), the increase of \ac{NAR} is quite limited. Therefore, 
 if there are at least two messages in the observed time slot $t$, i.e., $Z\geq2$ in eq.~\eqref{eq:Z}, the results do not change considerably (e.g., see results for 200~ms to 2 seconds time slots in Fig.~\ref{fig:NARDifferentT}). 
 These results go in line with the conclusion that the CAM transmissions succeed (and fail) in bursts due to the communication being dominated by shadowing; if there are at least two messages sent in a time period $t$, sending additional messages results in little benefit (two as opposed to one, in order to counter: i) no messages reaching the receiver on time due to queuing or processing delay at either the transmitter or receiver; and ii) sudden message loss due to small-scale fading). We further explore this topic through simulations in Section~\ref{sec:Simulation}.

In terms of the accuracy of the model, Fig.~\ref{fig:NARModelAll} shows that the model matches the measurements better when the data is separated according to environments (e.g., urban, suburban, highway). The main reason is that the behavior of \ac{PDR} for separate environments is less variable than when \ac{PDR} results are combined (see, for example, Fig.~\ref{fig:PDRV2VOverall} and related figures). Figs.~\ref{fig:NARMFinHw}-\ref{fig:NARMI}
 show that \ac{NAR} generated by the model is very close to measured values. On the other hand, results for combined environments (e.g., Fig.~\ref{fig:NARMS} and~\ref{fig:NARMN}), the estimate is not as accurate, particularly at distances larger than 200~m. The main reason for this is that 
combining results from different environments increases the variation of \ac{PDR} used for calculating \ac{NAR} in eq.~\eqref{eq:Z}, particularly
at larger distances. Furthermore, the range of values for $Z$ is relatively large (2-8); this confirms the results shown in Fig.~\ref{fig:NARDifferentT}: while a single message sent in a time period is not sufficient, the difference between having two or more transmitted messages per time period $t$ is comparatively small. 

\begin{figure}
\centering
\includegraphics[width=0.65\linewidth]{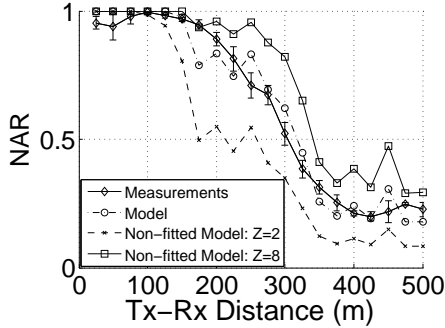}
\caption{\ac{NAR} results for Test Site Finland: measurements, results from model generated using best-fit $Z$ value (4.2768), $Z$=2, and $Z$=8.}
\label{fig:NARBounds}
\end{figure}

A practical application of the model is providing upper and lower bound for awareness. For instance, for Test Site Finland, 
 Fig.~\ref{fig:NARBounds} shows the fitted ($Z=4.2768$) and non-fitted model results for $Z$ equal to 2 and 8. The curves for non-fitted model encompass the measured \ac{NAR} curve, while not being overly wide to render them obvious. Therefore, when actual measurements of \ac{NAR} are not available to fit the parameter $Z$, the theoretical model can be used to give a relatively confident range of \ac{NAR} values based on \ac{PDR} measurements only.

To conclude, the simple model we developed in this section can estimate \ac{NAR} by knowing \ac{PDR} behavior over distance for a given environment. While it is not able to account for all the effects that impact cooperative awareness, the model can give an insight into the behavior of awareness in an environment for which \ac{PDR} is available.

%% file: simulation.tex
The measurement results presented in Section~\ref{sec:Evaluation} provide valuable insights into the performance of cooperative awareness under realistic conditions, by considering different environments, V2V and V2I communication, different vehicle setups (antenna, effective power), etc. 
However, the measurements analyzed in Section~\ref{sec:Evaluation} 
are limited in scale (e.g., the number of communicating nodes is below 10 on all test sites) and scope (e.g., CAM transmit rate and power were fixed). 

In this section, we resort to realistic simulations to study the achievable performance of cooperative awareness by varying the transmit rate and transit power of CAM messages in scenarios containing thousands of vehicles in different environments. The main questions we aim to answer in this section are: 1) to increase awareness in a given environment, is it better to transmit more CAM messages at lower power or fewer messages at higher power? 2) how many CAM messages do we need to transmit before gains are diminished? 3) given realistic transmit power limitations, what is the largest distance at which high levels of awareness can be achieved for a specific environment? 4) for the same transmit power and rate settings, how significant are the differences between urban and highway environments?

\subsection{Simulation Platform}
Measurements described in Section~\ref{sec:Evaluation} showed that \ac{PDR}, \ac{NAR}, and \ac{RNAR} are highly dependent on the propagation environment where \ac{V2V} communication occurs. Therefore, simulating cooperative awareness requires a simulation tool that is able to represent distinct propagation environments (e.g., urban intersection, rural highway, urban canyon). For that reason, we used \acf{GEMV2}, a freely available \ac{V2V} propagation model and simulation framework (see~\cite{boban14TVT}) to perform a realistic assessment of cooperative awareness on a large-scale. 
\ac{GEMV2} is an efficient geometry-based propagation model for \ac{V2V} communications, 
which explicitly accounts for surrounding objects (buildings, foliage and other vehicles). The model considers three \ac{V2V} links categories, depending on the \ac{LOS} conditions between transmitter and receiver, to deterministically calculate large-scale signal variations (i.e., path-loss and shadowing):
\begin{itemize}
\item \acf{LOS}: links that have an obstructed optical path between the transmitting and receiving antennas;
\item \acf{NLOSv}: links whose \ac{LOS} is obstructed by other vehicles;
\item \acf{NLOSb}: links whose \ac{LOS} is obstructed by buildings or foliage.
\end{itemize}
Additionally, \ac{GEMV2} determines small-scale signal variations using a simple geometry-based stochastic model
that takes into account the the number and size of surrounding objects. 
The simulator allows importing realistic mobility data from \ac{SUMO}~\cite{SUMO2012} and building/foliage outlines from OpenStreepMap (\url{http://www.openstreetmap.org/}). 

\begin{figure}[!t]
    \includegraphics[width=0.98\columnwidth]{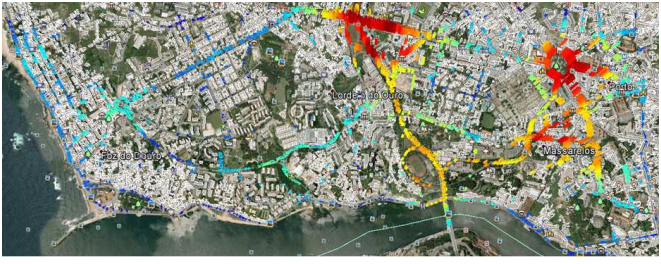}
      \caption{Urban environment (Porto, Portugal) used for simulations. The map overlay shows the number of directly reachable vehicles (``neighbors'') for each vehicle, as generated by the \ac{GEMV2} model. For each vehicle, the colored bar represents the number of neighbors. Warmer and taller bar colors indicate more neighbors.      
      }
      \label{fig:Neighborhood}
\end{figure}

\subsection{Simulated environments}\label{sec:simEnvironments}
To evaluate the behavior of cooperative awareness in different environments, we performed simulations in \ac{GEMV2} using roadways and geographic data from highway and urban locations in and near the city of Porto (i.e., the same locations as those in the measurements reported in~\cite{boban14TVT}) 
\footnote{Received power level measurements were not available for the Test Sites of the empirical evaluation study described in Section~\ref{subsec:Scenarios}.}.
Received power distributions for cooperative messages was based on the measurements in the same location (as explained in Section.~\ref{subsec:CAMPower}).
Specifically, we simulated two distinct environments: 
\begin{itemize}
\item Urban environment, a core part of the city of Porto containing 2410 vehicles delimited by a rectangle with the following coordinates: (41.1426,-8.6850),(41.1624,-8.6203). The area is shown in Fig.~\ref{fig:Neighborhood} and contains 2410 vehicles. 
\item Highway environment, comprising a 12.5~km stretch of A28 Highway with approximate center coordinates at (41.2327, -8.6954) and containing 404 vehicles.
\end{itemize} 
To ensure credible locations of vehicles, we used vehicle locations collected through aerial photography (details on the datasets are available in Ferreira et al.~\cite{ferreira09}).

\subsection{\ac{CAM} message received power and inter reception times}\label{subsec:CAMPower}
To generate realistic small-scale signal variation and inter reception times for cooperative messages,
we used the data collected during \ac{V2V} measurements. Specifically, we used the standard deviation of received power for cooperative messages collected in the study by Boban et al.~\cite{boban14TVT}. The measurements were made using ITS-G5 compliant radios (NEC LinkBird MX) operating in the 5.9 GHz frequency band. The messages were generated at 10~Hz frequency and exchanged by passenger cars traveling in Porto (urban environment) and on surrounding highways.

\begin{figure}[!t]
\centering
    \subfigure[\scriptsize Highway Porto.]{\includegraphics[width=0.48\linewidth]{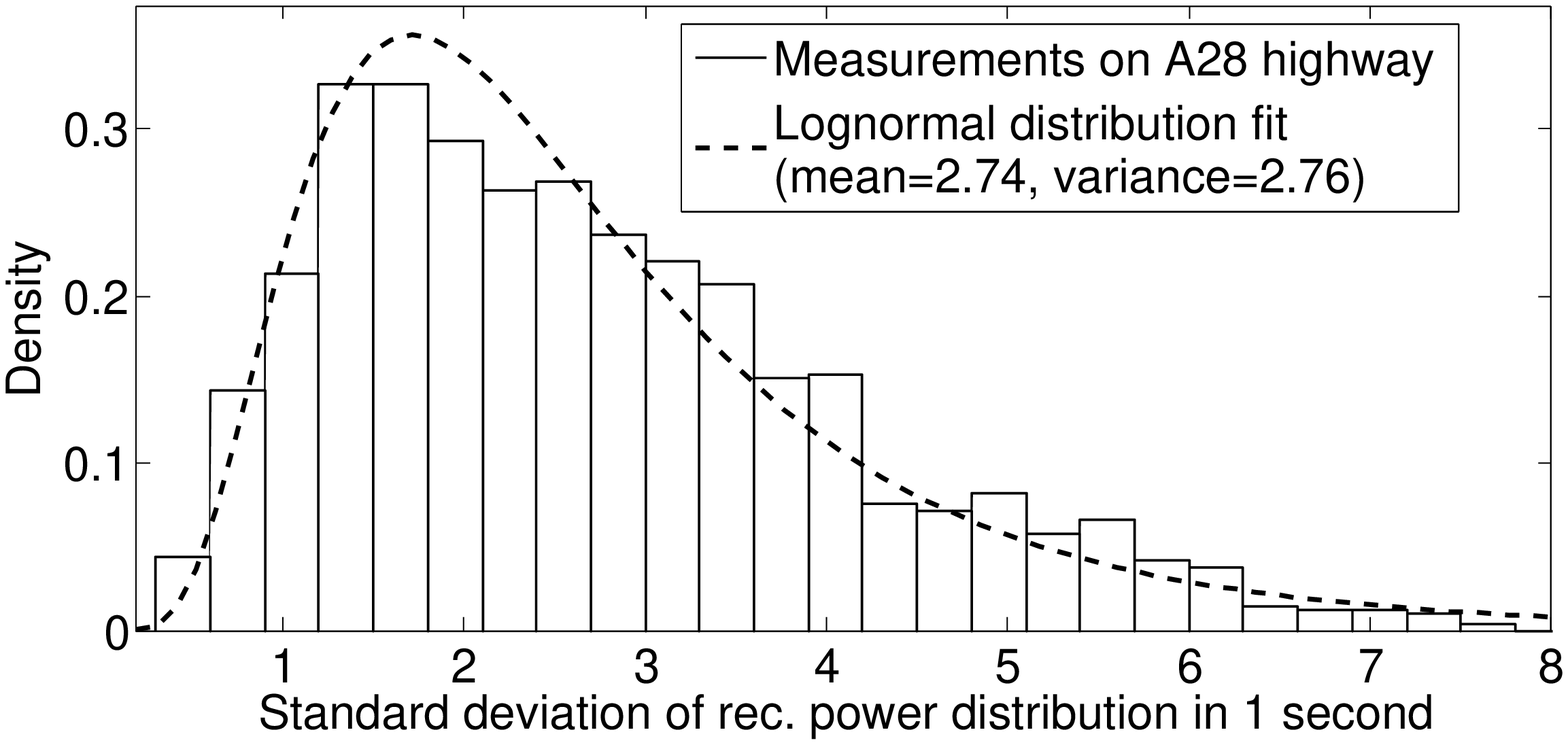}}\label{fig:fitPortoHighway}
        \subfigure[\scriptsize Urban Porto.]{\includegraphics[width=0.48\linewidth]{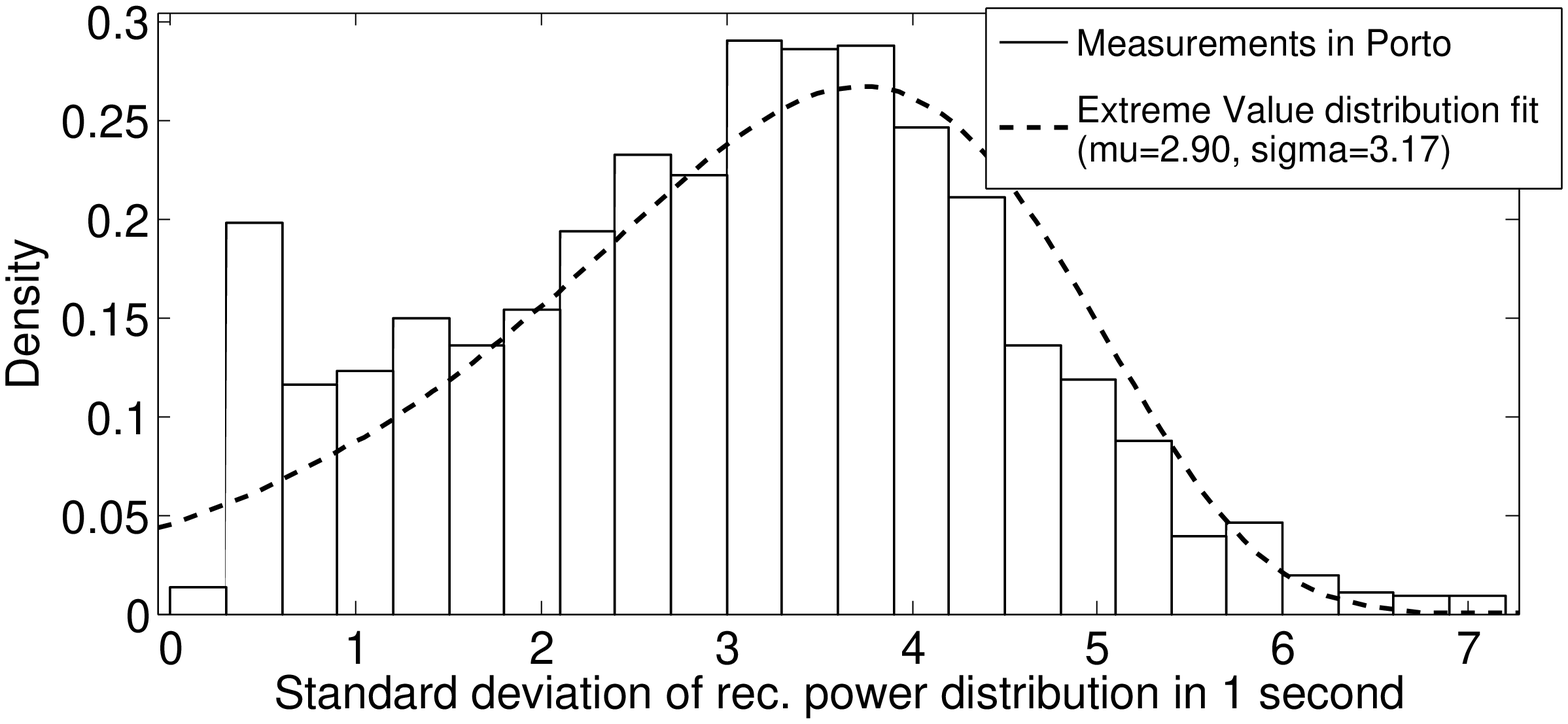}}\label{fig:fitPortoUrban}
     \caption{Std. dev. of 10 Hz CAM distribution.}
      \label{fig:10HzCAMDistribution}
\end{figure}

Since message inter-reception times are directly dependent on the successful packet decoding at the receiver,
we used the received power variation measured in V2V experiments to generate realistically simulated inter-reception times. 
Specifically, for the two environments (urban and highway), we divided the measurement data 
into one second bins and calculated the standard deviation of received power for each bin. 
We excluded the bins with \ac{PDR} below 50\% to ensure a minimum of five messages per bin.
We fit the measured standard deviation across the entire measurement dataset to the theoretical distribution functions available in MATLAB Distribution Fitting tool.
The measurement data and the corresponding fits are shown in Fig.~\ref{fig:10HzCAMDistribution}. Finally, we modified the small-scale signal variation model in \ac{GEMV2} so that it draws a random number from the corresponding best-fit theoretical distributions (Fig.~\ref{fig:10HzCAMDistribution}), representing the standard deviation of received power for each one-second bin. The generated small-scale variation is then added on top of the received power calculated by large-scale signal variation model.

\begin{figure}[!t]
\centering
    \subfigure[\scriptsize Test Site Finland -- Highway.]{\includegraphics[width=0.4\linewidth]{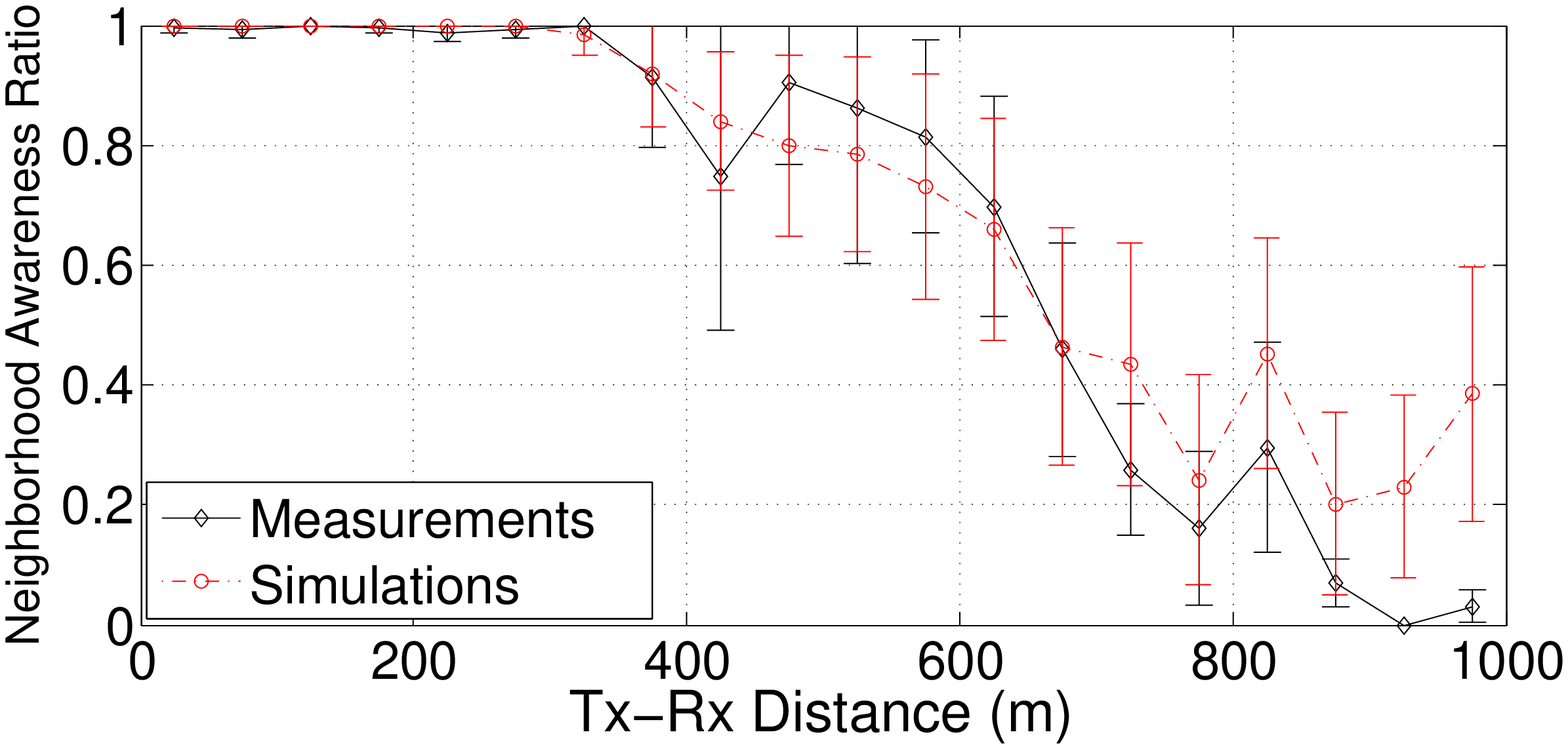}}
	\subfigure[\scriptsize Test Site Finland -- Urban.]{\includegraphics[width=0.4\linewidth]{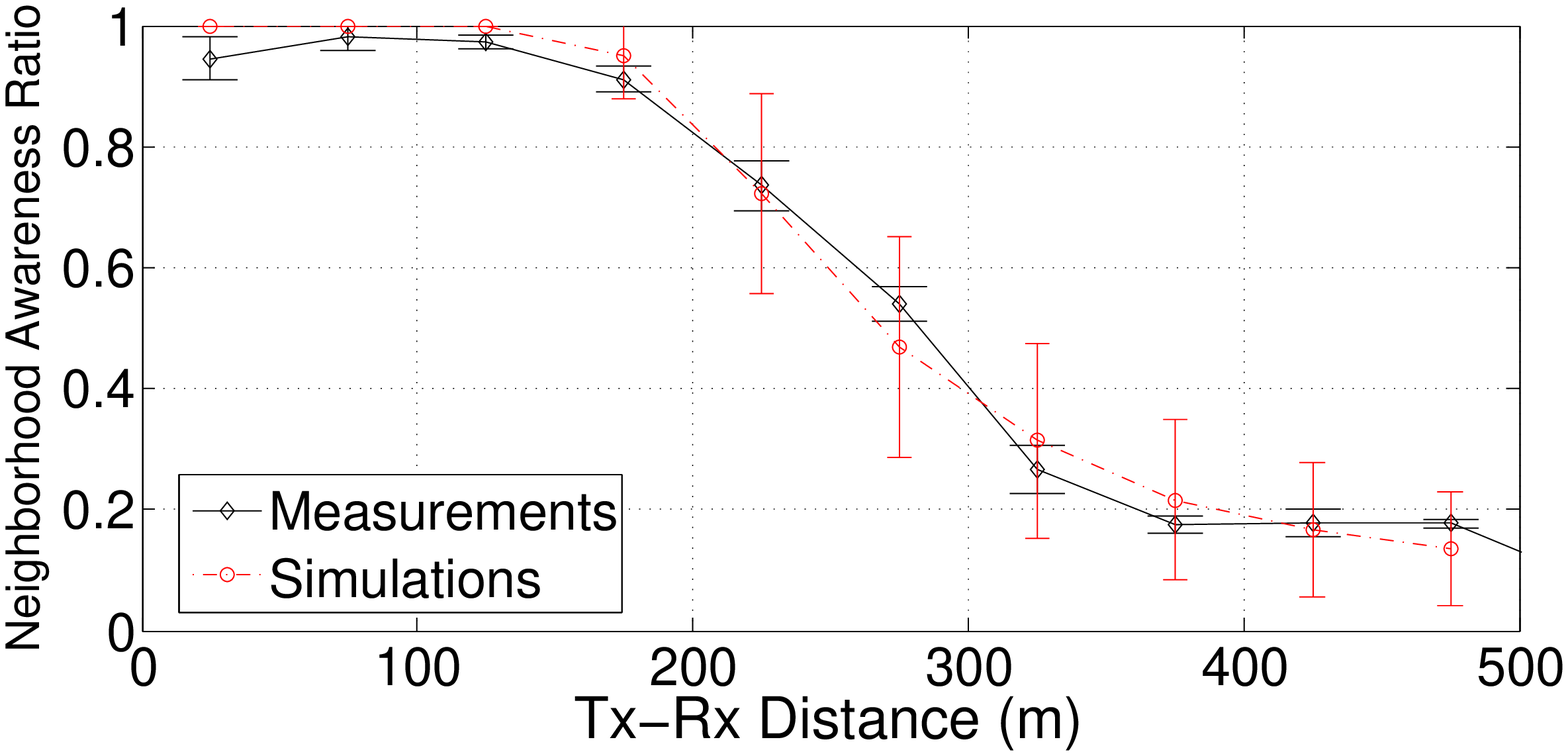}}
     \caption{Comparison of \ac{NAR} results: measurements vs. simulations. The error bars represent one standard deviation around the mean.}
      \label{fig:comparison}
\end{figure}

\subsection{Validation of simulation results against measurements}\label{sec:simValidation}
Before performing the large-scale simulations, we wanted to validate our simulation model against the small-scale measurements.
For that purpose, we used comparable simulated environment (including area size and road layout), effective transmit power, vehicle types, and cooperative message generation rates to those where measurements were performed in Test Site Finland. Fig.~\ref{fig:comparison} shows the results of the comparison in terms of \ac{NAR}. The results for both highway and urban comparison are quite similar. The larger standard deviation for simulated urban environment can be explained by simulated vehicles taking more diverse routes (the number of simulated vehicles was 2000 compared to 3 vehicles used for measurements), thus experiencing a larger number of distinct propagation environments. The reason for the discrepancy above 800 meters in highway environment is that the simulated location, while similar to the Test Site Finland  Highway, is not completely identical to it.

\subsection{Simulation Results \& Discussion}
In this section, we study the behavior of \ac{NAR} by varying the transmit power and transmit rate of cooperative messages in urban and highway environments. We note that, in simulations, we do not consider interference generated by CAM exchange; therefore, the results in this section are an upper bound of awareness performance for the given transmit power and transmit rate. Also, note that the simulation results were generated with the assumption of -95~dBm receiver sensitivity threshold, in line with the sensitivity of devices used for DRIVE C2X measurements; given different receiver sensitivity thresholds of radios, the results we show in this section would be equivalent to changing the transmit power level by the same amount (i.e., \ac{NAR} would be increased by the same amount by increasing sensitivity by 1~dB or by increasing transmit power by 1~dBs).

\begin{figure*}[!tb]
\centering
    \subfigure[\scriptsize 5~dBm.]{\includegraphics[width=0.29\linewidth]{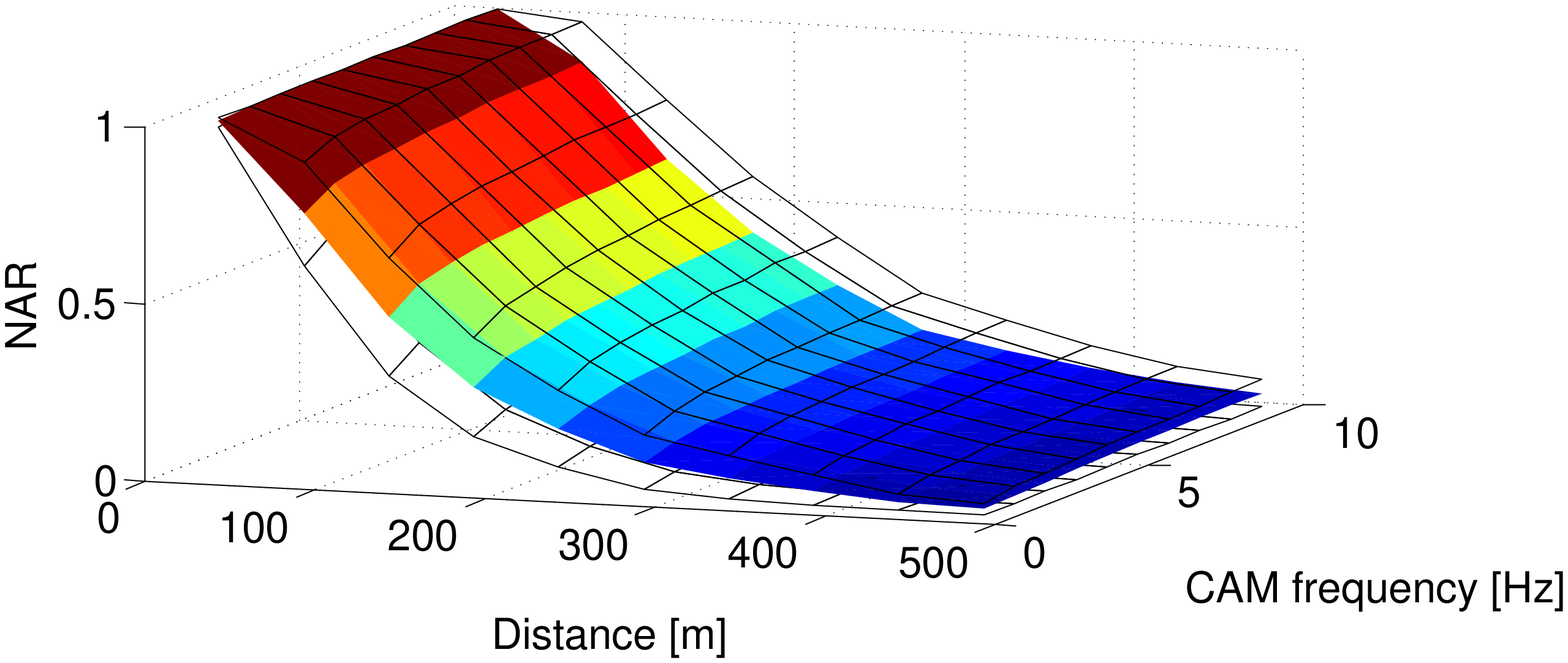}\label{subfig:CAM5dBm}}
    \subfigure[\scriptsize 15~dBm.]{\includegraphics[width=0.29\linewidth]{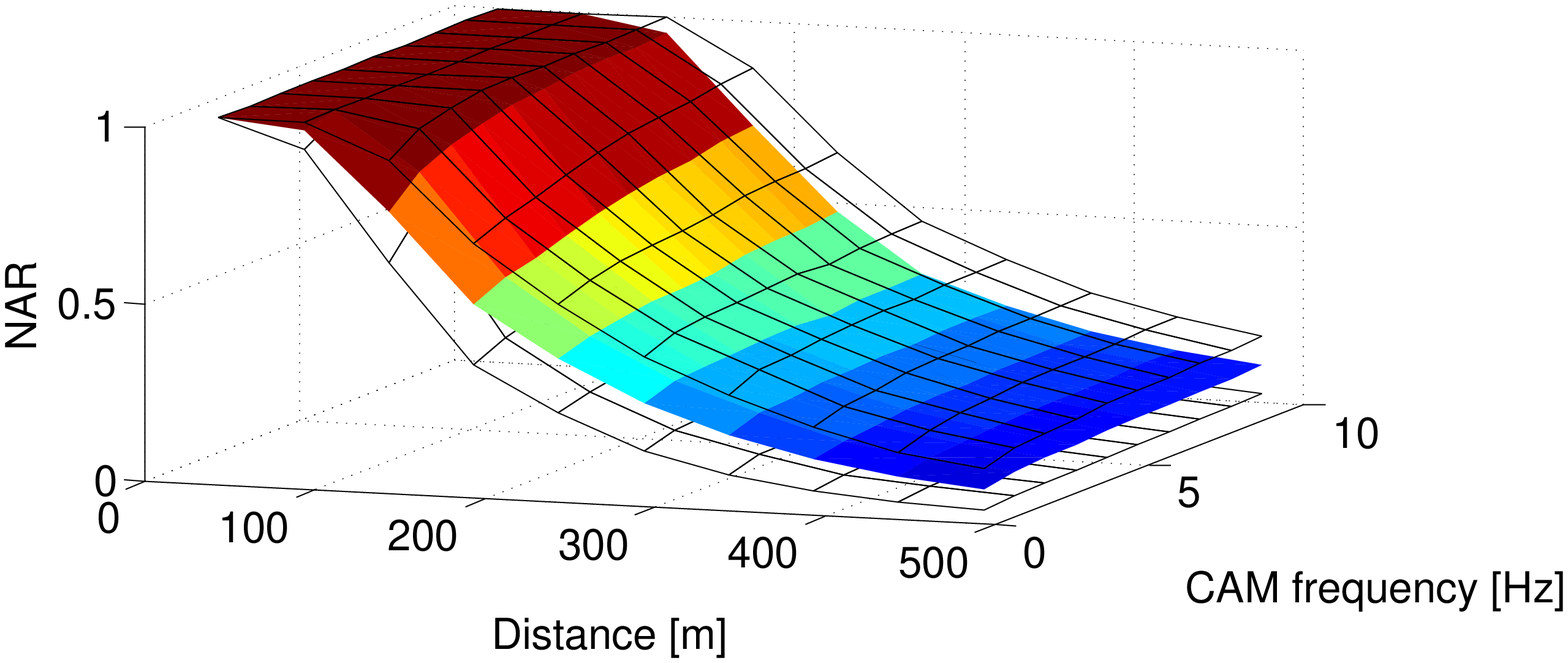}\label{subfig:CAM15dBm}}
        \subfigure[\scriptsize 23~dBm.]{\includegraphics[width=0.29\linewidth]{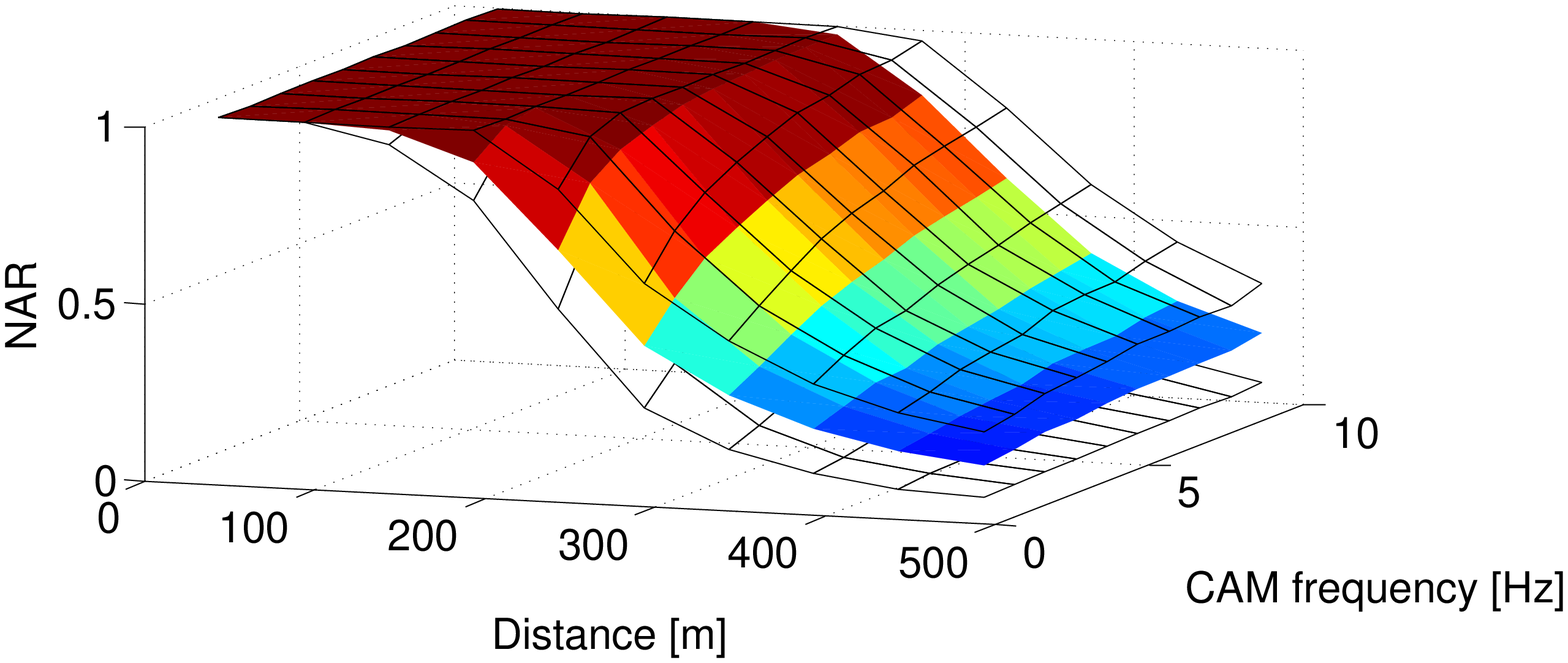}\label{subfig:CAM23dBm}}
     \caption{\ac{NAR} simulations in urban environment: CAM rate varied, Tx power fixed to 5, 15, and 23~dBm. Mean value is represented by the surface; standard deviation is represented by the black grid.}
      \label{fig:CAMRateUrban}
\end{figure*}

\begin{figure*}[!tb]
\centering
    \subfigure[\scriptsize 1~Hz.]{\includegraphics[width=0.29\linewidth]{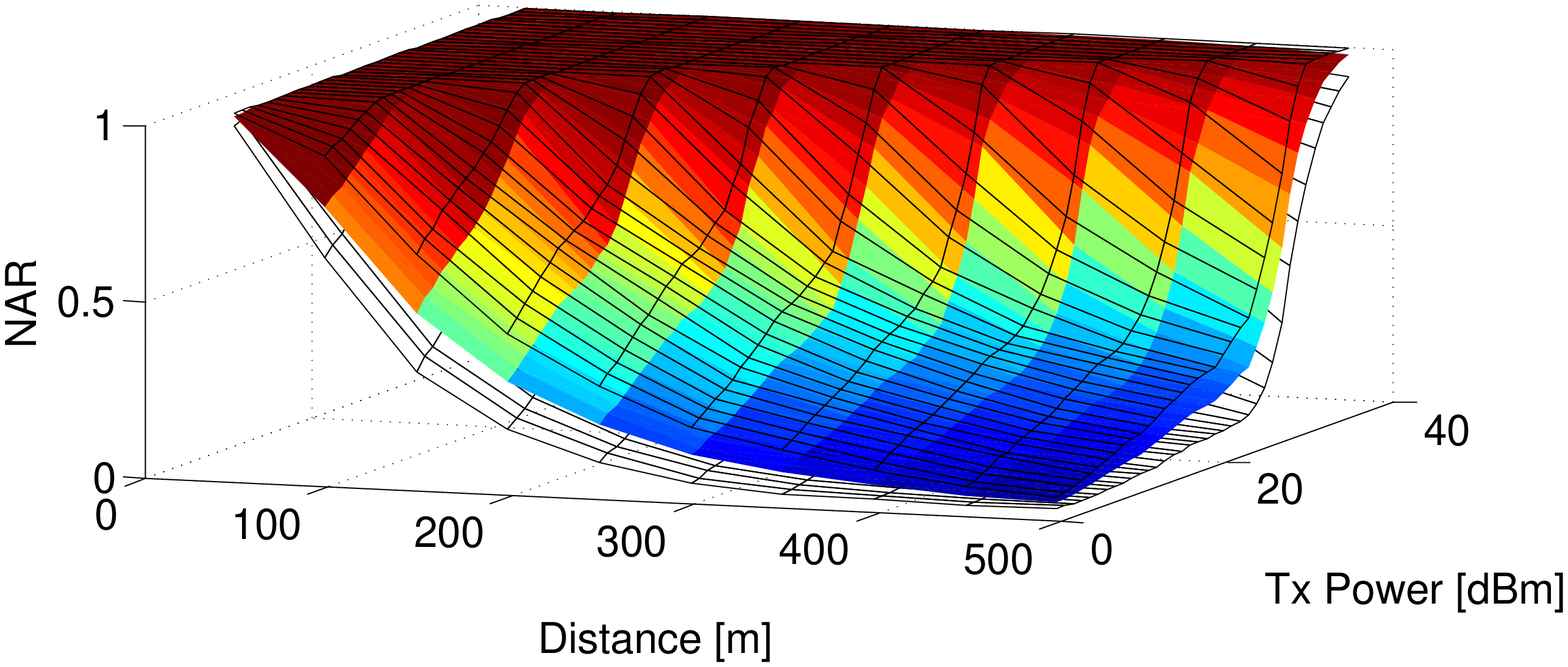}\label{fig:TxPwrUrban1Hz}}
	\subfigure[\scriptsize 5~Hz.]{\includegraphics[width=0.29\linewidth]{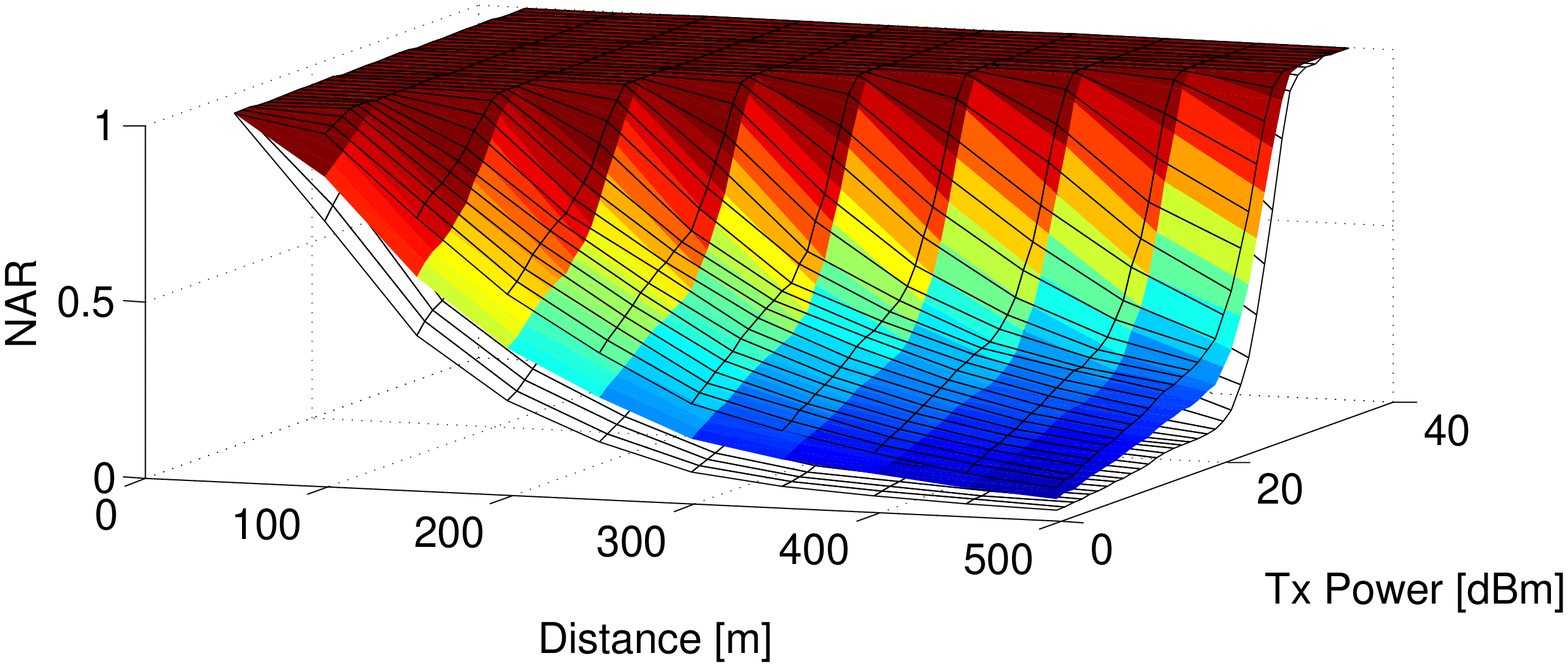}\label{fig:TxPwrUrban5Hz}}
    \subfigure[\scriptsize 10~Hz.]{\includegraphics[width=0.29\linewidth]{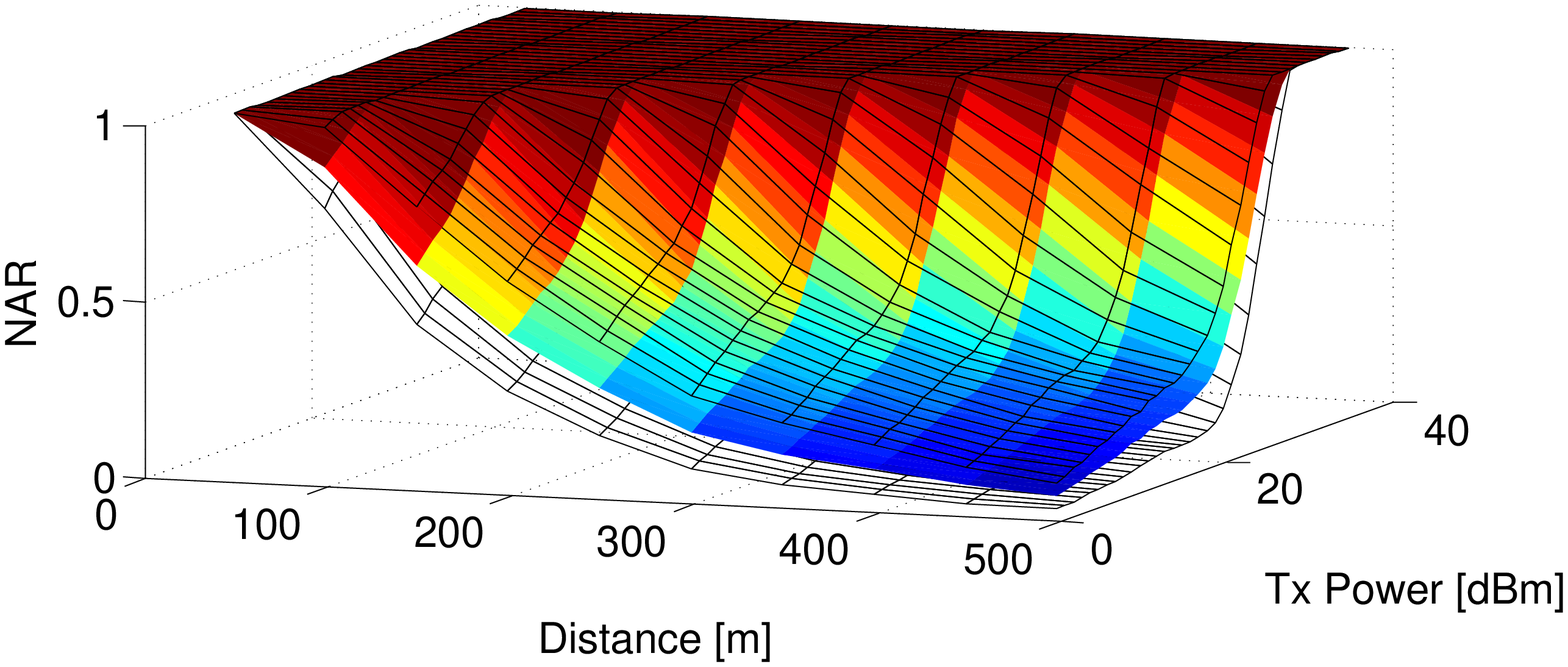}\label{fig:TxPwrUrban10Hz}}	
     \caption{\ac{NAR} simulations in urban environment: Tx power varied, CAM rate fixed to 1, 5, and 10~Hz. Mean value is represented by the surface; standard deviation is represented by the black grid.}
      \label{fig:TxPwrUrban}
\end{figure*}

\subsubsection{Urban Environment}

Figs.~\ref{fig:CAMRateUrban} and~\ref{fig:TxPwrUrban} show \ac{NAR} as a function of CAM transmit rate and transmit power, respectively, over distance. Fig.~\ref{subfig:CAM5dBm} shows that, at 5~dbm effective transmit power, awareness above 90\% can be achieved only up to 50~m, irrespective of the CAM rate. For 15~dBm and 23~dBm, 90\% awareness is achievable at 200~m and 300~m, respectively (Fig.~\ref{subfig:CAM15dBm},~\ref{subfig:CAM23dBm}). 
Similar to what we observed in measurements (Fig.~\ref{fig:NARDifferentT}), \ac{NAR} increase can be noticed when going from the rate of one CAM per time period to two and three; increasing the rate further results in minimal benefits. On the other hand, increasing power has a direct influence on the awareness. Fig.~\ref{fig:TxPwrUrban} shows how \ac{NAR} increases with power; it is interesting to see that, for each distance, there is a relatively narrow range of transmit powers at which the awareness ``transition'' occurs, i.e., where \ac{NAR} increases rapidly with each 1~dB power increase. As noted before, increasing the CAM frequency has little effect on the modification of the transition zone.

\begin{figure*}[!t]
\centering
    \subfigure[\scriptsize 5~dBm.]{\includegraphics[width=0.29\linewidth]{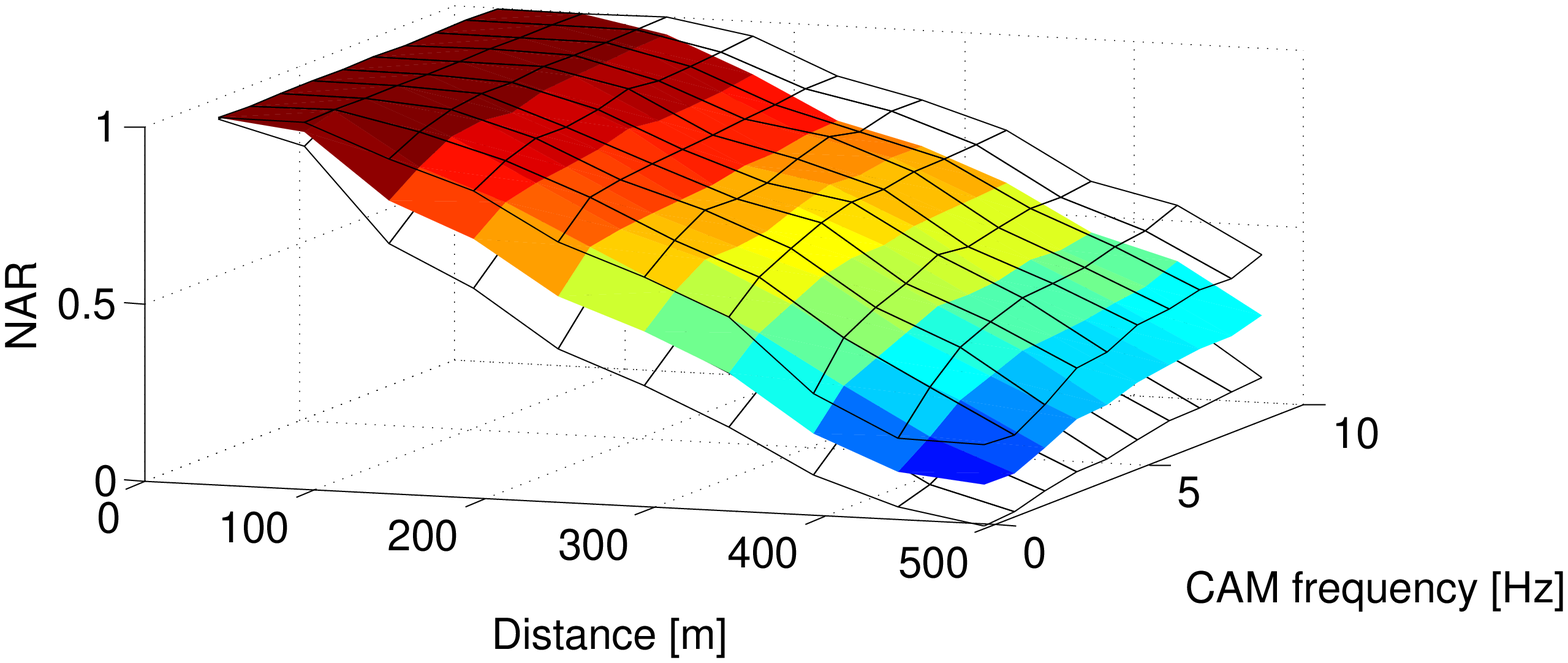}}
    \subfigure[\scriptsize 15~dBm.]{\includegraphics[width=0.29\linewidth]{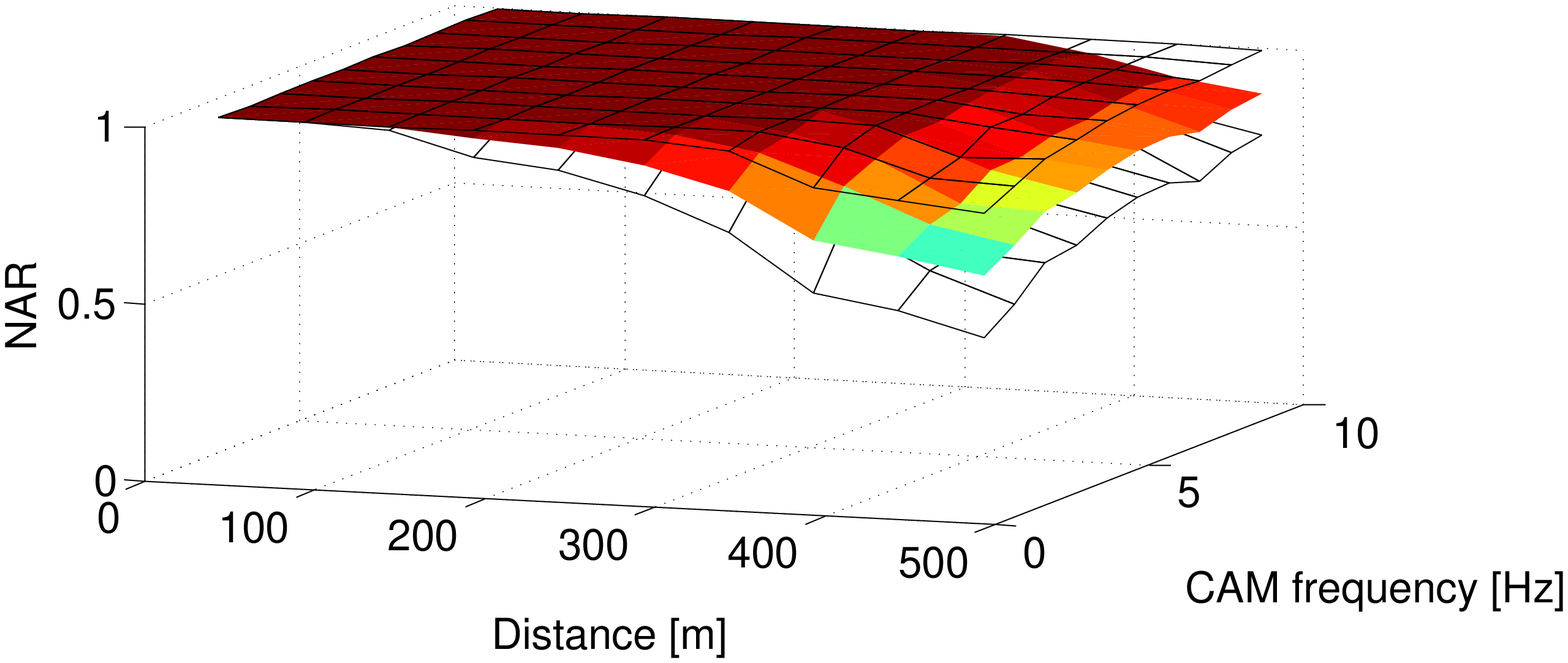}}
        \subfigure[\scriptsize 23~dBm.]{\includegraphics[width=0.29\linewidth]{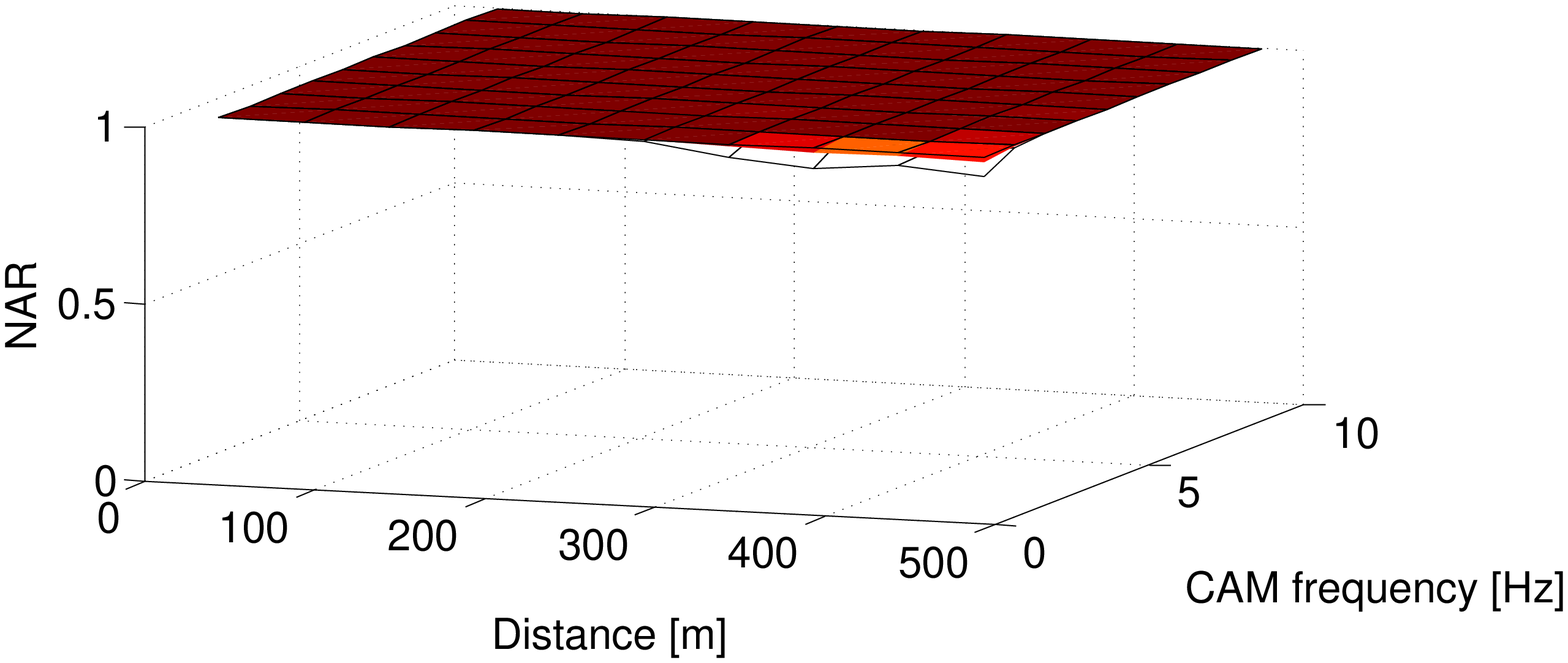}}
     \caption{\ac{NAR} simulations in highway environment: CAM rate varied, Tx power fixed to 5, 15, and 23~dBm. Mean value is represented by the surface; standard deviation is represented by the black grid.}
      \label{fig:CAMRateHighway}
\end{figure*}

\begin{figure*}[!t]
\centering
    \subfigure[\scriptsize 1~Hz.]{\includegraphics[width=0.29\linewidth]{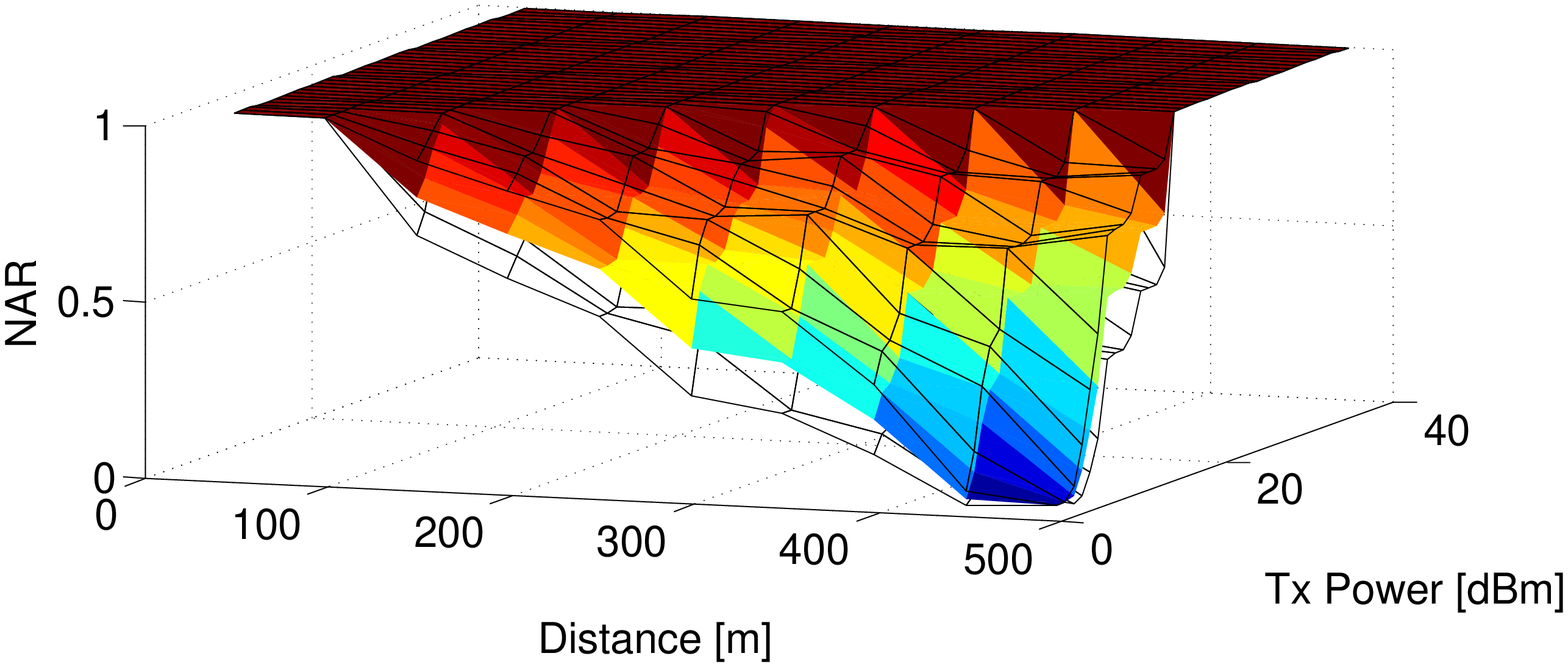}  \label{fig:TxPwrHighway1Hz}}
	\subfigure[\scriptsize 5~Hz.]{\includegraphics[width=0.29\linewidth]{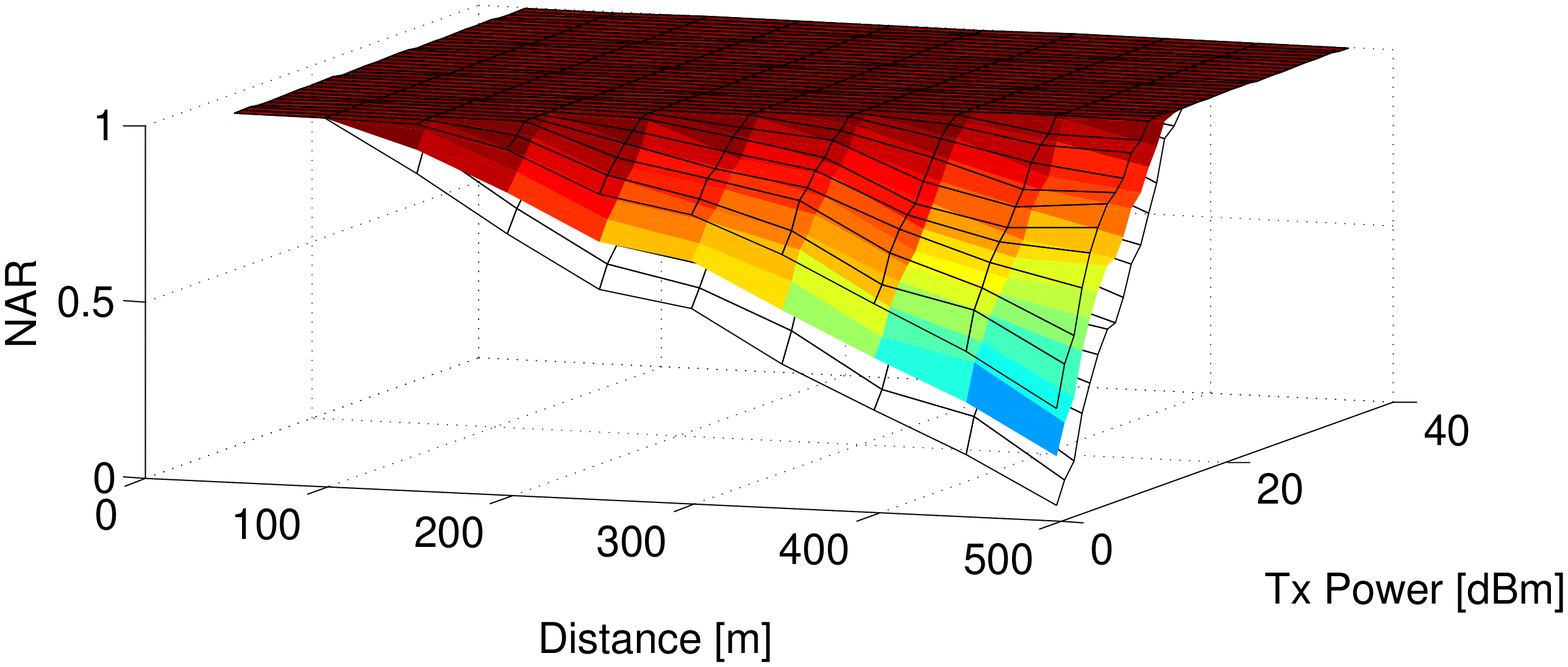}  \label{fig:TxPwrHighway5Hz}}
    \subfigure[\scriptsize 10~Hz.]{\includegraphics[width=0.29\linewidth]{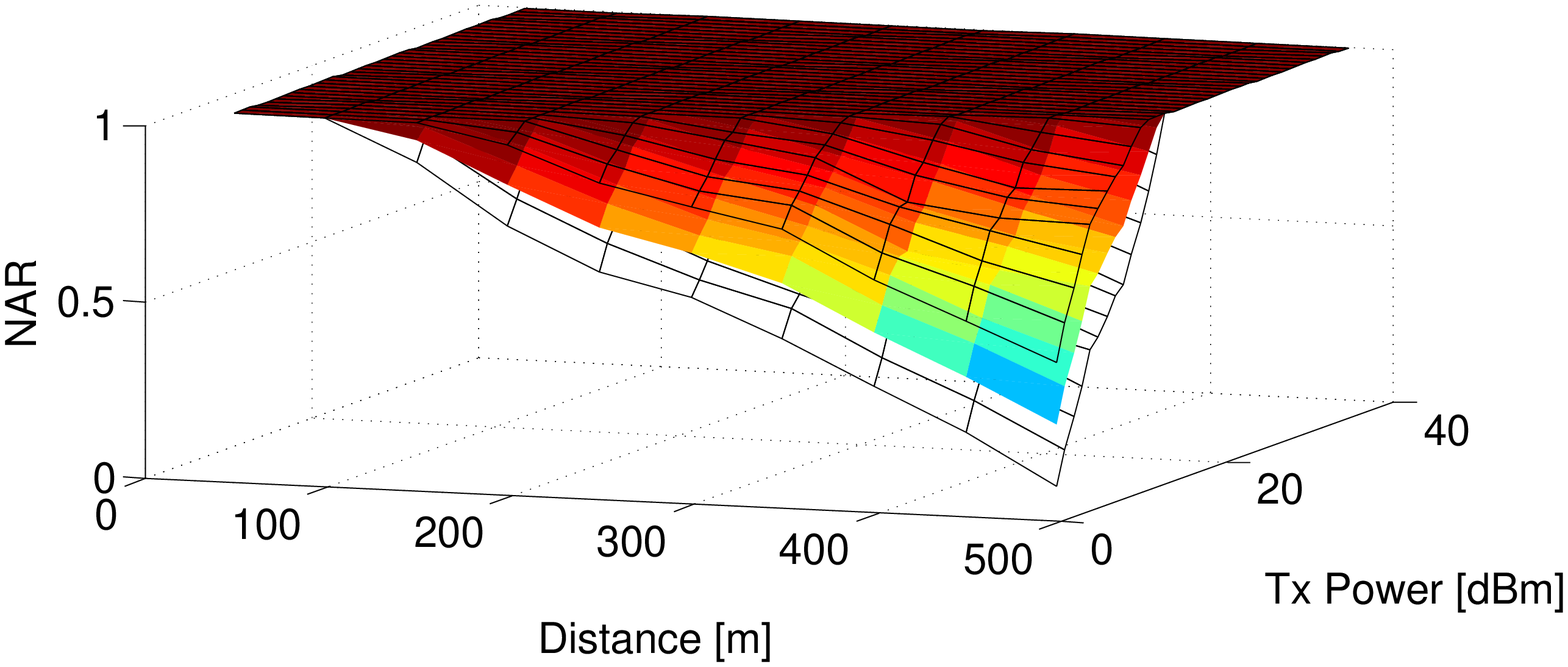} \label{fig:TxPwrHighway10Hz}}	
     \caption{\ac{NAR} simulations in highway environment: Tx power varied, CAM rate fixed to 1, 5, and 10~Hz. Mean value is represented by the surface; standard deviation is represented by the black grid.}
      \label{fig:TxPwrHighway}
\end{figure*}

\subsubsection{Highway Environment}
In highway environment, compared to urban, we observe a notably higher \ac{NAR} for the same distance and CAM transmit power and rate\footnote{Somewhat higher variability in of \ac{NAR} in highway is a result of a smaller number of simulated vehicles compared to urban (404 vs. 2410).}. While this is to be expected, it is interesting to note that the transmit power and rate required to reach 90\% \ac{NAR} at 400~m is approximately 20~dB and two CAM transmissions per period (Fig.~\ref{fig:TxPwrHighway}). This shows that production-ready DSRC radios, often limited to 23~dBm EIRP, have the ability to provide high awareness in highways.
To achieve the same performance in urban would require over 33~dB EIRP (Fig.~\ref{fig:TxPwrUrban}), which is not allowed according to the current transmit power limits in the US and EU~\cite{etsi202663}. With realistic limits in mind, our results show that high awareness (above 90\%) for urban environment can be achieved up to 250~m. 
Furthermore, similar to results for urban, the transition from low (sub-20\%) to high (above 90\%) \ac{NAR} on highway requires a limited range of transmit power values (8-10~dB difference in Fig.~\ref{fig:TxPwrHighway} compared to 5-7~dB in Fig.~\ref{fig:TxPwrUrban}), albeit at  
different absolute transmit powers (up to 15~dB in highway, compared to 35~dB in urban). 
These results indicate that, for each location, there is a specific transmit power level that is sufficient for achieving high awareness for a given distance. However, since each location has a distinct propagation pattern (compare, for example, Fig.~\ref{fig:TxPwrUrban} and Fig.~\ref{fig:TxPwrHighway}), determining the correct power for a given environment requires adaptive power control algorithms (e.g.,~\cite{Tielert2013JPC,aygun2015ecpr}). 

In both environments, there is virtually no difference in terms of \ac{NAR} for 5~Hz and 10~Hz CAM rate (Figs.~\ref{fig:TxPwrUrban5Hz} and~\ref{fig:TxPwrUrban5Hz} for urban, Figs.~\ref{fig:TxPwrHighway5Hz} and~\ref{fig:TxPwrHighway10Hz} for highway); this agrees with the measurement results shown in Fig.~\ref{fig:NARDifferentT}. Therefore, we conclude that transmitting more than approximately two to four messages per time period 
does not result in improved awareness, while at the same time increasing the 
channel load\footnote{Note that the message rate required to reach a certain \ac{NAR} is not tied to a time period of specific duration, but to the number of messages per time period $t$ (eq.~\ref{eq:Z}): if, for example, $t$ is one second (as is the case in our measurements and simulations), then the rate is considered per one second; if an application requires awareness within 100~ms, the rate should be considered per 100~ms. This relationship holds for sufficiently small time periods (e.g., up to a few seconds).}. 

In our simulations, we did not observe any significant impact of vehicle density on \ac{NAR} for a given distance. This is in line with the assumptions that the model makes in Section~\ref{sec:Model}, as well as the comparison between simulations and measurements (Fig.~\ref{fig:comparison}), where \ac{NAR} within a given distance range did not depend on the number of the vehicles; on the other hand, \ac{NAR} does depend on the environment and distance between vehicles. However, we should highlight that we do not account for interference in our simulations; 
when considering interference between nodes, we expect vehicle density to have a considerable impact on the awareness level at high channel load values.